\documentclass{jfm}

\usepackage{amsmath,amssymb}
\usepackage{nccmath}
\usepackage{mathrsfs}
\usepackage{xcolor}
\usepackage{subfigure}
\usepackage{bbm}
\usepackage{bm}
\usepackage{pict2e} 
\usepackage{booktabs}
\usepackage{tabularx}
\usepackage{dcolumn}
\usepackage{setspace}
\usepackage{mathtools}
\usepackage{tikz}
\usepackage{textcomp}

\usepackage[colorlinks=true, linkcolor=blue, urlcolor=blue, citecolor=blue]{hyperref}
\usepackage{natbib}
\usepackage{soul}

\usepackage{multirow,bigdelim}
\usepackage{makecell}

\usepackage[absolute]{textpos}

\newcounter{subeq}
\renewcommand{\thesubeq}{\theequation\alph{subeq}}
\newcommand{\newsubeqblock}{\setcounter{subeq}{0}\refstepcounter{equation}}
\newcommand{\subeqn}{\refstepcounter{subeq}\tag{\thesubeq}}

\newcolumntype{d}[1]{D{.}{.}{#1}}

\title{Rarefied gas flow past a liquid droplet: Interplay between internal and external flows}

\shorttitle{Rarefied gas flow past a liquid droplet} 
\shortauthor{R. Bhattacharjee, S. Saini, V. K. Gupta and A. S. Rana} 

\author{
Rahul Bhattacharjee\aff{1},
Sonu Saini\aff{1},
Vinay Kumar Gupta\aff{2},
\and
{Anirudh S. Rana\aff{1}
\corresp{\email{anirudh.rana@pilani.bits-pilani.ac.in}}
}
}

\affiliation
{
\aff{1}
Department of Mathematics, Birla Institute of Technology \& Science Pilani, Rajasthan 333031, India
\aff{2}
Department of Mathematics, Indian Institute of Technology Indore, 
Indore 453552, India
}

\begin{document}
\maketitle
\begin{textblock*}{\textwidth}(20mm, 8mm) 
\centering
\noindent
{\color{green!50!black}
Published in the 
\href{https://doi.org/10.1017/jfm.2023.994}{\ul{\emph{J.~Fluid Mech.}, {\bf 980}, A4 (2024)}}
\\[0.5ex]
A copy of the published version may also be obtained by mailing to the authors at 
\href{mailto:vkg@iiti.ac.in}{vkg@iiti.ac.in} or \href{mailto:anirudh.rana@pilani.bits-pilani.ac.in}{anirudh.rana@pilani.bits-pilani.ac.in}.}
\end{textblock*}

\begin{abstract} 
Experimental and theoretical studies on millimetre-sized droplets suggest that at low Reynolds number the difference between the drag force on a circulating water droplet and that on a rigid sphere is very small (less than 1\%) \citep{leclair1972theoretical}. 
While the drag force on a spherical liquid droplet at high viscosity ratios (of the liquid to the gas), is approximately the same as that on a rigid sphere of the same size, the other quantities of interest (e.g., the temperature) in the case of a rarefied gas flow over a liquid droplet differ from the same quantities in the case of a rarefied gas flow over a rigid sphere.
The goal of this article is to study the effects of internal motion within a spherical micro/nano droplet---such that its diameter is comparable to the mean free path of the surrounding gas---on the drag force and its overall dynamics. 
To this end, the problem of a slow rarefied gas flowing over an incompressible liquid droplet is investigated analytically by considering the internal motion of the liquid inside the droplet and also by accounting for kinetic effects in the gas.
Detailed results for different values of the Knudsen number, the ratio of the thermal conductivities and the ratio of viscosities are presented for the pressure and
temperature profiles inside and outside the liquid droplet. 
The results for the drag force obtained in the present work are in good agreement with the theoretical and experimental results existing in the literature.  
\end{abstract}
\section{\label{Sec:Intro}Introduction}


Liquid droplets moving in a gaseous medium are frequently encountered in nature; for instance, in rainfalls, in coughing or sneezing, in irrigation mist, etc.; they also have tremendous industrial applications, such as in aerosol spray, spray cooling, thermal spray coating, agricultural spraying, spray painting, food processing, fuel injection, fuel combustion, etc. 
Therefore, a clear understanding of the motion of liquid droplets in a gaseous medium (or, in other words, gas flow over a liquid droplet) is crucial in designing devices involving droplet motion in a gaseous medium and/or in improving their performance.   


For a gas flow over a liquid droplet, the ratio of the viscosities of the liquid inside the droplet to that of the surrounding fluid---referred to as the inside-to-outside viscosity ratio or, simply, the viscosity ratio and denoted by $\Lambda_\mu$---is a nonzero finite number.
The two limiting cases of the problem are (i) fluid flow over a solid sphere [case of infinite inside-to-outside viscosity ratio ($\Lambda_\mu \to \infty$)] and (ii) liquid flow over a gas bubble [case of zero inside-to-outside viscosity ratio ($\Lambda_\mu \approx 0$)]. 
In the former case, there is no question of internal motion and in the latter case, the internal fluid motion has negligible effect on the shape of the gas bubble as well as on the dynamics of the external flow \citep{oliver1985steady, oliver1987flow, Pozrikidis1989}. 
Thus, it is not surprising that these two limiting cases have been explored extensively in the literature [see the references given in chapters 3 and 5 of the textbook \cite{CGW1978}, which present a comprehensive review of these two limiting cases],
and they are now seemingly well understood. 
The internal fluid motion in the case of liquid droplets, however, has a significant impact on the dynamics of the external flow and, hence, should not be disregarded \citep{oliver1987flow, Pozrikidis1989}.
From a mathematical standpoint, the coupling of the external flow in the case of liquid droplet with the internal flow is through complicated boundary conditions on the droplet interface that makes the theoretical and computational methods of analysis considerably involved.
From an experimental point of view, it is very challenging to measure the flow inside the droplet without disturbing the shape of the droplet or without changing the physical properties of the fluids.
Consequently, only a handful of studies have looked into the problem of gas flow over a liquid droplet considering the internal fluid motion so far.  
Indeed, the authors could not find any paper on a rarefied gas flow over a \emph{micro-/nano-size} liquid droplet that accounts for the internal flow dynamics, although the problem of rarefied gas flow past a micro-/nano-size evaporating/non-evaporating droplet without internal circulation has been a subject of some recent works \citep[see, e.g.,][]{rana2019lifetime, RGST2021, RSCLS2021, TKR2021, de2022efficient, RLS2018}. 
%
%
Some open-source software, like OpenFOAM, in combination with methods, such as the volume-of-fluid, level-set and the direct simulation Monte Carlo (DSMC) methods, have also been utilised to investigate gas-liquid multiphase flow problems \citep{MR2015, CRYTL2019, C2019}. 
Nevertheless, the problems studied with these software are typically in a somewhat different direction than the one considered in this paper; for instance, the above references focus on bubble/droplet formation and its dynamics. 


The existence of internal circulation in liquid droplets falling in air (for which $\Lambda_\mu \approx 56$) was speculated already in the beginning of the last century by \citet{Lenard1904} and was confirmed later through wind tunnel experiments by \citet{GL1959} for large drops and by \citet{PB1970} for small drops.
To the best of authors' knowledge, the first attempt to explain the effects of almost all the factors, including internal circulation, on the shape and dynamics of large raindrops falling in air was made by \citet{mcdonald1954shape}. 
Through this study,  \cite{mcdonald1954shape} concluded that, for large drops, the internal circulation plays only a negligible role in controlling the shape of the drop.
To investigate the internal circulation in water drops falling at terminal velocity in air, \citet{leclair1972theoretical} proposed four theoretical approaches based on (i) creeping flow assumption for both internal and external flows, (ii) the assumptions of irrotational external flow and inviscid internal flow, (iii) the boundary layer theory and (iv) solving the vorticity-streamfunction formalism of the Navier--Stokes equations numerically for both internal and external flows together. 
In the same paper, they also presented a wind tunnel experimental study, similarly to that of \citet{PB1970}, to gauge the validity of the results from their theoretical approaches.
By comparing the results obtained from the theoretical approaches with those obtained from the wind tunnel experiment, \citet{leclair1972theoretical} found that the first approach markedly underestimated the internal velocity while the second approach markedly overestimated it and that the results from the third and fourth approaches were in reasonably good agreement with the experimental data for drops of diameters smaller than 1 millimetre. 
However, for large drops (of diameters bigger than 1 millimetre), they found that even their third approach overestimated the internal velocity significantly showing a completely wrong trend
and that their numerical approach, although overestimating the internal velocity slightly, was able to capture the trend of the internal velocity qualitatively.
Furthermore, \citet{leclair1972theoretical} also concluded that, for small values of the Reynolds number $\Rey$, the drag force on the drop 
is practically the same as the drag force on a solid sphere of the same Reynolds number.
Following the numerical approach of \cite{leclair1972theoretical}, 
\citet{AH1975} investigated the effect of internal circulation on the drag on a spherical droplet falling at terminal velocity and presented an empirical formula---obtained by fitting their numerical results---for the drag coefficient as a function of the viscosity ratio and external Reynolds number.   
In another similar numerical study,
\citet{rivkind1976flow} also solved the vorticity-streamfunction formalism of the Navier--Stokes equations numerically via the method of finite differences to determine the drag on a spherical fluid drop falling in another fluid for viscosity ratios $0 \leq \Lambda_\mu < \infty$ and for external Reynolds numbers $0.5 \leq \Rey \leq 100$ that cover flow over a solid sphere, over a liquid drop and over a small gas bubble. 
For $\Rey \ll 1$, they found that the drag coefficient of the drop can be expressed as a convex combination of the drag coefficients of the solid sphere and that of the gas bubble, with the coefficients in the combination being functions of the viscosity ratio.
Their formula for the drag coefficient of the drop turned out to yield a fairly accurate drag coefficient for $\Rey \ll 1$. 
\citet{RR1976} furthered the study to moderate Reynolds numbers and also gave another (empirical) formula for the drag coefficient of the drop as a function of the viscosity ratio and external Reynolds number.
However, a comparison of the drag coefficients obtained from the formulae of \cite{AH1975} and \cite{RR1976} reveals that there could be differences up to 20\% (for $\Rey \leq 20$) in the values of the drag coefficients obtained from them.
Moreover, the drag coefficients computed from neither of the formulae of \cite{AH1975} and \cite{RR1976} could approach the drag coefficient obtained from the Hadamard and Rybczynski relation \citep{CGW1978} in the vanishing Reynolds number limit.
Aiming to decipher the discrepancies arising from the formulae of \cite{AH1975} and \cite{RR1976}, Oliver and Chung performed two studies on flows inside and outside of a fluid sphere---first one for low Reynolds numbers and the second one for moderate Reynolds numbers. 
In their first study, \citet{oliver1985steady} employed a hybrid semi-analytical method comprising of the series-truncation technique and the finite-difference method to study the effect of internal circulation on bubble and droplet dynamics at low Reynolds numbers. 
They found that the density difference has no significant effect on the drag coefficient at low
Reynolds numbers and that the drag coefficient increases with increasing viscosity ratio.
In their second study, \citet{oliver1987flow} employed another hybrid semi-analytical method comprising of the series-truncation technique and the finite-element method to predict the flows inside and outside a fluid droplet at low to moderate Reynolds numbers. 
In this study, they found that the formula of the drag coefficient from \cite{AH1975} is actually dubious while that from \cite{RR1976} is good in predicting the drag coefficient for $2 \leq \Rey \leq 20$.
Since the formulae of both \cite{AH1975} and \cite{RR1976} for the drag coefficient are inadequate in the zero Reynolds number limit, \citet{oliver1987flow} also gave a predictive formula for the drag coefficient valid for $0 < \Rey < 2$. 
They also concluded that the strength of the internal circulation increases with increasing Reynolds number.

As aforementioned, we have found neither any theoretical work nor any experimental work on rarefied gas flow around a micro-/nano-size liquid droplet---especially when accounting for the internal circulation---in the literature.
Given that setting up an experiment at such a small scale is even more challenging, the objective of this paper is to investigate the aforesaid problem theoretically.
The liquid phase inside the droplet can be modelled with the Navier--Stokes equations.
However, it is important to note that the Navier--Stokes--Fourier (NSF) equations are not adequate for describing rarefied gas flow  \citep{Sone2002, Struchtrup2005}---outside the droplet.
Any fluid flow---including a rarefied gas flow---can, in principle, be described by the Boltzmann equation; nevertheless, its numerical solutions are computationally very expensive in general and particularly for flows in the so-called transition regime \citep{Struchtrup2005}.
The main source of problems in dealing with the Boltzmann equation is the Boltzmann collision operator appearing on the right-hand side of the Boltzmann equation.
Thus there has been a significant amount of research in developing ways alternative to directly solving the Boltzmann equation for investigating rarefied gas flows.
One of the most commonly used numerical techniques for investigating rarefied gas flows is the DSMC method, which is a probabilistic particle-based method developed by \cite{Bird1994} to solve the Boltzmann equation numerically. 
Since its development, the DSMC method has been ameliorated and employed to investigate several canonical rarefied gas flow problems; see, e.g., \cite{RMS2015, SRS2022, TRS2022, SH2023} and references therein.
Nevertheless, the DSMC method also demands a very high computational cost for processes in the transition regime.
Aiming to substitute for the involved Boltzmann collision operator in the Boltzmann equation, some simplified models---generically referred to as kinetic models---have also been proposed. 
Some widely used kinetic models are the Bhatnagar--Gross--Krook (BGK) model \citep{BGK1954}, the ellipsoidal statistical BGK (ES-BGK) model \citep{Holway1966} and the Shakhov model \citep{Shakhov1968}.
These kinetic models have also been utilised with the DSMC method. 
However, each of these kinetic models has its own shortcomings/difficulties; the reader is referred to \cite{Struchtrup2005} for details of these kinetic models.
The widely accepted models for describing transition-regime flows are the extended macroscopic equations derived from the Boltzmann equation predominantly through two asymptotic-expansion based approaches, namely the Chapman--Enskog expansion method \citep{CC1970} and the Grad moment method \citep{Grad1949b}.
The models resulting from both methods again have their own merits and demerits. 
Let us skip the details of them for the sake of succinctness; the interested reader may refer to \cite{Struchtrup2005} and \cite{TorrilhonARFM} for details.
To circumvent the demerits associated with the above two methods, Struchtrup and Torrilhon regularised the equations resulting from the Grad moment method (referred to as the Grad moment equations) by performing a Chapman--Enskog-like expansion on the Grad moment equations and derived the so-called regularised 13-moment (R13) equations \citep{StruchtrupTorrilhon2003, Struchtrup2004}.
The R13 equations, since their derivation, have been remarkably successful in describing rarefied gas flows in the transition regime; see \cite{TorrilhonARFM} and reference therein.
Since the R13 equations have been derived via an asymptotic expansion in the powers of a dimensionless parameter the Knudsen number, which is defined as the ratio of the mean free path of the gas to a characteristic length scale in the problem, it is not surprising that the R13 equations yield meaningful results mostly for small Knudsen numbers (i.e.~for flows in the early transition regime). 
Aiming to cover more of the transition regime, \cite{GuEmerson2009} derived the regularised 26-moment (R26) equations by extending the method proposed by \cite{StruchtrupTorrilhon2003}.
The R26 equations, in general, describe transition-regime flows better than the R13 equations, especially for relatively large Knudsen numbers.
Notwithstanding, the R26 equations also have limitations due to their derivation also through an asymptotic expansion in powers of the Knudsen number.
It can be stated empirically that the R26 equations yield very good results for transition-regime flows up to the Knudsen number close to unity \citep{GuEmerson2009, RLS2018, RGST2021} but may yield quantitatively different results for certain processes beyond the Knudsen number unity; see, e.g.~\cite{RLS2018, RGST2021}. 
Despite this, the system comprised of  the R26 equations \citep{GuEmerson2009} is the best known macroscopic model till date for describing transition-regime rarefied gas flows.
Therefore, we shall model the gas phase (outside the droplet) with the system of the R26 equations \citep{GuEmerson2009} and the liquid phase inside the droplet with the Navier--Stokes equations.
For comparison purpose, we shall also include the analytic solution obtained by solving the NSF equations for the gas phase. 
It is worthwhile noting that although the surface tension force is an important force that ought to be accounted for while investigating gas-liquid multiphase flows, considering the effect of the surface tension forces is beyond the scope of this paper and  will be considered elsewhere in the future.  
Here, we shall assume that the surface tension forces on the droplet are strong enough to maintain its spherical shape.
This assumption is justified at least for droplets made of some commonly used liquids---as discussed in \S\,\ref{Subsec:discussion}.


To find appropriate boundary conditions concomitant to the R13 and R26 equations is another challenging task; nevertheless, remarkable progress has been made in this direction since the pioneering work of \cite{Gu&Emerson2007} on deriving the boundary conditions for the R13 equations. 
\citet{ST2008} noticed some inconsistencies in the boundary conditions derived by \cite{Gu&Emerson2007} and presented improved boundary conditions for the R13 equations based on physical and mathematical requirements for the problem under consideration.
The boundary conditions of \cite{ST2008} may generically be referred to as the macroscopic boundary conditions (MBC). 
Following the approach of \cite{ST2008}, \cite{GuEmerson2009} derived the MBC for the R26 equations.
Recently, the MBC for the R13 equations have also been combined with the discrete velocity method in a hybrid approach by  \cite{YGWEZT2020} to make the computations faster in the near-wall region.
Notwithstanding, \citet{RanaStruchtrup2016} and \citet{RGST2021} showed that the MBC for the linearised R13 (LR13) equations as well as for the linearised R26 (LR26) equations are thermodynamically inconsistent and violate the Onsager reciprocity relations \citep{Onsager37, Onsager38, BRTS2018} for some boundary value problems, and proposed a new set of phenomenological boundary conditions (PBC) for the LR13 equations in \cite{RanaStruchtrup2016} and for the LR26 equations in \citet{RGST2021}.
As a next step, the PBC valid for processes involving phase change were also derived by \citet{BRTS2018} for the R13 equations and by \citet{RGST2021} for the R26 equations.
In this paper, we shall employ the PBC derived in \citet{RGST2021} for the external flow. 
In summary, we solve the LR26 equations---and also the NSF equations for comparison purposes---for the gas phase (outside the droplet) along with the PBC and the linearised Navier--Stokes equations for the liquid phase (inside the droplet) along with the coupled boundary conditions to obtain the analytic solution for the flow fields.
To validate the analytic solution obtained in the present work, we also compare the drag force on the liquid droplet computed analytically in the present work with that obtained from Millikan's famous oil-drop experiment.
A comparison of the drag force obtained from the present theory with the results obtained from Millikan's oil-drop experiment reveals that the drag force on a spherical liquid droplet at high viscosity ratios is nearly the same as that on a rigid sphere of the same size. 
Hence, in many practical applications (wherein the viscosity ratio is usually large), e.g.,~a water droplet moving through air, the droplet can be treated as a rigid sphere for the drag force computations.
However, we show that the internal motion of the liquid in the droplet does have effects on the other quantities of interest (e.g., the temperature).

The remainder of this paper is organised as follows. 
The governing equations in spherical coordinates along with the boundary conditions are presented in \S\,\ref{Sec:Problem}. 
The methodology for solving the problem analytically is outlined in \S\,\ref{Sec:method}.
The main results on the effect of the internal flow inside the liquid droplet on the motion of the rarefied gas flow are illustrated in \S\,\ref{Sec:results}.
Finally, concluding remarks are made in \S\,\ref{Sec:concl}.

\section{\label{Sec:Problem} Problem formulation}
\begin{figure}
    \centering
    \includegraphics[scale=0.5]{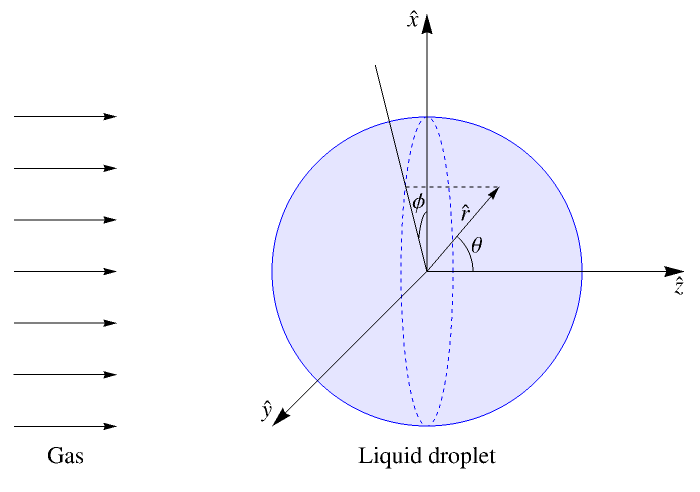}
    \caption{Schematic of a rarefied gas flow past a spherical liquid droplet.}
    \label{Fig:schematic}
\end{figure}

We consider a slow steady uniform flow of a monatomic rarefied gas approaching from the negative $\hat{z}$-direction with a uniform velocity $\hat{u}_\infty$ over a spherical droplet made of an incompressible liquid and centred at origin, as depicted in figure~\ref{Fig:schematic}.
Since incompressible liquid flows can be described accurately by the most celebrated equations of fluid dynamics---the Navier--Stokes equations, we model the flow inside the liquid droplet with the Navier--Stokes equations.
However, as stated in \S\,\ref{Sec:Intro}, the Navier--Stokes equations are not adequate for describing rarefied gas flows; therefore, we model the gas flow using the R26 equations, which describe rarefied gas flows remarkably well.
To exploit the spherical symmetry of the droplet, we shall express all the equations in the spherical coordinate system $(\hat{r},\theta,\phi)$, which is related to the Cartesian coordinate system $(\hat{x},\hat{y},\hat{z})$ via $(\hat{x},\hat{y},\hat{z}) \equiv (\hat{r} \sin{\theta} \cos{\phi}, \hat{r} \sin{\theta} \sin{\phi}, \hat{r} \cos{\theta})$. Here, $\hat{r} \in [0,\infty)$, $\theta \in [0,\pi]$ and $\phi \in [0,2\pi)$.
The spherical symmetry of the droplet implies that the flow parameters are independent of the direction $\phi$. Consequently, all the field variables pertaining to the problem are functions of $\hat{r}$ and $\theta$ only.

For the aforesaid problem, an analytic solution to the full Navier--Stokes equations and the fully nonlinear R26 equations is seemingly impossible.
Therefore, we restrict the present study to Stokes flows, i.e.~to small Reynolds number flows ($\Rey \ll 1$), and to slow flows, i.e.~to small Mach number flows ($\Ma \ll 1$), so that the linearised equations and linearised boundary conditions can be utilised in order to obtain an analytic solution of the problem.
Such an analytic solution is valid for slow flows ($\Ma \ll 1$) and in for low Reynolds numbers ($\Rey \ll 1$).
Given that rarefied gas flows encountered in micro and nano devices are usually slow flows, an analytic solution obtained by solving the linearised equations along with the linearised boundary conditions is   very plausible for all practical purposes.

To obtain the linearised equations, the governing equations and boundary conditions are linearised around a reference state, given by a constant density $\hat{\rho}_0$, a constant temperature $\hat{T}_0$ and all other field variables as zero.
For simplicity, we shall work with the dimensionless equations and boundary conditions, which are obtained by introducing the dimensionless deviations from their reference state values. 
Here, the deviations are assumed to be sufficiently small so that flow description with the linearised equations remains valid.
\subsection{Modelling of the gas phase}
The gas phase in the problem is modelled with the linear, dimensionless, steady-state R26 equations. 
The (fully nonlinear) R26 equations in the Cartesian coordinate system have been propounded in \cite{GuEmerson2009}. 
The field variables in the R26 equations are the density $\hat{\rho}$, velocity $\hat{v}_i$, temperature $\hat{T}$, stress tensor $\hat{\sigma}_{ij}$, heat flux $\hat{q}_i$, (tracefree) third velocity moment $\hat{m}_{ijk}$, partially contracted  (tracefree) fourth  velocity moment $\hat{R}_{ij}$ and fully contracted fourth velocity moment (also referred to as the scalar fourth moment) $\hat{\Delta}$. 
The field variables with hats are the usual quantities with dimensions, like the ones taken in \cite{GuEmerson2009} but without hats.
The R26 equations are linearised around the reference state described above and are made dimensionless by introducing the dimensionless deviations in the field variables from their respective reference state values:
\begin{align}
\label{deviations}
\left.
\begin{gathered}
\rho := \frac{\hat{\rho} - \hat{\rho}_0}{\hat{\rho}_0},
\quad
v_i := \frac{\hat{v}_i}{\sqrt{\hat{R}\hat{T}_0}},
\quad
T := \frac{\hat{T} - \hat{T}_0}{\hat{T}_0},
\quad
p := \frac{\hat{p} - \hat{p}_0}{\hat{p}_0},
\quad
\sigma_{ij} := \frac{\hat{\sigma}_{ij}}{\hat{p}_0},
\\
q_i := \frac{\hat{q}_i}{\hat{p}_0 \sqrt{\hat{R}\hat{T}_0}},
\quad
m_{ijk} := \frac{\hat{m}_{ijk}}{\hat{p}_0 \sqrt{\hat{R}\hat{T}_0}},
\quad
R_{ij} := \frac{\hat{R}_{ij}}{\hat{p}_0 \hat{R}\hat{T}_0},
\quad
\Delta := \frac{\hat{\Delta}}{\hat{p}_0 \hat{R}\hat{T}_0},
\end{gathered}
\right\}
\end{align}
where $\hat{R}$ is the gas constant; $\hat{p}_0 = \hat{\rho}_0\hat{R}\hat{T}_0$ is the pressure in the reference state; $\rho$, $v_i$, $T$, $p$, $\sigma_{ij}$ and $q_i$ are the dimensionless deviations in the density, velocity, temperature, pressure, stress and heat flux of the gas, respectively; similarly, $m_{ijk}$, $R_{ij}$ and $\Delta$ are the dimensionless deviations in the corresponding quantities. 
In addition, the droplet radius $\hat{R}_0$ is taken as the length scale for making the space variable $\hat{r}$ dimensionless, i.e.~$r := \hat{r}/\hat{R}_0$. 
We insert the dimensionless deviations \eqref{deviations} in the original R26 equations and drop all the nonlinear terms in  deviations along with the time derivative terms.
Finally, on transforming the resulting equations from the Cartesian coordinate system to the spherical coordinate system, we obtain the linear, dimensionless, steady-state R26 equations, which read \citep{RGST2021}
\begingroup
\begin{align}
\label{mass_bal_g}
\frac{\partial v_r}{\partial r}
+\frac{2 v_r}{r}
+\frac{\mathscr{D} v_\theta}{r} 
&= 0,
\\
\newsubeqblock
\subeqn
\label{r_mom_bal_g}
\frac{\partial p}{\partial r}
+\frac{\partial \sigma_{rr}}{\partial r}
+\frac{3 \sigma_{rr}}{r}
+\frac{\mathscr{D} \sigma_{r\theta}}{r}
&= 0,
\\
\subeqn
\label{theta_mom_bal_g}
\frac{\partial \sigma_{r\theta}}{\partial r}
+\frac{3 \sigma_{r\theta}}{r}
-\frac{1}{2 r} \frac{\partial \sigma_{rr}}{\partial \theta}
+\frac{1}{r} \frac{\partial p}{\partial \theta}
&= 0,
\\
\label{energy_bal_g}
\frac{\partial q_r}{\partial r}
+\frac{2 q_r}{r}
+\frac{\mathscr{D} q_\theta}{r}
&= 0,
\end{align}
\begin{subequations}
\begin{align}
\label{rr_stress_bal}
\frac{\partial m_{rrr}}{\partial r}
+\frac{4 m_{rrr}}{r}
+\frac{4}{5} \frac{\partial q_r}{\partial r}
+2 \frac{\partial v_r}{\partial r}
+\frac{\mathscr{D} m_{rr\theta}}{r}
&=-\frac{1}{\mathrm{Kn}} \sigma_{rr},
\\
\label{rtheta_stress_bal}
\frac{\partial m_{rr\theta}}{\partial r}
+\frac{4 m_{rr\theta}}{r}
+\frac{2}{5} \left(\frac{\partial q_\theta}{\partial r}
-\frac{q_\theta}{r}
\right)
+\frac{\partial v_\theta}{\partial r}
-\frac{v_\theta}{r}
\nonumber\\
-\frac{1}{2 r} \frac{\partial m_{rrr}}{\partial \theta}
+\frac{1}{r} \frac{\partial v_r}{\partial \theta}
+\frac{2}{5r} \frac{\partial q_r}{\partial \theta}
&= - \frac{1}{\mathrm{Kn}} \sigma_{r\theta},
\end{align}
\end{subequations}
\begin{subequations}
\begin{align}
\label{r_HF_bal}
\frac{1}{2} \left(\frac{\partial R_{rr}}{\partial r}
+\frac{3 R_{rr}}{r}
\right)
+\frac{1}{2} \frac{\mathscr{D} R_{r\theta}}{r}
+\frac{1}{6} \frac{\partial \Delta}{\partial r}
-\frac{\partial p}{\partial r}
+\frac{5}{2} \frac{\partial T}{\partial r}
&= - \frac{\mathrm{Pr}}{\mathrm{Kn}} q_r,
\\
\label{theta_HF_bal}
\frac{1}{2} \left(\frac{\partial R_{r\theta}}{\partial r}
+\frac{3 R_{r\theta}}{r}\right)
+\frac{1}{6r} \frac{\partial \Delta}{\partial \theta}
-\frac{1}{4 r} \frac{\partial R_{rr}}{\partial \theta}
-\frac{1}{r}\frac{\partial p}{\partial \theta}
+\frac{5}{2r} \frac{\partial T}{\partial \theta}
&= - \frac{\mathrm{Pr}}{\mathrm{Kn}} q_\theta,
\end{align}
\end{subequations}
\begin{subequations}
\label{m_bal}
\begin{align}
\label{mrrr_bal}
\frac{1}{r}\mathscr{D} 
\left(-\frac{6}{5}\sigma_{r\theta}+\Phi_{rrr\theta} - \frac{6}{35}R_{r\theta}\right) +\frac{9}{5}\left(\frac{\partial \sigma_{rr}}{\partial r}
-\frac{2 \sigma_{rr}}{r}\right)&
\nonumber \\
+ \frac{\partial \Phi_{rrrr}}{\partial r}+\frac{5 \Phi_{rrrr}}{r}
+\frac{9}{35}\left( \frac{\partial R_{rr}}{\partial r}-\frac{2 R_{rr}}{r}\right) &= -\frac{\mathrm{Pr}_{m}}{\mathrm{Kn}}m_{rrr}, 
\\
\label{mrrtheta_bal}
\frac{6}{5r}\frac{\partial \sigma_{rr}}{\partial \theta} 
+\frac{6}{35r} \frac{\partial R_{rr}}{\partial \theta }-\frac{1}{2r}\frac{\partial \Phi_{rrrr}}{\partial \theta} +\frac{8}{5}\left( \frac{\partial \sigma_{r\theta}}{\partial r}-\frac{2 \sigma_{r\theta}}{r}\right)&
\nonumber\\
+\frac{8}{35} \left( \frac{\partial R_{r\theta}}{\partial r}-\frac{2 R_{r\theta}}{r}\right) 
+\frac{\partial \Phi_{rrr\theta}}{\partial r}+\frac{5 \Phi_{rrr\theta}}{r} &=-\frac{\mathrm{Pr}_{m}}{\mathrm{Kn}}m_{rr\theta},
\end{align}
\end{subequations}
\begin{subequations}
\label{R_bal}
\begin{align}
\label{Rrr_bal}
\frac{1}{r}\mathscr{D}
\left( 2m_{rr\theta}
-\frac{2}{15}\Omega_{\theta}+\psi_{rr\theta}-\frac{28}{15}q_\theta \right) 
+\frac{56}{15}
\left(\frac{\partial q_r}{\partial r}-\frac{q_r}{r}\right)\hspace*{9.5mm} &
\nonumber\\
+2\left(\frac{\partial m_{rrr}}{\partial r}+\frac{4 m_{rrr}}{r}\right)
+\frac{\partial \psi_{rrr}}{\partial r}+\frac{4\psi_{rrr}}{r}
+\frac{4}{15}\left( \frac{\partial \Omega_{r}}{\partial r}-\frac{\Omega_{r}}{r}\right) &=-\frac{\mathrm{Pr}_{R}}{\mathrm{Kn}}R_{rr},  
\\
\label{Rrtheta_bal}
2\left(\frac{\partial m_{rr\theta}}{\partial r}+\frac{4 m_{rr\theta}}{r}\right) 
+ \frac{\partial \psi_{rr\theta}}{\partial r}+\frac{4 \psi_{rr\theta}}{r}
+\frac{1}{5}\left( \frac{\partial \Omega _{\theta}}{\partial r}-\frac{\Omega _{\theta}}{r}
\right)\hspace*{6mm} &
\nonumber\\
+ \frac{14}{5}\left(\frac{\partial q_\theta}{\partial r}-\frac{q_\theta}{r}\right)  
-\frac{1}{r}\frac{\partial m_{rrr}}{\partial \theta }+\frac{14}{5r}\frac{\partial q_r }{\partial \theta }-\frac{1}{2r}\frac{\partial 
\psi_{rrr}}{\partial \theta }+\frac{1}{5r}\frac{\partial 
\Omega_{r}}{\partial \theta }&=-\frac{\mathrm{Pr}_{R}}{\mathrm{Kn}}R_{r\theta},
\end{align}
\end{subequations}
\begin{align}
\label{Delta_bal}
\frac{1}{r}\mathscr{D}
\left( 8 q_{\theta}+\Omega _{\theta }\right) 
+ 8 \left( \frac{\partial 
q_r}{\partial r}+\frac{2 q_r}{r}\right) 
+ \frac{\partial 
\Omega_{r}}{\partial r}+\frac{2\Omega_{r}}{r} 
=-\frac{\mathrm{Pr}_{\Delta }}{\mathrm{Kn}}\Delta
%
\end{align}
\endgroup
with the additional unknowns in \eqref{m_bal}--\eqref{Delta_bal} being
\begin{subequations}
\label{Phi_closure}
\begin{align}
\Phi_{rrrr} &= - 4 \frac{\mathrm{Kn}}{\mathrm{Pr}_{\Phi}} \left[\frac{4}{7}\left( \frac{\partial m_{rrr}}{\partial r}-\frac{3 m_{rrr}}{r}\right)  
-\frac{3}{7}\frac{\mathscr{D} m_{rr\theta}}{r}\right],
\\
\Phi_{rrr\theta} &= - 4 \frac{\mathrm{Kn}}{\mathrm{Pr}_{\Phi}} \left[\frac{15}{28}\left( \frac{\partial m_{rr\theta}}{\partial r}-\frac{3 m_{rr\theta}}{r}\right) 
+\frac{5}{14 r} \frac{\partial m_{rrr}}{\partial \theta}\right],
\end{align}
\end{subequations}
\begin{subequations}
\label{psi_closure}
\begin{align}
\psi_{rrr} &= -\frac{27}{7} \frac{\mathrm{Kn}}{\mathrm{Pr}_{\psi}} \left[\frac{3}{5}\left( \frac{\partial R_{rr}}{\partial r}-\frac{2 R_{rr}}{r}\right)  
-\frac{2}{5} \frac{\mathscr{D} R_{r\theta}}{r}\right],
\\
\psi_{rr\theta} &= -\frac{27}{7} \frac{\mathrm{Kn}}{\mathrm{Pr}_{\psi}} \left[\frac{8}{15}\left( \frac{\partial R_{r\theta}}{\partial r}-\frac{2 R_{r\theta}}{r}\right) + \frac{2}{5 r} \frac{\partial R_{rr}}{\partial \theta}\right],
\end{align}
\end{subequations}
\begin{subequations}
\label{Omega_closure}
\begin{align}
\label{Omega_r_closure}
\Omega_{r} &= -\frac{7}{3} \frac{\mathrm{Kn}}{\mathrm{Pr}_{\Omega}} \left[\frac{\partial \Delta }{\partial r}+\frac{12}{7}\frac{\mathscr{D} R_{r\theta}}{r} + \frac{12}{7} \left(\frac{\partial R_{rr}}{\partial r}+\frac{3 R_{rr}}{r}\right) \right], 
\\
\label{Omega_theta_closure}
\Omega_{\theta} &= -\frac{7}{3} \frac{\mathrm{Kn}}{\mathrm{Pr}_{\Omega}} \left[\frac{1}{r}\frac{\partial \Delta}{\partial \theta} + \frac{12}{7}\left( \frac{\partial R_{r\theta}}{\partial r}+\frac{3 R_{r\theta}}{r}\right) -\frac{6}{7 r}\frac{\partial R_{rr}}{\partial \theta}\right].
\end{align}
\end{subequations}
Here $\mathscr{D}\equiv \cot{\theta}+\frac{\partial}{\partial\theta}$. Equations \eqref{mass_bal_g}--\eqref{Omega_closure}, henceforth, will be referred to as the linearised R26 (LR26) equations.
The coefficients $\mathrm{Kn}$, $\mathrm{Pr}$, $\mathrm{Pr}_m$, $\mathrm{Pr}_R$, $\mathrm{Pr}_\Delta$, $\mathrm{Pr}_{\Phi}$, $\mathrm{Pr}_{\psi}$, $\mathrm{Pr}_{\Omega}$ in the LR26 equations are the dimensionless numbers arising from the non-dimensionalisation of the equations. 
In particular, the numbers 
\begin{align}
\label{KnandPr}
\mathrm{Kn} = \frac{\hat{\mu}}{\hat{\rho}_0 \sqrt{\hat{R}\hat{T}_0} \hat{R}_0}
\quad\text{and}\quad
\mathrm{Pr} = \frac{5}{2} \frac{\hat{\mu}}{\hat{\kappa}} \hat{R}
\end{align}
are referred to as the Knudsen number and the Prandtl number, respectively, with $\hat{\mu}$ being the viscosity of the gas and $\hat{\kappa}$ being the thermal conductivity of the gas. 
It should be noted that, owing to the linearisation, the viscosity $\hat{\mu}$ and thermal conductivity $\hat{\kappa}$ in \eqref{KnandPr} are the viscosity and thermal conductivity of the gas at the reference state temperature $\hat{T}_0$; and hence both are constant.
Let us denote the viscosity and thermal conductivity of the gas at the reference state temperature $\hat{T}_0$ by $\hat{\mu}_0$ and $\hat{\kappa}_0$, respectively. 
Thus, owing to the linearisation, $\hat{\mu} = \hat{\mu}_0$ and $\hat{\kappa} = \hat{\kappa}_0$ throughout this work.  
The values of the numbers $\mathrm{Pr}$, $\mathrm{Pr}_m$, $\mathrm{Pr}_R$, $\mathrm{Pr}_\Delta$, $\mathrm{Pr}_{\Phi}$, $\mathrm{Pr}_{\psi}$, $\mathrm{Pr}_{\Omega}$ depend on the choice of the interaction potential between two gas molecules. For the Maxwell interaction potential used in the present work, the values of these numbers are $\mathrm{Pr} = 2/3$, $\mathrm{Pr}_m = 3/2$, $\mathrm{Pr}_R = 7/6$, $\mathrm{Pr}_\Delta = 2/3$, $\mathrm{Pr}_{\Phi} = 2.097$, $\mathrm{Pr}_{\psi} = 1.698$, $\mathrm{Pr}_{\Omega} = 1$ \citep{GuEmerson2009}.
The subscripts $r$ and $\theta$ with the vectors/tensors in the LR26 equations denote their respective components; for instance, $v_r$ is the $r$-component of the deviation in the velocity vector and $\sigma_{r\theta}$ is the $r\theta$-component of the deviation in the stress tensor.
Equation \eqref{mass_bal_g} can be identified as the equation of continuity for the gas, equations \eqref{r_mom_bal_g} and \eqref{theta_mom_bal_g} as the momentum balance equations in the $r$- and $\theta$-directions, respectively, and equation \eqref{energy_bal_g} as the energy balance equation, equations \eqref{rr_stress_bal} and \eqref{rtheta_stress_bal} as the balance equations for the $rr$- and $r\theta$-components of the stress, equations \eqref{r_HF_bal} and \eqref{theta_HF_bal} as the heat flux balance equations in the $r$- and $\theta$-directions, and so on.

For comparison purposes, we shall also model the gas phase with the linearised NSF equations. 
In this case, the linear, dimensionless, steady-state NSF equations are \eqref{mass_bal_g}--\eqref{energy_bal_g} along with the closure 
\begin{gather}
\label{closure_sigma}
\sigma_{rr} = - 2 \mathrm{Kn}\frac{\partial v_r}{\partial r},
\quad
\sigma_{r\theta} = - \mathrm{Kn}\left(\frac{\partial v_\theta}{\partial r} - \frac{v_\theta}{r} + \frac{1}{r} \frac{\partial v_r}{\partial \theta}\right),
\\
\label{closure_q}
q_r = - \frac{5}{2} \frac{\mathrm{Kn}}{\mathrm{Pr}} \frac{\partial T}{\partial r}
\quad\text{and}\quad
q_\theta = - \frac{5}{2} \frac{\mathrm{Kn}}{\mathrm{Pr}} \frac{1}{r} \frac{\partial T}{\partial \theta}.
\end{gather}
\subsection{Modelling of the liquid phase inside the droplet}
The liquid phase inside the spherical droplet is modelled with the linearised, steady-state, incompressible NSF equations.
The complete (fully nonlinear and unsteady) NSF equations in the spherical coordinate system can be found in some standard textbooks on fluid dynamics; see, e.g., the textbook \cite{Batchelor1967}.
The linear, dimensionless, steady-state NSF equations are obtained by dropping the time derivative terms in the full NSF equations, linearising the field variables around the reference state defined above and making them dimensionless using the density $\hat{\rho}_0$ and temperature $\hat{T}_0$ in the reference state and the droplet radius $\hat{R}_0$ as the length scale. 
After simplification, the linear, dimensionless steady-state NSF equations for modelling the liquid phase of the problem under consideration read 
%
\begin{align}
\label{mass_bal_l}
\frac{\partial v_r^{(\ell)}}{\partial r}+\frac{2 v_r^{(\ell)}}{r}+\frac{\mathscr{D} v_\theta^{(\ell)}}{r}=0,
\\
\newsubeqblock
\label{r_mom_bal_l}
\subeqn
\frac{\partial p^{(\ell)}}{\partial r}+\frac{\partial\sigma_{rr}^{(\ell)}}{\partial r}+\frac{3\sigma_{rr}^{(\ell)}}{r}+\frac{\mathscr{D} \sigma_{r\theta}^{(\ell)}}{r}=0,
\\
\subeqn
\label{theta_mom_bal_l}
\frac{\partial\sigma_{r\theta}^{(\ell)}}{\partial r}+\frac{3\sigma_{r\theta}^{(\ell)}}{r}-\frac{1}{2r}\frac{\partial\sigma_{rr}^{(\ell)}}{\partial \theta}+\frac{1}{r}\frac{\partial p^{(\ell)}}{\partial \theta}=0,
\\
\label{energy_bal_l}
\frac{\partial q_r^{(\ell)}}{\partial r}+\frac{2q_r^{(\ell)}}{r}+\frac{\mathscr{D}q_{\theta}^{(\ell)}}{r}=0\hphantom{,}
\end{align}
with
\begingroup
\allowdisplaybreaks
\begin{gather}
\label{closure_sigmal}
\sigma_{rr}^{(\ell)} = - 2 \Lambda_\mu \mathrm{Kn} \frac{\partial v_r^{(\ell)}}{\partial r},
\quad
\sigma_{r\theta}^{(\ell)} = - \Lambda_\mu \mathrm{Kn} \left(\frac{\partial v_\theta^{(\ell)}}{\partial r} - \frac{v_\theta^{(\ell)}}{r} + \frac{1}{r} \frac{\partial v_r^{(\ell)}}{\partial \theta}\right),
\\
\label{closure_ql}
q_r^{(\ell)} = - \frac{5}{2} \Lambda_\kappa \frac{\mathrm{Kn}}{\mathrm{Pr}} \frac{\partial T^{(\ell)}}{\partial r}
\quad\text{and}\quad
q_\theta^{(\ell)} = - \frac{5}{2} \Lambda_\kappa \frac{\mathrm{Kn}}{\mathrm{Pr}} \frac{1}{r} \frac{\partial T^{(\ell)}}{\partial \theta}.
\end{gather}
\endgroup
The superscript `$(\ell)$' in \eqref{mass_bal_l}--\eqref{closure_ql} has been used to indicate that the variables with the superscript `$(\ell)$' belong to the liquid phase (i.e.~to the liquid droplet). 
The variables in \eqref{mass_bal_l}--\eqref{closure_ql} are as follows. 
The variables 
\begin{align}
v_r^{(\ell)} = \frac{\hat{v}_r^{(\ell)}}{\sqrt{\hat{R}\hat{T}_0}}
\quad\text{and}\quad
v_\theta^{(\ell)} = \frac{\hat{v}_\theta^{(\ell)}}{\sqrt{\hat{R}\hat{T}_0}}
\end{align}
are the dimensionless deviations in the $r$- and $\theta$-components of the velocity of the liquid, respectively;
\begin{align}
p^{(\ell)} = \frac{\hat{p}^{(\ell)}-\hat{p}^{(\ell)}_0}{\hat{p}_0}
\label{p_liquid}
\end{align}
is the dimensionless deviation in the pressure of the liquid droplet from its pressure in the equilibrium state $\hat{p}_0^{(\ell)}$, which is given by the pressure inside a stationary droplet in a quiescent environment, i.e.~$\hat{p}_0^{(\ell)} = \hat{p}_0 + 2\hat{\gamma}/\hat{R}_0$. Note that the reference pressure  inside a stationary droplet is higher than the reference pressure outside the droplet because of the surface tension $\hat{\gamma}$. 
Furthermore, in \eqref{mass_bal_l}--\eqref{closure_ql}, 
\begin{align}
\sigma_{rr}^{(\ell)} = \frac{\hat{\sigma}_{rr}^{(\ell)}}{\hat{p}_0}
\quad\text{and}\quad
\sigma_{r\theta}^{(\ell)} = \frac{\hat{\sigma}_{r\theta}^{(\ell)}}{\hat{p}_0}
\end{align}
are the dimensionless deviations in the $rr$- and $r\theta$-components of the stress tensor for the liquid, respectively; 
\begin{align}
q_r^{(\ell)} = \frac{\hat{q}_r^{(\ell)}}{\hat{p}_0\sqrt{\hat{R}\hat{T}_0}}
\quad\text{and}\quad
q_\theta^{(\ell)} = \frac{\hat{q}_\theta^{(\ell)}}{\hat{p}_0\sqrt{\hat{R}\hat{T}_0}}
\end{align}
are the dimensionless deviations in the $r$- and $\theta$-components of the heat flux of the liquid, respectively;
\begin{align}
T^{(\ell)} = \frac{\hat{T}^{(\ell)}-\hat{T}_0}{\hat{T}_0}
\end{align}
is the dimensionless deviation in the temperature of the liquid droplet from the temperature in the reference state $\hat{T}_0$;
$\Lambda_\mu = \hat{\mu}^{(\ell)} / \hat{\mu}$ is  the ratio of the viscosity of the liquid to the viscosity of the gas
and $\Lambda_\kappa = \hat{\kappa}^{(\ell)} / \hat{\kappa}$ is the ratio of the thermal conductivity of the liquid to the thermal conductivity of the gas.
%
Similarly to the above, owing to the linearisation, $\hat{\mu}^{(\ell)}$ and $\hat{\kappa}^{(\ell)}$ are the viscosity and thermal conductivity of the liquid at the reference temperature $\hat{T}_0$; consequently, $\hat{\mu}$, $\hat{\mu}^{(\ell)}$, $\hat{\kappa}$, $\hat{\kappa}^{(\ell)}$, $\Lambda_\mu$ and $\Lambda_\kappa$ are all constant in the present work.
%
The field variables with hats and superscript `${(\ell)}$' can be identified as the field variables of the original NSF equations. Furthermore, equation \eqref{mass_bal_l} can be identified as the equation of continuity, equations \eqref{r_mom_bal_l} and \eqref{theta_mom_bal_l} as the momentum balance equations in the $r$- and $\theta$-directions, respectively, and equation \eqref{energy_bal_l} as the energy balance equation.
%
%
\subsection{Boundary conditions}\label{BCON}
The physically admissible boundary conditions, which respect the second law of thermodynamics and satisfy the Onsager reciprocity relations, for the LR26 equations have been derived in \cite{RGST2021}.
It may be noted that the boundary conditions derived in \cite{RGST2021,RGS2018} are general---they consider the motion of both gas and boundary in the normal direction along with evaporation. 
In the problem under consideration, evaporation has been ignored for simplicity, the interface between the liquid and gas has been assumed to be fixed, and that neither the liquid nor the gas can penetrate the interface is assumed.
Therefore, we would take the evaporation/condensation coefficient $\vartheta = 0$ and the normal component of the flow velocity relative to the interface velocity $\mathcal{V}_n = 0$ in the boundary conditions given in \cite{RGST2021}. 
We skip more details of the boundary conditions for the sake of succinctness and present them here directly; interested readers are referred to \citep{RGST2021, RSCLS2021} for more details. 
For the problem under consideration, the $r$-direction is the normal direction while the $\theta$- and $\phi$-directions are the two tangential directions; nevertheless, the present problem is independent of $\phi$ due to spherical symmetry. 
Therefore, we shall replace the subscripts $n$ with $r$ and $i$ with $\theta$ in the boundary conditions derived in \cite{RGST2021}.
Consequently, the boundary conditions complementing the LR26 equations \eqref{mass_bal_g}--\eqref{Omega_closure}---for the problem under consideration---at the interface (i.e.~at $r=1$) read \citep{RGST2021}
\begingroup 
\allowdisplaybreaks\begin{subequations}
\label{BCgas}
\begin{align}
\label{Interface_cond}
v_r &= v_r^{I} = 0,
\\
\label{For_finding_NSF_jump_cond}
q_r &= - \frac{\chi}{2 - \chi} \sqrt{\frac{2}{\pi}} \left(2 \mathcal{T} + \frac{1}{2} \sigma_{rr} + \frac{5}{28} R_{rr} + \frac{1}{15} \Delta -\frac{1}{6} \Phi_{rrrr} \right),
\\
m_{rrr} &= \frac{\chi}{2 - \chi}\sqrt{\frac{2}{\pi}} \left(\frac{2}{5} \mathcal{T} - \frac{7}{5} \sigma_{rr} - \frac{1}{14} R_{rr} + \frac{1}{75} \Delta - \frac{13}{15} 
\Phi_{rrrr}\right),
\\
\Psi_{rrr} &= \frac{\chi}{2 - \chi}\sqrt{\frac{2}{\pi}} \left(\frac{6}{5} \mathcal{T} + \frac{9}{5} \sigma_{rr} - \frac{93}{70} R_{rr} + \frac{1}{5} \Delta + \frac{11}{15} \Phi_{rrrr}\right),
\\
\Omega_r &= \frac{\chi}{2 - \chi}\sqrt{\frac{2}{\pi}} \left(8 \mathcal{T} + 2 \sigma_{rr} - R_{rr} - \frac{4}{3} \Delta
- \frac{2}{3} \Phi_{rrrr}\right),
\\
\label{For_finding_NSF_slip_cond}
\sigma_{r\theta} &=- \frac{\chi}{2 - \chi}\sqrt{\frac{2}{\pi}}
\left( \mathscr{V}_\theta
+\frac{1}{5} q_\theta +\frac{1}{2}m_{rr\theta} - \frac{1}{14} \Psi_{rr\theta} - \frac{1}{70} \Omega_r\right),
\\
R_{r\theta} &=- \frac{\chi}{2 - \chi}\sqrt{\frac{2}{\pi}} \left( -\mathscr{V}_\theta 
+ \frac{11}{5} q_\theta + \frac{1}{2} m_{rr\theta} + \frac{13}{14} \Psi_{rr\theta} + \frac{13}{70} \Omega_r\right), 
\\
\Phi_{rrr\theta} &=- \frac{\chi}{2 - \chi}\sqrt{\frac{2}{\pi}} \left( -\frac{4}{7} \mathscr{V}_\theta
-\frac{12}{35} q_\theta +\frac{9}{7} m_{rr\theta} - \frac{2}{49} \Psi_{rr\theta}
- \frac{2}{245} \Omega_r\right),
\end{align}
\end{subequations}
\endgroup
where $v_r^{I}$ is the velocity of the interface and is zero for the present problem, $\chi$ is the accommodation coefficient, $\mathscr{V}_\theta = v_\theta - v_\theta^{(\ell)}$ is the velocity slip and $\mathcal{T} = T - T^{(\ell)}$ is the temperature jump.
Note that the remaining boundary conditions for the rank-2 and rank-3 tensors given in \cite{RGST2021} are not needed due to the fact that the present problem is independent of $\phi$.

While solving the gas phase with the linearised NSF equations (for comparison purposes), the appropriate boundary conditions are \eqref{Interface_cond} and the ones obtained by ignoring the higher-order moments in boundary conditions \eqref{For_finding_NSF_jump_cond} and \eqref{For_finding_NSF_slip_cond}. 
Thus, for the linearised NSF equations, the appropriate boundary conditions are \eqref{Interface_cond} and 
\begin{align}
\label{Jump1}
\frac{\chi}{2 - \chi} \sqrt{\frac{2}{\pi}} \left(2 \mathcal{T} + \frac{1}{2} \sigma_{rr}\right) + q_r = 0,
\\
\label{Slip1}
\frac{\chi}{2 - \chi}\sqrt{\frac{2}{\pi}}
\left( \mathscr{V}_\theta
+\frac{1}{5} q_\theta\right) + \sigma_{r\theta} = 0.
\end{align}

In order to solve the equations corresponding to the liquid and gas phases together, we need additional boundary conditions, which are as follows.
(i) Similarly to the gas, the liquid cannot penetrate the interface. This means that the normal component of the velocity of the liquid should also vanish at the interface, i.e.
\begin{align}
\label{vn_cond}
v_r^{(\ell)} = 0
\quad\text{at}\quad
r=1.
\end{align}
(ii) The heat flux and shear stress are continuous at the interface; this implies that
\begin{align}
\label{cont_cond}
q_r = q_r^{(\ell)}
\quad\text{and}\quad
\sigma_{r\theta} = \sigma_{r\theta}^{(\ell)}
\quad\text{at}\quad
r=1.
\end{align}  
It is worthwhile noting that the density ratio (of the liquid to the gas)---despite being an important parameter in gas-liquid interfacial flows---appear neither in the governing equations \eqref{mass_bal_g}--\eqref{Delta_bal} and \eqref{mass_bal_l}--\eqref{energy_bal_l} nor in the constitutive relations \eqref{Phi_closure}--\eqref{Omega_closure}, \eqref{closure_sigmal} and \eqref{closure_ql} nor in boundary conditions \eqref{BCgas}, \eqref{vn_cond} and \eqref{cont_cond} due to our assumptions of no phase change and of the surface tension force being strong enough to keep the spherical shape and size of the droplet unchanged. 
Hence, the density ratio does not play any role in the present work. 
Nevertheless, when taking the effect of the surface tension force into account (i.e.~when the droplet is allowed to change its shape) and/or taking the phase-change into account (i.e.~when the droplet is allowed to change its size), boundary conditions \eqref{Interface_cond} and \eqref{vn_cond} need to be modified appropriately.
The modified boundary conditions will consist of the densities of both liquid and gas, and hence the density ratio.
In addition, when accounting for the effects of the surface tension forces, the stress boundary condition \eqref{cont_cond}$_2$ also needs to be changed. 
A general force balance condition at an interface between two fluids (labelled `1' and `2') is given by \citep{LL1959}
\begin{align}
\label{gen_stress_cond}
(\hat{p}_1-\hat{p}_2+\hat{\gamma}_c \hat{\varkappa}) n_i = -\left[\hat{\sigma}_{ij}^{(1)} - \hat{\sigma}_{ij}^{(2)}\right] n_j + \frac{\partial \hat{\gamma}_c}{\partial \hat{x}_i},
\end{align}
where $\hat{p}_1$ and $\hat{p}_2$ are the pressures exerted on the interface by the fluids `1' and `2', respectively; $\hat{\gamma}_c$ is the surface tension coefficient; $\hat{\varkappa}$ is the local surface curvature; $n_i$ is the unit normal at the interface; and $\hat{\sigma}_{ij}^{(1)}$ and $\hat{\sigma}_{ij}^{(2)}$ are the  stress tensors of the fluids `1' and `2', respectively.
In the present work, we assume that the droplet remains spherical without any growth or shrinkage and that it does not deform too; in other words, the radius of the droplet $\hat{R}_0$ is assumed to remain constant. 
Therefore, we neglect the nonequilibrium force balance in the normal direction as described in equation \eqref{gen_stress_cond}. 
It is justified since the equilibrium aspect has already been considered when defining $\hat{p}_0^{(\ell)}$ above.
Additionally, for the force balance in the tangential direction, we multiply equation \eqref{gen_stress_cond} by the unit tangent vector $t_i$ at the interface. This multiplication makes the left-hand side of the resulting equation vanish. By disregarding the effect of surface tension forces, specifically the term $\partial \hat{\gamma}_c / \partial \hat{x}_i$ (commonly referred to as the Marangoni surface tension gradient) in equation \eqref{gen_stress_cond}, the boundary condition simplifies to $\hat{\sigma}_{nt}^{(1)} = \hat{\sigma}_{nt}^{(2)}$, which in the dimensionless form is the same as boundary condition \eqref{cont_cond}$_2$ considered in the present work. 
\section{\label{Sec:method}Analytic solution methodology}
We solve the equations derived in \S\,\ref{Sec:Problem} analytically by following a method proposed in~\citep{Torrilhon2010, RLS2018}.
The key idea of this method is to convert the system of partial differential equations to a system of ordinary differential equations by presuming the dependence of the field variables on the azimuthal angle $\theta$ through the sine and cosine functions alone. 
In this method, the scalar variables and vectorial/tensorial components of a field variable having an even number of $\theta$ indices are taken to be proportional to $\cos{\theta}$, the vectorial/tensorial components having an odd number of $\theta$ indices are taken to be proportional to $\sin{\theta}$ and the proportionality constants are taken to be functions of $r$ alone \citep{Torrilhon2010, RGST2021}. 
Using this idea, we assume that the field variables are given by

\begin{align}
\label{field_ansatz_gas}
\left.
\begin{aligned}
v_r(r,\theta) &= \mathbbm{v}_1(r) \cos{\theta},&
\quad
v_\theta(r,\theta) &= - \mathbbm{v}_2(r) \sin{\theta}
\\
p(r,\theta) &= \mathbbm{p}(r) \cos{\theta},
&\quad
T(r,\theta) &= \mathbbm{T}(r) \cos{\theta},
\\
\sigma_{rr}(r,\theta) &= \mathbbm{s}_1(r) \cos{\theta},
&\quad
\sigma_{r\theta}(r,\theta) &= \mathbbm{s}_2(r) \sin{\theta},
\\
q_r(r,\theta) &= \mathbbm{q}_1(r) \cos{\theta},
&\quad
q_\theta(r,\theta) &= -\mathbbm{q}_2(r) \sin{\theta},
\\
m_{rrr}(r,\theta) &= \mathbbm{m}_1(r) \cos{\theta},
&\quad
m_{rr\theta}(r,\theta) &= \mathbbm{m}_2(r) \sin{\theta},
\\
R_{rr}(r,\theta) &= \mathbbm{R}_1(r) \cos{\theta},
&\quad
R_{r\theta}(r,\theta) &= \mathbbm{R}_2(r) \sin{\theta},
\\
\Delta(r,\theta) &= \mathbbm{d}(r) \cos{\theta},
\end{aligned}
\right\}
\end{align}
\begin{align}
\label{field_ansatz_liquid}
\left.
\begin{aligned}
v_r^{(\ell)}(r,\theta) &= \mathbbm{v}_1^{(\ell)}(r) \cos{\theta},
&\quad
v_\theta^{(\ell)}(r,\theta) &= - \mathbbm{v}_2^{(\ell)}(r) \sin{\theta},
\\
p^{(\ell)}(r,\theta) &= \mathbbm{p}^{(\ell)}(r) \cos{\theta},
&\quad
T^{(\ell)}(r,\theta) &= \mathbbm{T}^{(\ell)}(r) \cos{\theta},
\end{aligned}
\right\}
\end{align}
where the functions $\mathbbm{v}_1, \mathbbm{v}_2, \mathbbm{p}, \mathbbm{T}, \mathbbm{s}_1, \mathbbm{s}_2, \mathbbm{q}_1, \mathbbm{q}_2, \mathbbm{m}_1, \mathbbm{m}_2, \mathbbm{R}_1, \mathbbm{R}_2, \mathbbm{d}, \mathbbm{v}_1^{(\ell)}, \mathbbm{v}_2^{(\ell)}, \mathbbm{p}^{(\ell)}$ and $\mathbbm{T}^{(\ell)}$ are the functions of $r$ alone.

The above ansatzes for the field variables are inserted in \eqref{mass_bal_g}--\eqref{Omega_closure} and in \eqref{mass_bal_l}--\eqref{closure_ql}. 
After simplification ($\cos{\theta}$ and $\sin{\theta}$ in each equation get canceled), one obtains two systems of ordinary differential equations---one for the liquid and the other for the gas. 
These systems of ordinary differential equations are solved independently and analytically. 
The analytic solution for the system associated with the liquid phase is easy to obtain. 
It turns out to be 
\begin{align}
\label{sol_l}
\mathbbm{v}_1^{(\ell)}(r) = b_1 + \frac{b_2 r^2}{2},
\quad
\mathbbm{v}_2^{(\ell)}(r) = b_1 + b_2 r^2,
\quad
\mathbbm{p}^{(\ell)}(r) = 5 b_2 \Lambda_\mu \mathrm{Kn} \, r,
\quad
\mathbbm{T}^{(\ell)}(r) = b_3 r,
\end{align}  
where $b_1$, $b_2$ and $b_3$ are the integration constants, which are computed using the interface conditions \eqref{vn_cond} and \eqref{cont_cond}. 
While obtaining solution \eqref{sol_l}, we have also used the fact that the solution should remain bounded as $r$ approaches zero. 
The analytic solution for the system associated with the gas phase is, however, not so straightforward to obtain.
We have used the computer algebra software {\textsc{Mathematica}}\textsuperscript{\textregistered} to obtain the analytic solution for the system associated with the gas phase. 
For better readability, the solution has, however, been relegated to appendix~\ref{app:sol}. 
It may be noted that this solution contains eight integration constants, namely $C_1$, $C_2$, $C_3$ and $K_1, K_2, \dots, K_5$, which  are computed using the eight boundary conditions \eqref{BCgas}.
After applying the boundary conditions, the solution for all field variables for both the liquid and gas phases become known.
\section{\label{Sec:results}Results and discussion}
To access the validity of the findings of this work, we first present the results on the drag force and compare them with those obtained from an experiment. 
After validating the drag force, we shall present the results on the physical field variables, which are often difficult to measure through experiments.
\subsection{\label{Subsec:drag}Drag force}
Before computing the drag force with the analytic solution obtained in the present work, let us first comment on the experimental data, which will be used to validate our analytical findings on the drag force.
The experimental data is actually from the famous oil-drop experiment conducted in 1909 by R.~A.~Millikan, the Nobel laureate in Physics 1923, to measure the electric charge carried by a single electron.
Through this experiment, he also computed the drag force on oil drops of different radii falling in the air at terminal velocity \citep{Millikan1923}.
Millikan's experimental data have been fitted by several researchers to obtain an empirical formula for the drag force in the Knudsen--Weber form \citep{KW1911}
\begin{align}
\label{KW_emp_formula}
F= F_\mathrm{Stokes}\left[\frac{1}{1+\mathrm{Kn}\left(\mathfrak{a} + \mathfrak{b} \, \mathrm{e}^{-\mathfrak{c}/\mathrm{Kn}}\right)}\right], 
\end{align}
where $F_\mathrm{Stokes}=6\pi \,\mathrm{Kn} \, u_{\infty}$ is the Stokes drag with $u_{\infty} = \hat{u}_\infty / \sqrt{\hat{R}\hat{T}_0}$ being the far-field dimensionless velocity of gas approaching the droplet (for simplicity, we have taken $u_{\infty} = 1$ in this article), and $\mathfrak{a}$, $\mathfrak{b}$ and $\mathfrak{c}$ are the experimentally-determined constants. 
It turns out that for an oil drop of size comparable with the mean free path of the air (i.e.~in the transition regime), the experimental data for oil drops in air from Millikan's oil-drop experiment requires these constants to be \citep{Kennard1938}%
\begin{align}
\label{Kennard_coeff}
\mathfrak{a} = 1.23,
\quad
\mathfrak{b} = 0.41
\quad\text{and}\quad
\mathfrak{c} = 0.88.
\end{align}
The most accurate raw data from Millikan's experiment were reviewed later by \cite{AR1982} using very precise values of the physical constants known at that time and the nonlinear least-squares fitting technique. 
With these, the new values of the constants $\mathfrak{a}$, $\mathfrak{b}$ and $\mathfrak{c}$ obtained by \cite{AR1982} are
\begin{align}
\label{AR1982_coeff}
\mathfrak{a} = 1.155 \pm 0.008,
\quad
\mathfrak{b} = 0.471 \pm 0.011
\quad\text{and}\quad
\mathfrak{c} = 0.596 \pm 0.050.
\end{align}
An improved version of Millikan's experimental apparatus was designed and built by \cite{AR1985} aiming to replace the oil drops by solid spheres in Millikan's experiment and to determine new values of the constants $\mathfrak{a}$, $\mathfrak{b}$ and $\mathfrak{c}$ corresponding to the drag force on a solid sphere suspended in air. 
By taking three different types of solid spherical particles---namely,  polystyrene latex-divinylbenzene particles, polyvinyltoluene particles and polystyrene latex particles---with the Knudsen number ranging from $0.03$ to $7.2$ in this experiment, they found the values of the constants $\mathfrak{a}$, $\mathfrak{b}$ and $\mathfrak{c}$ to be
\begin{align}
\label{AR1985_coeff}
\mathfrak{a} = 1.142 \pm 0.0024,
\quad
\mathfrak{b} = 0.558 \pm 0.0024
\quad\text{and}\quad
\mathfrak{c} = 0.999 \pm 0.0212.
\end{align}
Using the modulated dynamic light scattering technique---a technique fundamentally different from the ones used by Millikan and by Allen and Raabe,
\cite{HHF1995} also measured the drag force on spherical polystyrene latex (solid) particles suspended in dry air for the Knudsen number values ranging from $0.06$ to $500$, and  found the values of the constants $\mathfrak{a}$, $\mathfrak{b}$ and $\mathfrak{c}$ to be
\begin{align}
\label{HHF1995_coeff}
\mathfrak{a} = 1.2310 \pm 0.0022, \quad
\mathfrak{b} = 0.4695 \pm 0.0037
\quad\text{and}\quad
\mathfrak{c} = 1.1783 \pm 0.0091.
\end{align}

To compare the drag force computed analytically in the present work, we use the data resulting from formula~\eqref{KW_emp_formula} with the constants $\mathfrak{a}$, $\mathfrak{b}$ and $\mathfrak{c}$ given in \eqref{Kennard_coeff} [taken from the textbook \cite{Kennard1938}] and in \eqref{AR1982_coeff} [determined by \cite{AR1982}], which give accurate fit to the experimental data from Millikan's oil-drop experiment.
In addition, we shall also include the data resulting from formula~\eqref{KW_emp_formula} on taking the constants $\mathfrak{a}$, $\mathfrak{b}$ and $\mathfrak{c}$ given in \eqref{AR1985_coeff} [determined by \cite{AR1985}] and in \eqref{HHF1995_coeff} [determined by \cite{HHF1995}], which give accurate fit to the data obtained from experiments performed with the (aforementioned) solid spheres suspended in air.
Although one may remonstrate that comparing the results on the drag force obtained in the present work, which is on a monatomic gas flow over a liquid droplet, with those from Millikan's oil-drop experiment, which was on oil drops falling in air, would not be fair due to the fact that air is not a monatomic gas, 
the viscosity of the monatomic gas in the present work is taken to be the same as that of air used in Millikan's oil-drop experiment---making the comparison sensible.
Furthermore, we are not aware of any theoretical/experimental work, which presents the data for monatomic gas flow over a liquid droplet. Therefore, it is reasonable to compare the results on the drag force obtained in the present work with the data from Millikan's oil-drop experiment.

In the present work, the drag force is computed analytically as follows. 
The drag force in the integral form can be written as \citep{Torrilhon2010}
\begin{align}
\label{int_stokes}
F = -2\pi\int_{0}^{\pi} \Vec{z}_{x}(\theta)\cdot(\boldsymbol{P}(r,\theta)\cdot \Vec{k})\sin{\theta}\,\mathrm{d}\theta,
\end{align}
where 
$\Vec{z}_{x}(\theta)=(\cos{\theta},-\sin{\theta},0)$, $\Vec{k}=(1,0,0)$ and $\boldsymbol{P}=p\boldsymbol{I}+\boldsymbol{\sigma}$ is the pressure tensor. The force on the surface of the liquid droplet is represented by $\boldsymbol{P}\cdot \Vec{k}$ and its scalar product with $\Vec{z}_{x}$ gives the component of the force in the direction of the flow.
Substituting ansatz \eqref{field_ansatz_gas} in \eqref{int_stokes}, we obtain
\begin{align}
    F=\frac{4\pi}{3}\left[-\mathbbm{p}(1)-\mathbbm{s}_{1}(1)+2\mathbbm{s}_{2}(1)\right].
\end{align}
This, on simplifying further, yields
\begin{align}
F=-2\pi C_{1} \mathrm{Kn},
\end{align}
where the constant $C_{1}$ is determined using the boundary conditions given in \eqref{BCON}.

\begin{figure}
    \centering
    \includegraphics[width=0.65\textwidth]{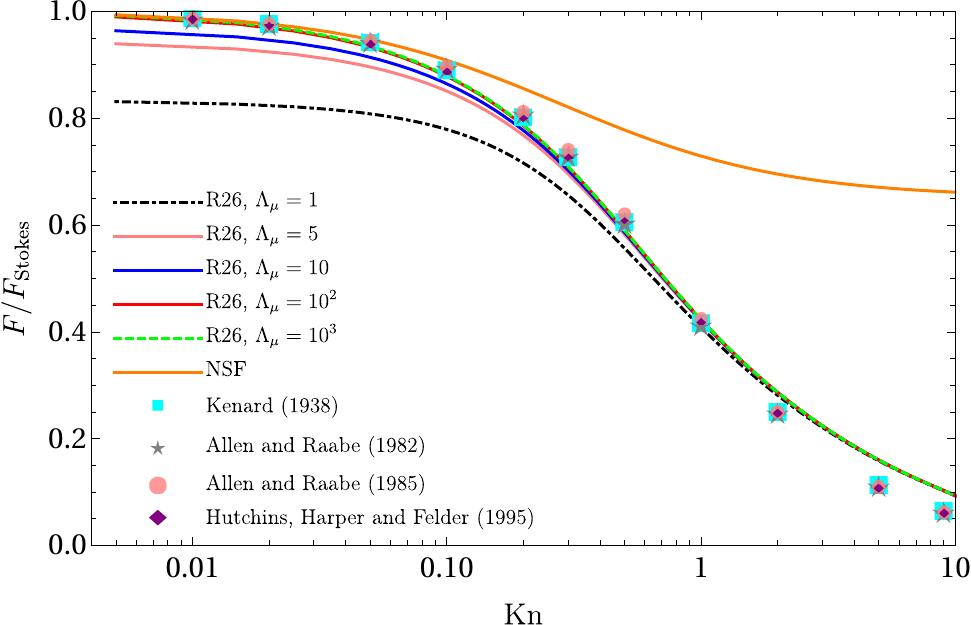} 
\caption{\label{Fig:Drag1}
Drag force (normalised with the Stokes drag $F_{\mathrm{Stokes}} =6\pi\,\mathrm{Kn}\,u_\infty$) on a liquid droplet from the linearised NSF and R26 theories as a function of the Knudsen number for a fixed thermal conductivity ratio $\Lambda_\kappa=100$. 
The square and star symbols denote the experimental data from Millikan's oil-drop experiment fitted by the empirical formulae of \cite{Kennard1938} [formula~\eqref{KW_emp_formula} with \eqref{Kennard_coeff}] and \cite{AR1982} [formula~\eqref{KW_emp_formula} with \eqref{AR1982_coeff}], respectively.
The disk and diamond symbols denote the data from experiments performed with solid spherical particles in air fitted by the empirical formulae of \cite{AR1985} [formula~\eqref{KW_emp_formula} with \eqref{AR1985_coeff}] and \cite{HHF1995} [formula~\eqref{KW_emp_formula} with \eqref{HHF1995_coeff}], respectively, and are included just for comparison.
}
\end{figure}
\renewcommand{\arraystretch}{2}
\begin{table}
\centering
\begin{tabular*}{\textwidth}{@{\extracolsep{\fill}} c c c c c c c @{}}
\multicolumn{1}{c}{\multirow{2}[3]{*}{\centering Knudsen number}} & \multicolumn{5}{c}{$F/F_{\mathrm{Stokes}}$ ($\Lambda_\kappa = 100$)} 
\\
\cmidrule{2-6}
& $\Lambda_\mu=1$ & $\Lambda_\mu=5$ & $\Lambda_\mu=10$ & $\Lambda_\mu=100$ & $\Lambda_\mu=1000$ 
\\
\midrule
0.01 & 0.826315 & 0.929546 & 0.952284 & 0.976313 & 0.978936 
\\ 
0.1 & 0.772654 & 0.843830 & 0.857370 & 0.870973 & 0.872416 
\\ 
0.5 & 0.544453 & 0.570468 & 0.574446 & 0.578199 & 0.578583 
\\ 
1.0 & 0.396739 & 0.408531 & 0.410211 & 0.411767 & 0.411926 
\\   
5.0 & 0.154088 & 0.155314 & 0.155473 & 0.155617 & 0.155631 
\\ 
10.0 & 0.091075 & 0.091425 & 0.091470 & 0.091510 & 0.091514 
\\
\bottomrule
\end{tabular*}
\caption{\label{Table-1}Drag force normalised with the Stokes drag, $F_{\mathrm{Stokes}}=6\pi\,\mathrm{Kn}\,u_\infty$, for different values of the Knudsen number and viscosity ratios at a fixed thermal conductivity ratio $\Lambda_\kappa=100$.
}
\end{table}

Figure \ref{Fig:Drag1} exhibits the drag force normalised with the Stokes drag $F_\mathrm{Stokes}$ plotted over the Knudsen number. 
The figure illustrates the drag force estimated by the LR26 equations for a fixed value of the thermal conductivity ratio $\Lambda_\kappa = 100$ and for different values of the viscosity ratio $\Lambda_\mu$. 
In addition, the explicit values of the normalised drag force for the same value of the thermal conductivity ratio $\Lambda_\kappa = 100$ and for different values of the Knudsen number and viscosity ratio are also given in table~\ref{Table-1}.

It is clear from figure~\ref{Fig:Drag1} and table~\ref{Table-1} that as the Knudsen number increases, the drag force decreases and approaches zero (in general, except for that from the NSF equations with which the drag force seemingly attains a nonzero positive value) as the Knudsen number approaches infinity. 
On the other hand, it can be seen from figure~\ref{Fig:Drag1} and table~\ref{Table-1} that with the increase in the viscosity ratio $ \Lambda_\mu$, the drag force approaches unity for smaller values of the Knudsen number.
Since the viscosity ratio of an oil to air is somewhere in between $10^2$ to $10^3$ (depending on the oil), the figure clearly shows that the drag force predicted by the LR26 equations for higher values of the viscosity ratios is in an excellent agreement with the experimental data from Millikan's oil-drop experiment for the Knudsen number values as big as $1$.
It is also worthwhile noting that the oil used in Millikan's experiment not only has a very high viscosity but also has a high thermal conductivity; therefore the results on the drag force from Millikan's experiment (and also those in the present work for high viscosity ratios) are very close to the drag force on a solid sphere (of the same size as the size of the liquid droplet)---as also manifested by the propinquity of symbols in figure~\ref{Fig:Drag1}.


\begin{figure}
    \centering
     \includegraphics[width=0.65\textwidth]{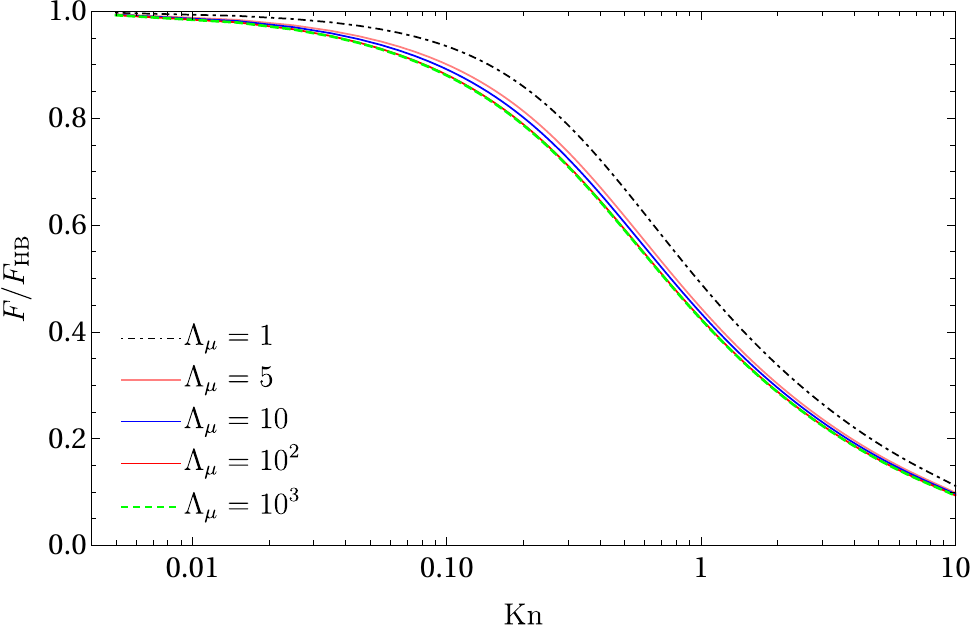} 
    \caption{\label{Fig:Drag2}
Drag force on the liquid droplet normalised with the drag force given by \eqref{HDrag}. 
The thermal conductivity ratio is $\Lambda_\kappa=100$.}
\end{figure}

\renewcommand{\arraystretch}{2}
\begin{table}
\centering
\begin{tabular*}{\textwidth}{@{\extracolsep{\fill}} c c c c c c c @{}}
\multicolumn{1}{c}{\multirow{2}[3]{*}{\centering Knudsen number}} & \multicolumn{5}{c}{$F/F_{\mathrm{HB}}$ ($\Lambda_\kappa = 100$)} 
\\
\cmidrule{2-6}
& $\Lambda_\mu=1$ & $\Lambda_\mu=5$ & $\Lambda_\mu=10$ & $\Lambda_\mu=100$ & $\Lambda_\mu=1000$ 
\\
\midrule
0.01 & 0.991579 & 0.984226 & 0.982044 & 0.979546 & 0.979262 
\\ 
0.1 & 0.927186 & 0.893468 & 0.884164 & 0.873857 & 0.872707 
\\ 
0.5 & 0.653344 & 0.604026 & 0.592398 & 0.580114 & 0.578777 
\\ 
1.0 & 0.476088 & 0.432563 & 0.423030 & 0.413131 & 0.412063 
\\ 
5.0 & 0.184906 & 0.164450 & 0.160332 & 0.156133 & 0.155684 
\\ 
10.0 & 0.109291 & 0.096803 & 0.094328 & 0.091813 & 0.091545 
\\
\bottomrule
\end{tabular*}
\caption{\label{Table-2}Drag force normalised with the drag force given by \eqref{HDrag} for different values of the Knudsen number and viscosity ratio with the thermal conductivity ratio being fixed at 100.}
\end{table}

Figure \ref{Fig:Drag2} depicts the drag force computed with the LR26 equations and normalised with the drag force given by \eqref{HDrag} again for a fixed value of the thermal conductivity ratio $\Lambda_\kappa = 100$ and for different values of the viscosity ratio $\Lambda_\mu$. 
In addition, the explicit values of the normalised drag force for the same value of the thermal conductivity ratio $\Lambda_\kappa = 100$ and for different values of the Knudsen number and viscosity ratio are also given in table~\ref{Table-2}.
We also compare the drag force computed analytically with the LR26 equations in the present work with the drag force computed with an explicit formula given by \cite{Happel1965}, which---in our notations---reads 
\begin{align}
\label{HDrag}
F_{\mathrm{HB}}=F_{\mathrm{Stokes}} \times \frac{1 + \dfrac{2}{3 \Lambda_\mu}}{1 + \dfrac{1}{\Lambda_\mu}}.
\end{align}
%
From figure~\ref{Fig:Drag2} and table~\ref{Table-2}, the drag force normalised with the drag force $F_{\mathrm{HB}}$ [given by \eqref{HDrag}] for very small Knudsen numbers is very close to unity for all values of the viscosity ratio $\Lambda_\mu$, which is not the case when the drag force normalised with the Stokes drag (cf.~figures \ref{Fig:Drag1} and \ref{Fig:Drag2} for small Knudsen numbers). 
However, for large Knudsen numbers, the drag force scaled with $F_{\mathrm{HB}}$ is more or less same as the drag force scaled with $F_{\mathrm{Stokes}}$.
Indeed, from \eqref{HDrag}, as $\Lambda_\mu \to \infty$, $F_{\mathrm{HB}} \to F_{\mathrm{Stokes}}$; therefore for large viscosity ratios the scaling with $F_{\mathrm{HB}}$ or $F_{\mathrm{Stokes}}$ does not matter. 
Hence the dotted green lines and solid red lines in figures \ref{Fig:Drag1} and \ref{Fig:Drag2} coincide with each other.
%
\subsection{\label{Subsec:Tq}Flow fields: temperature and heat flux}
\begin{figure}
    \centering
    \includegraphics[width = 0.48\textwidth]{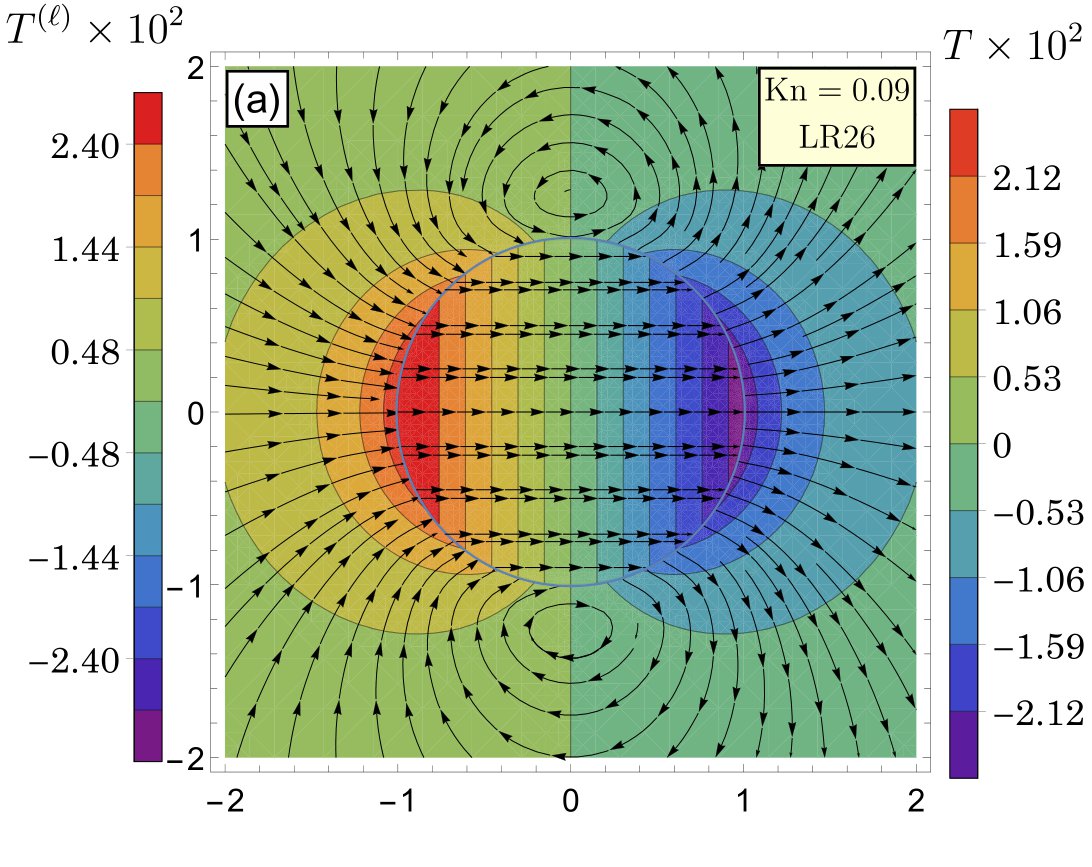}\hfill
     \includegraphics[width = 0.48\textwidth]{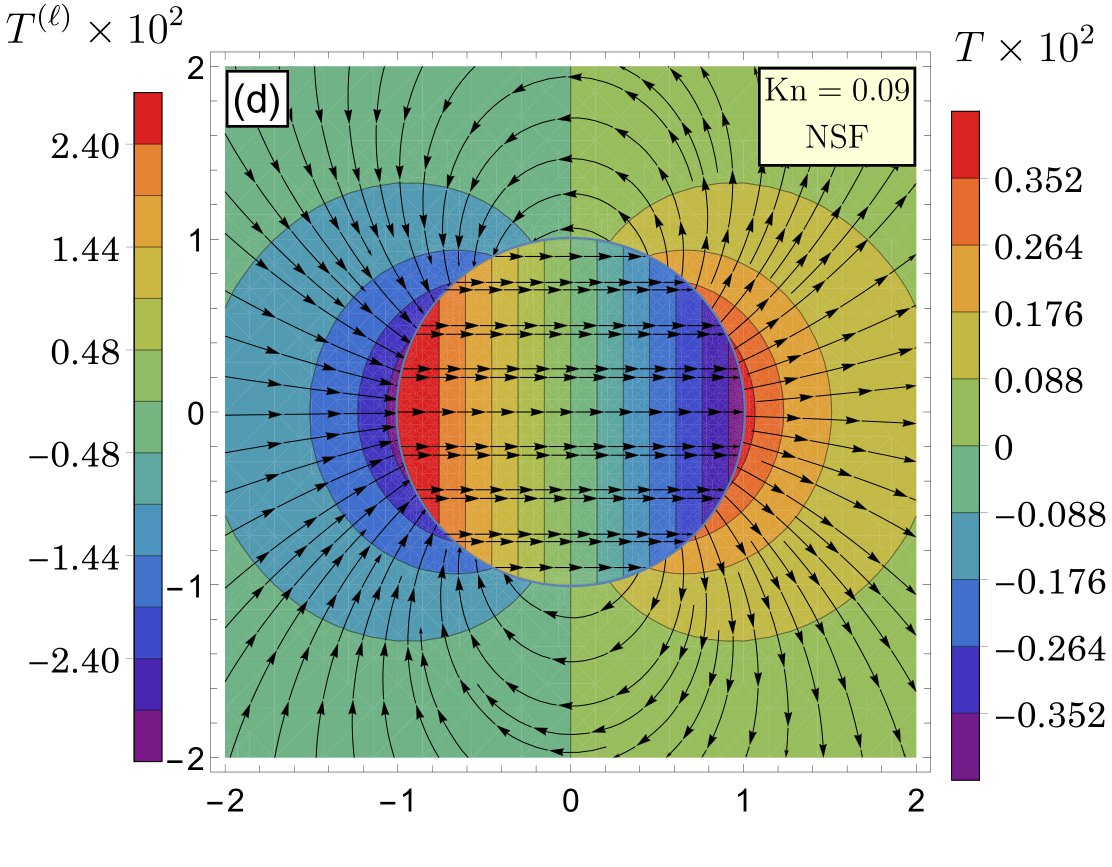}
     \includegraphics[width = 0.48\textwidth]{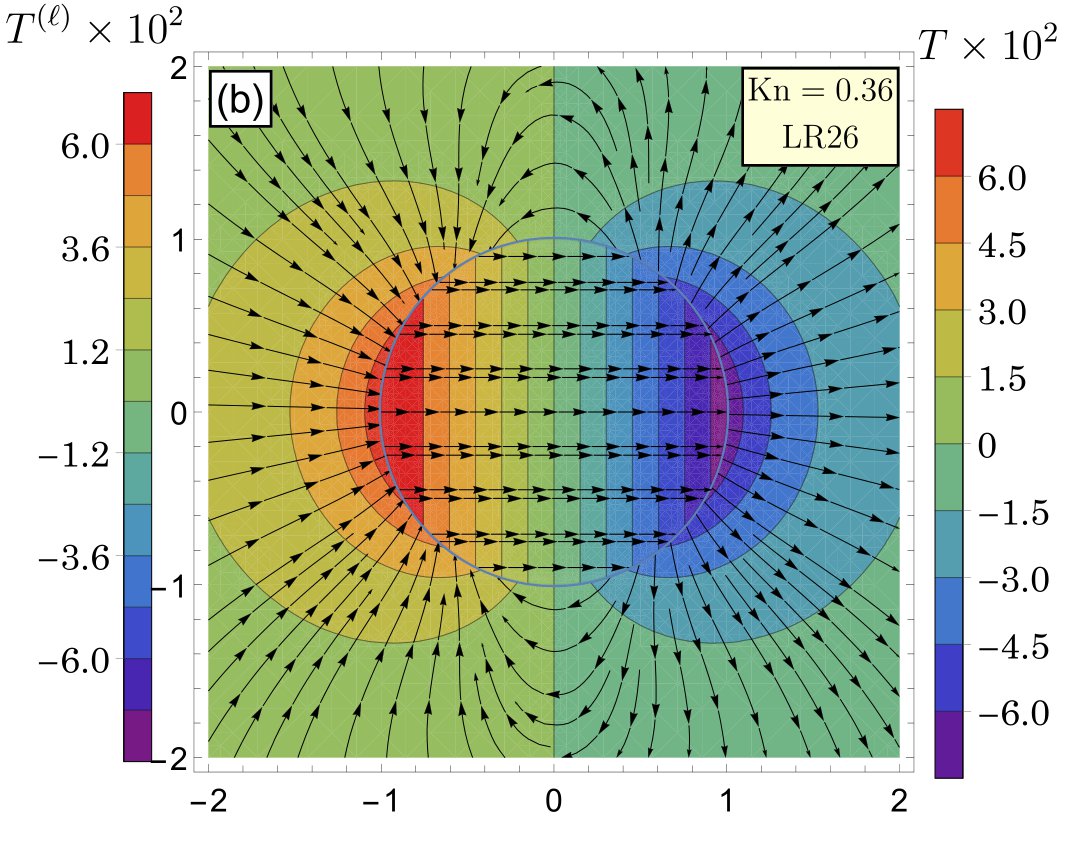}\hfill
     \includegraphics[width = 0.48\textwidth]{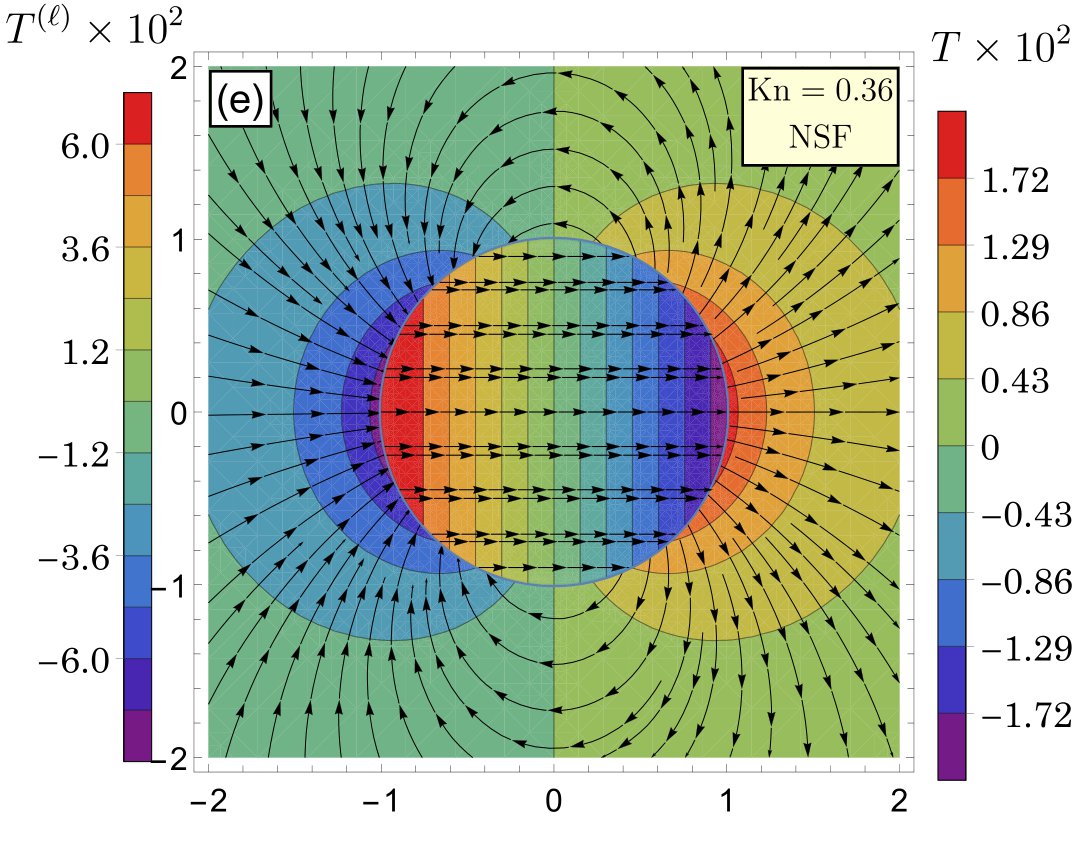}
     \includegraphics[width = 0.48\textwidth]{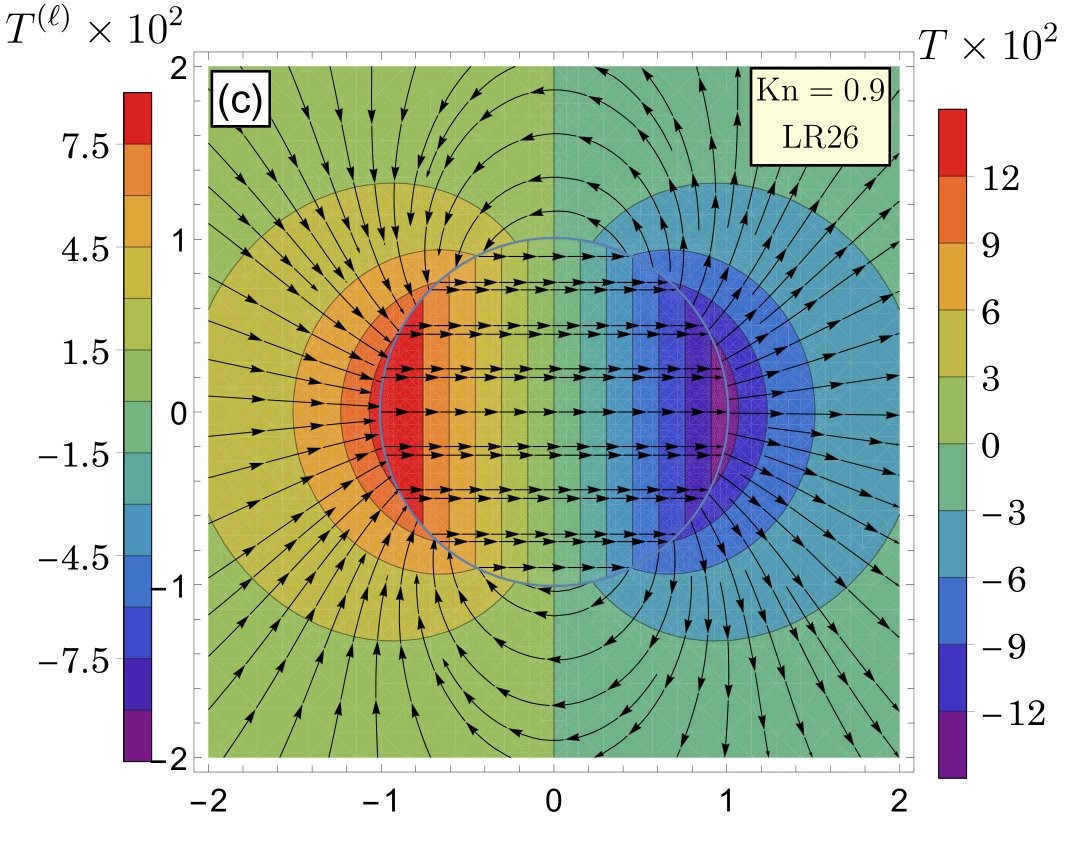}\hfill
     \includegraphics[width = 0.48\textwidth]{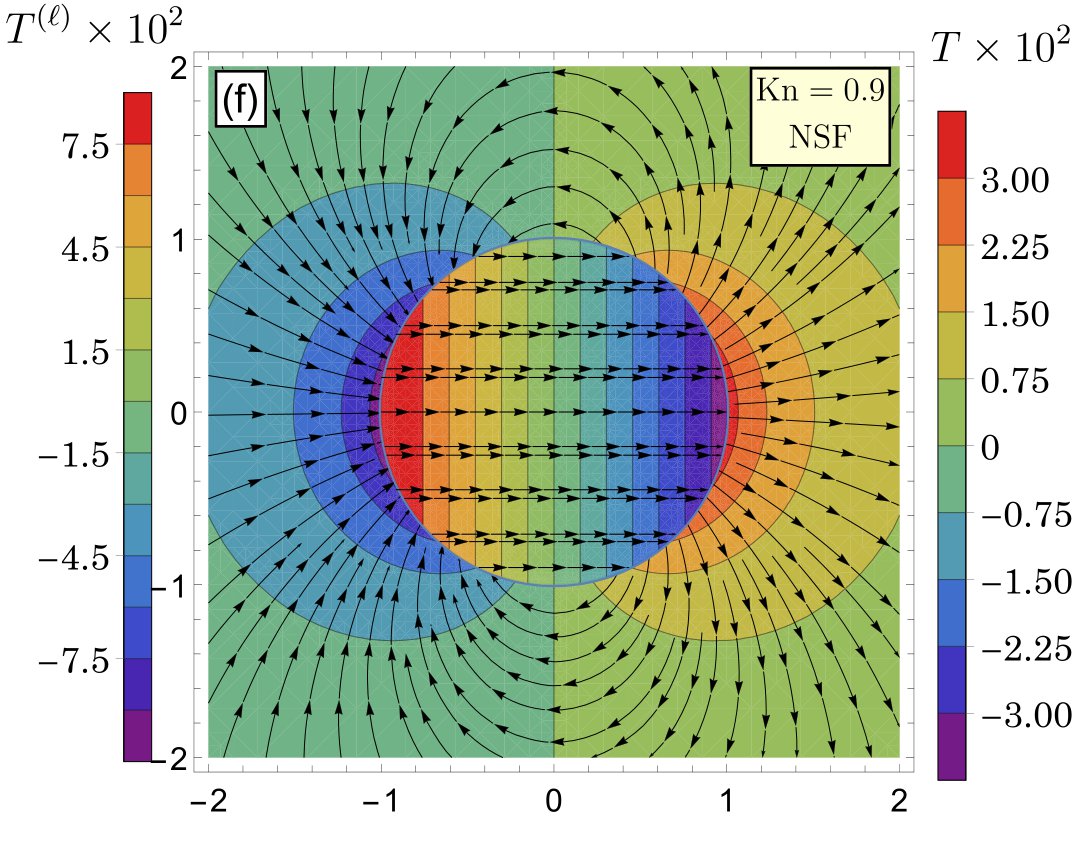}
\caption{\label{fig:tempcontLambdakappa1}Heat flux lines plotted on top of the temperature contours for the viscosity ratio $\Lambda_\mu=100$ and for different values of the Knudsen number: $\mathrm{Kn} = 0.09$ (top row), $\mathrm{Kn} = 0.36$ (middle row) and $\mathrm{Kn} = 0.9$ (bottom row).
The results for the liquid phase (internal flow) have been computed with the linear NSF equations in all the cases while those for the gas phase (external flow) have been computed with the LR26 equations for the panels in the left column and with the linear NSF equations for the panels in the right column. The thermal conductivity ratio is $\Lambda_\kappa=1$.}
\end{figure}%
\begin{figure}
     \centering
     \includegraphics[width = 0.48\textwidth]{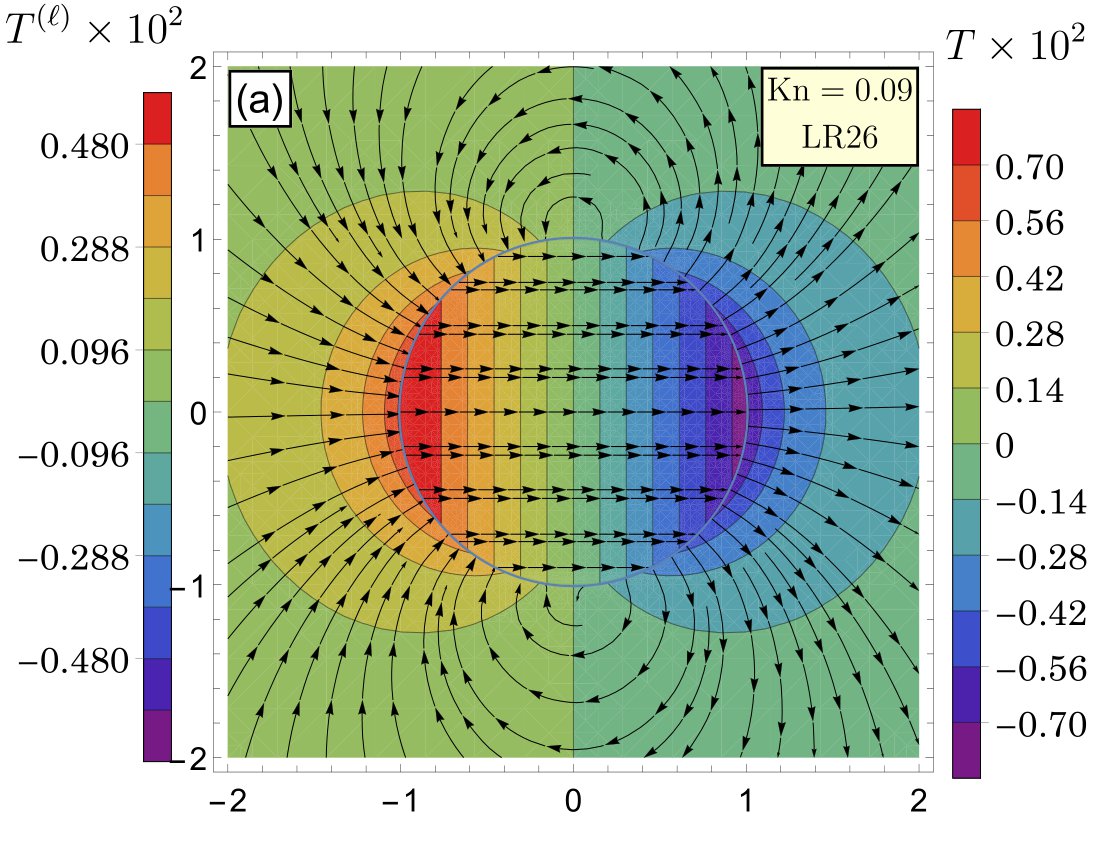}\hfill
      \includegraphics[width = 0.48\textwidth]{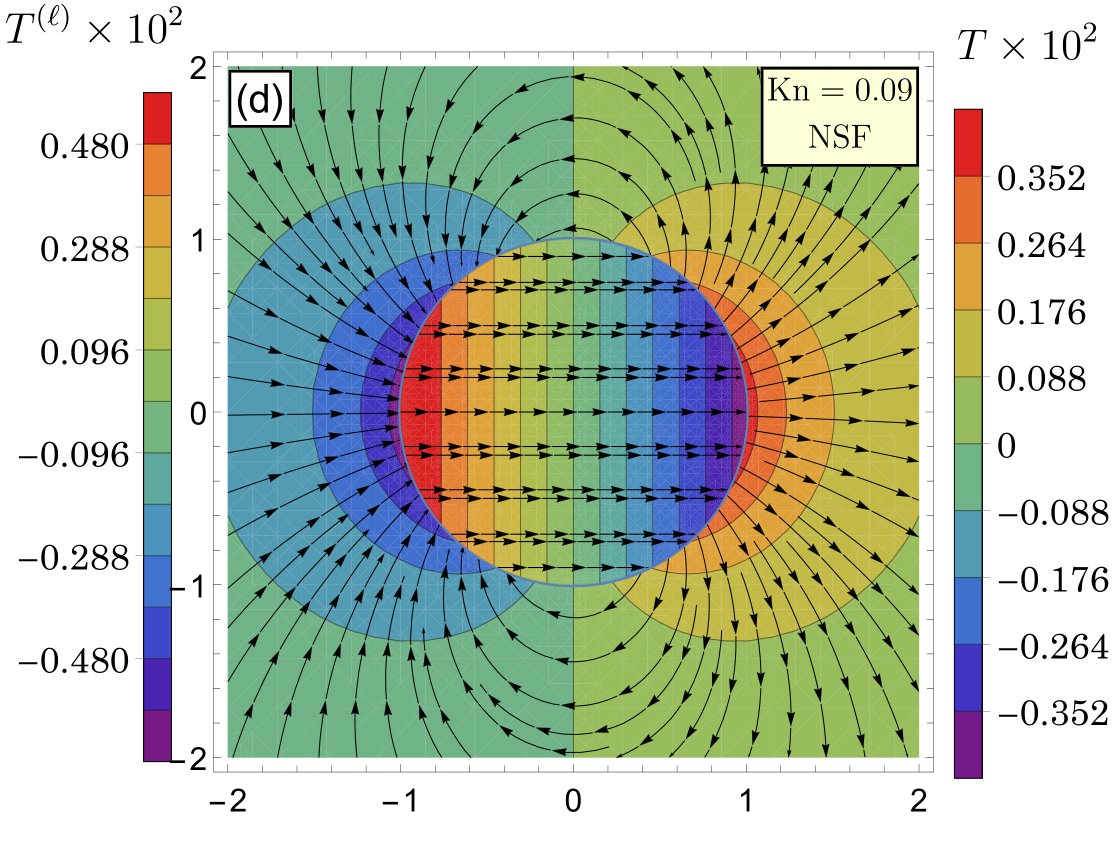}
      \includegraphics[width = 0.48\textwidth]{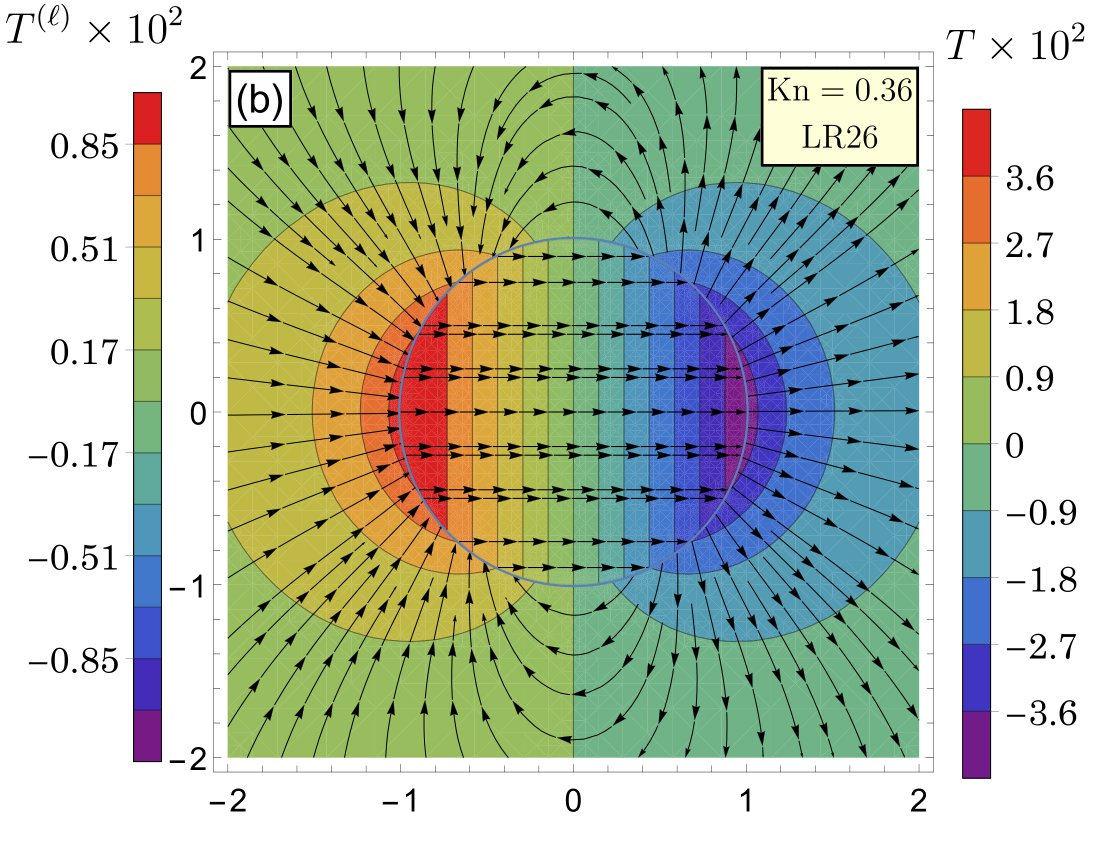}\hfill
      \includegraphics[width = 0.48\textwidth]{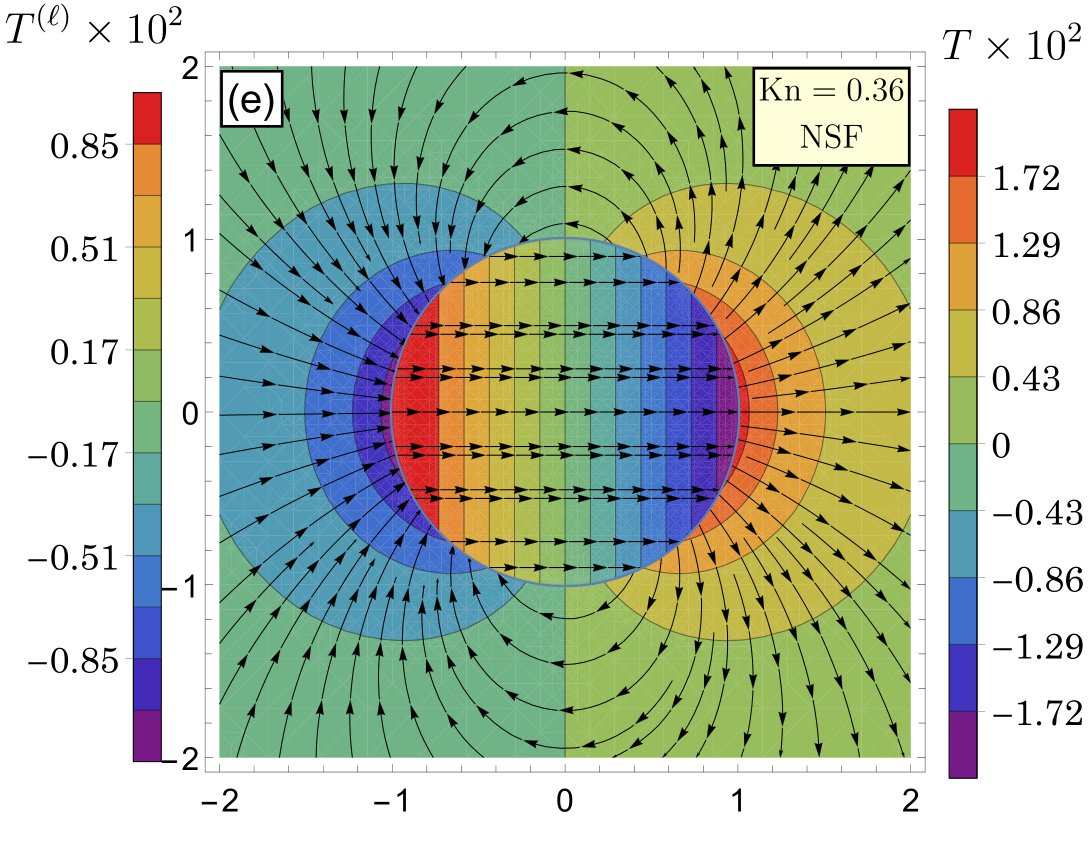}
      \includegraphics[width = 0.48\textwidth]{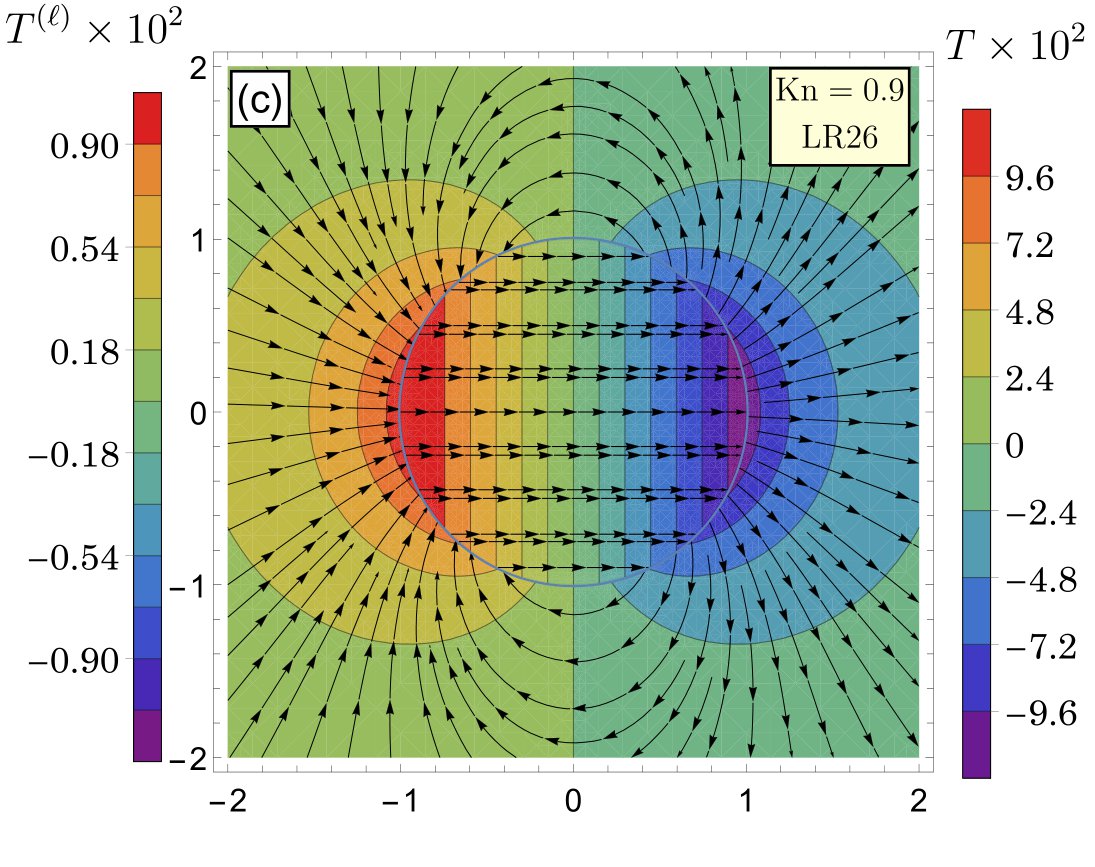}\hfill
      \includegraphics[width = 0.48\textwidth]{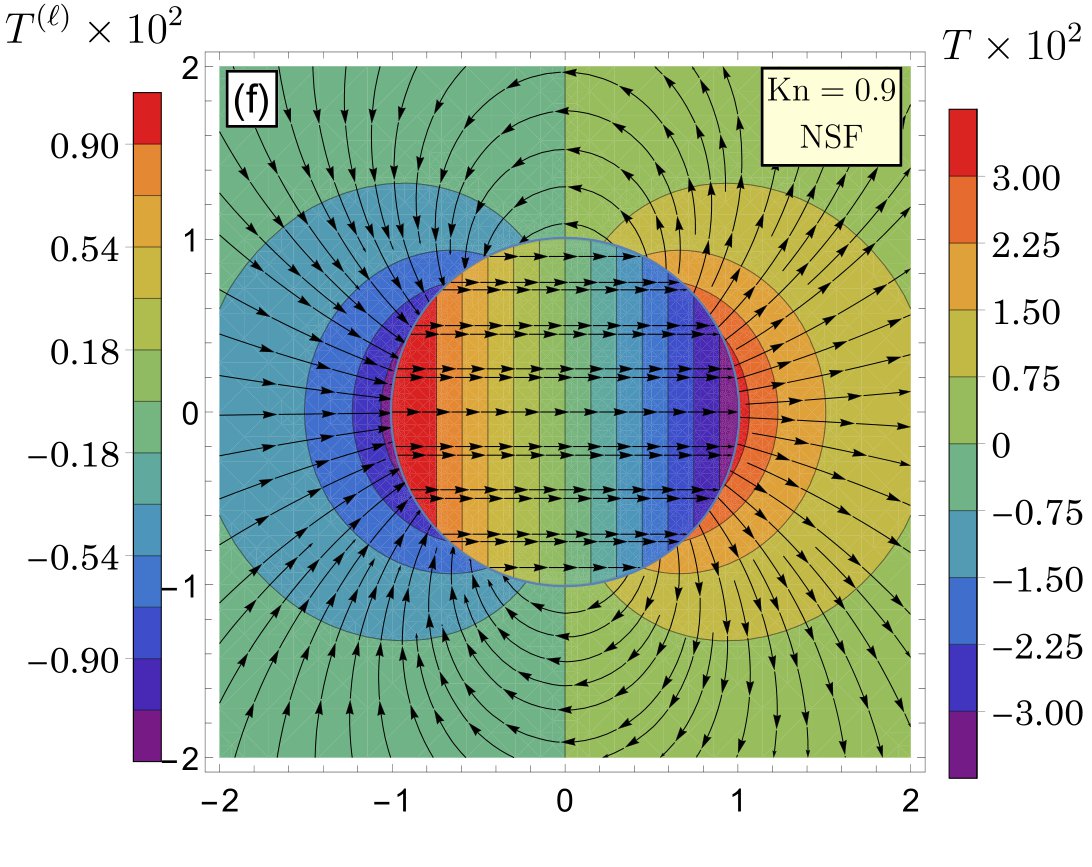}
 \caption{\label{fig:tempcontLambdakappa10}Same as figure \ref{fig:tempcontLambdakappa1} but for the thermal conductivity ratio $\Lambda_\kappa=10$.}
\end{figure}
\begin{figure}
     \centering
     \includegraphics[width = 0.48\textwidth]{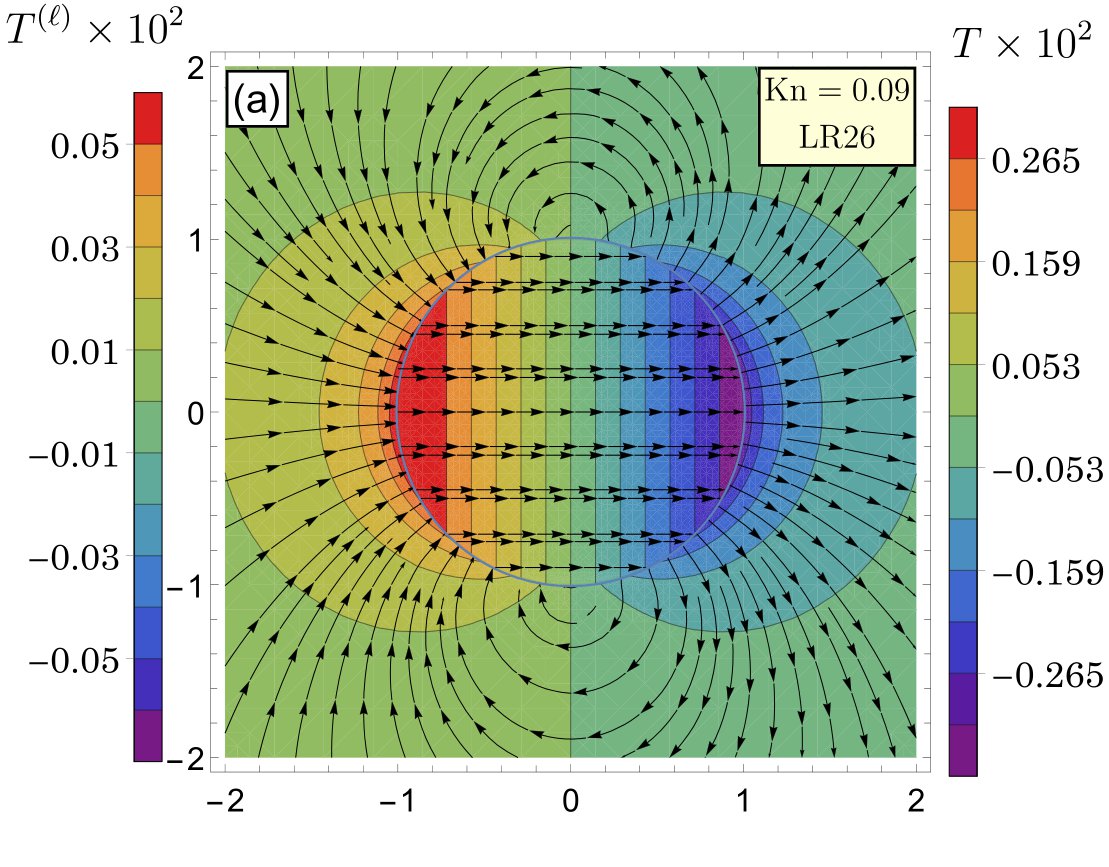}\hfill
      \includegraphics[width = 0.48\textwidth]{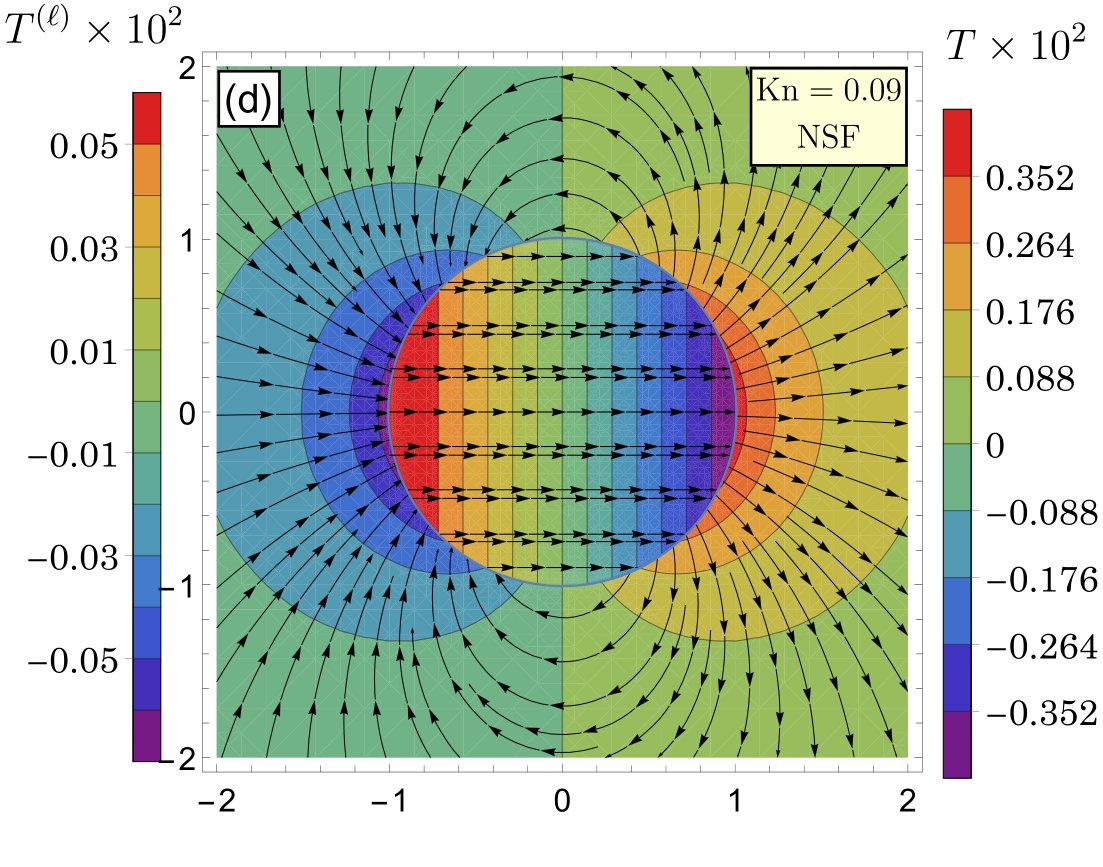}
      \includegraphics[width = 0.48\textwidth]{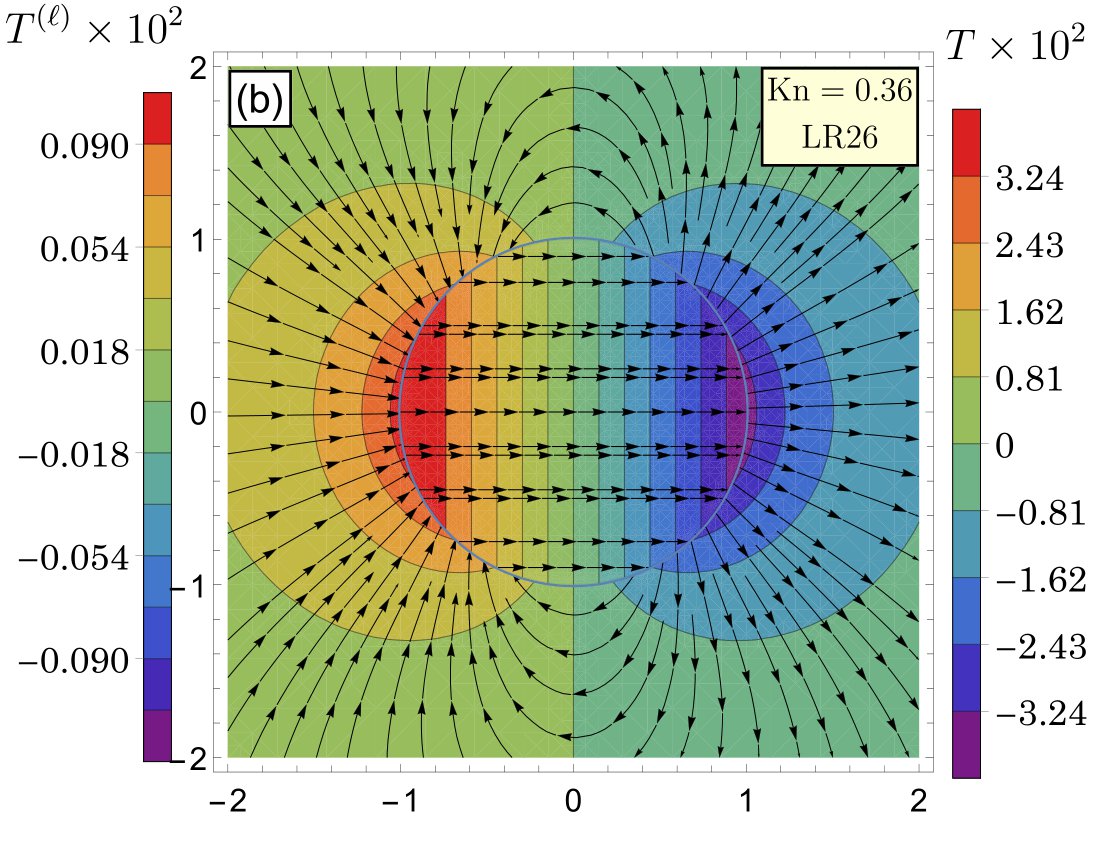}\hfill
      \includegraphics[width = 0.48\textwidth]{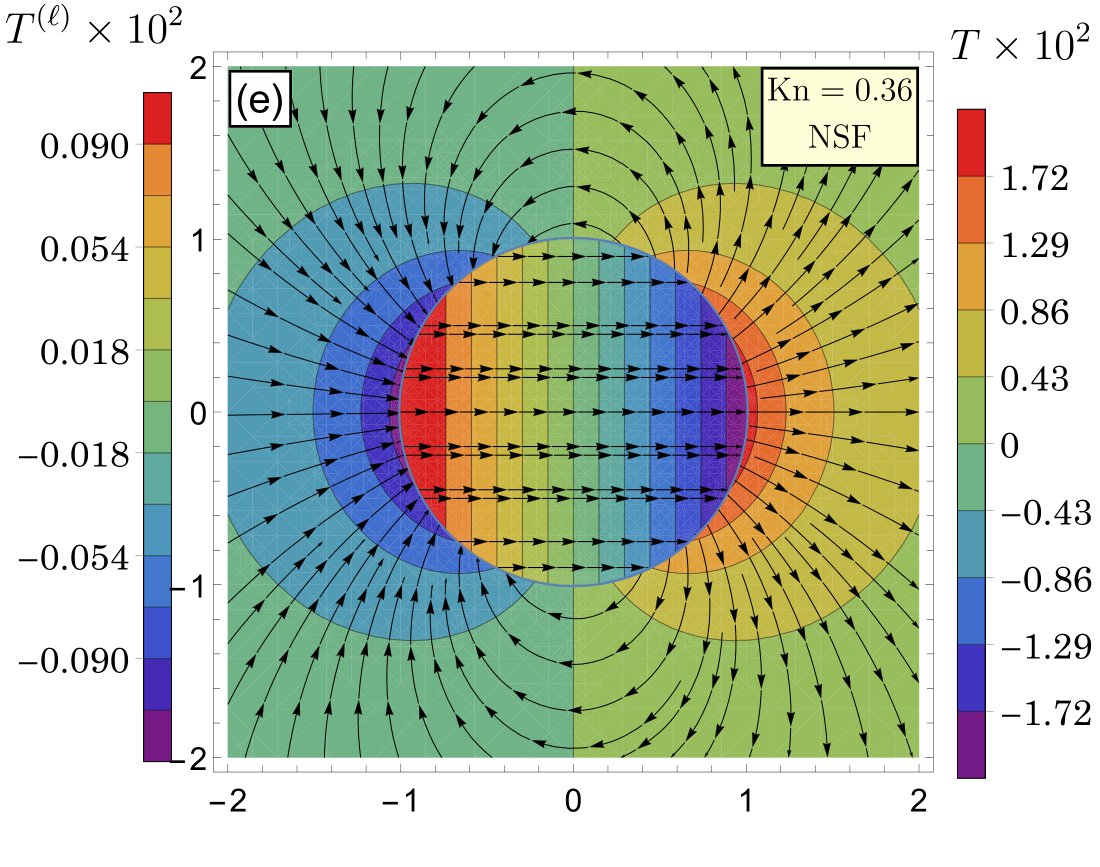}
      \includegraphics[width = 0.48\textwidth]{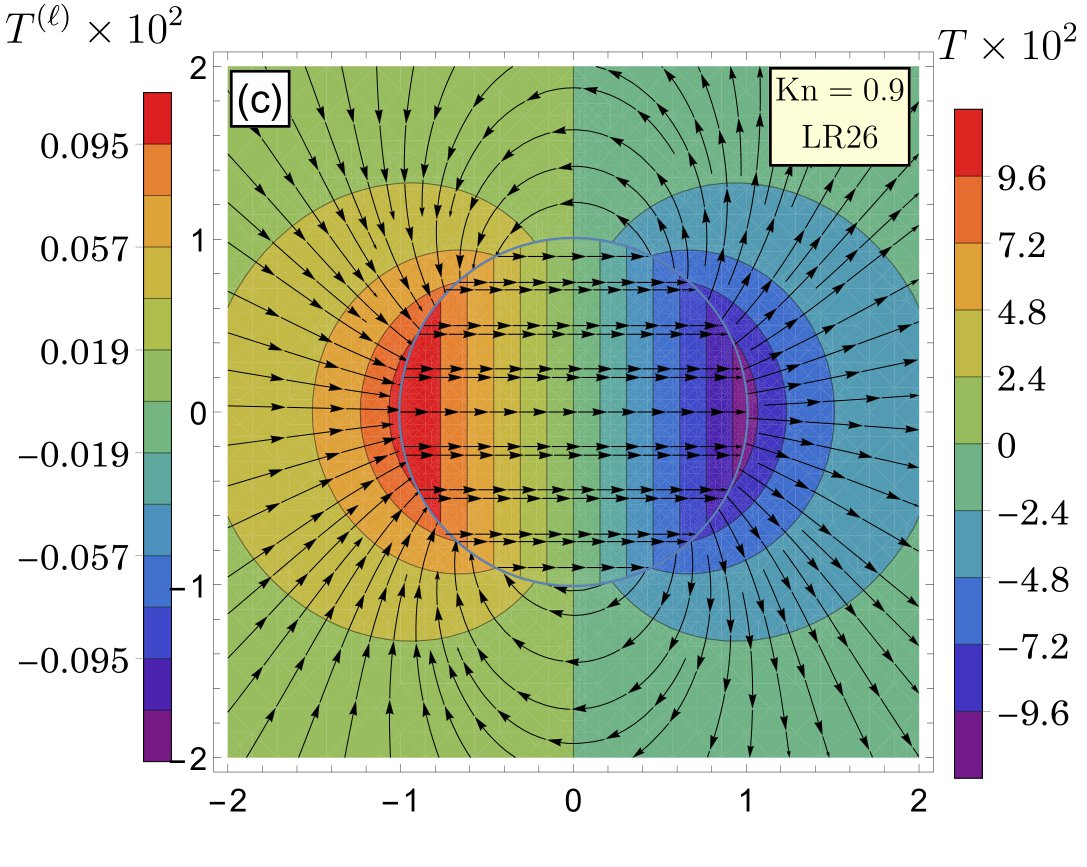}\hfill
      \includegraphics[width = 0.48\textwidth]{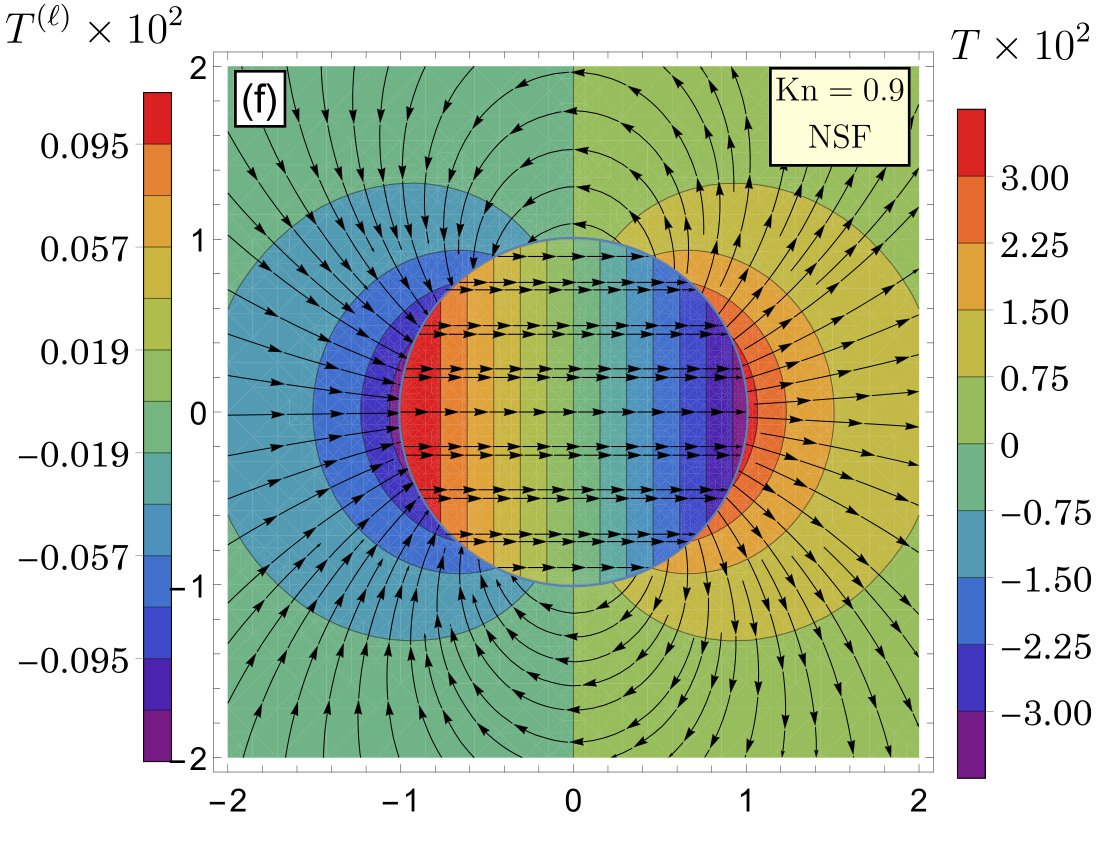}
 \caption{\label{fig:tempcontLambdakappa100}Same as figure \ref{fig:tempcontLambdakappa1} but for the thermal conductivity ratio $\Lambda_\kappa=100$.}
\end{figure}
Figures \ref{fig:tempcontLambdakappa1}, \ref{fig:tempcontLambdakappa10} and \ref{fig:tempcontLambdakappa100} illustrate the heat flux lines and temperature contours in the $\hat{y}=0$ plane for a fixed viscosity ratio $\Lambda_\mu = 100$ and for
the thermal conductivity ratios $\Lambda_\kappa=1$, $10$ and $100$, respectively.
The panels in top, middle and bottom rows of each figure display the results for the Knudsen numbers $\mathrm{Kn}=0.09, 0.36$ and $0.9$, respectively. 
The panels in the left columns of figures \ref{fig:tempcontLambdakappa1}, \ref{fig:tempcontLambdakappa10} and \ref{fig:tempcontLambdakappa100} illustrate the results obtained with the LR26 equations for the rarefied gas flow outside the droplet and with the linear NSF equations for the liquid inside the droplet while those on the right columns of figures \ref{fig:tempcontLambdakappa1}, \ref{fig:tempcontLambdakappa10} and \ref{fig:tempcontLambdakappa100} are obtained with the NSF equations for both the liquid inside the droplet and gas flow outside the droplet. 
Comparing the respective panels on the left and right columns of figures \ref{fig:tempcontLambdakappa1}, \ref{fig:tempcontLambdakappa10} and \ref{fig:tempcontLambdakappa100}, 
the R26 equations show that the heat in the external flow transfers from the cold region (back side of the liquid droplet) to the hot region (front side of the liquid droplet), which is a non-Fourier effect and for the liquid droplet inside as predicted by the NSF equations it is just the other way round, i.e.~the heat flows from the hot region to cold region inside the liquid droplet. 
On the other hand, panels in the right column of figures \ref{fig:tempcontLambdakappa1}, \ref{fig:tempcontLambdakappa10} and \ref{fig:tempcontLambdakappa100} show that the NSF equations cannot capture the non-Fourier heat transfer in the external (rarefied gas) flow. 
In fact, the R26 equations for the external flow give high (low) temperature on the front (back) side of the liquid droplet while the NSF equations for the external flow  predict the opposite of this (cf.~the corresponding panels in the left and right columns of figures \ref{fig:tempcontLambdakappa1}, \ref{fig:tempcontLambdakappa10} and \ref{fig:tempcontLambdakappa100}). 
In addition, comparing figures \ref{fig:tempcontLambdakappa1}, \ref{fig:tempcontLambdakappa10} and \ref{fig:tempcontLambdakappa100} with those obtained by \cite{RGST2021}, the difference in the values of temperature profiles is evident, which is caused due to the effect of the internal circulation inside the liquid droplet.

\begin{figure}
     \centering
     \includegraphics[width = 0.48\textwidth]{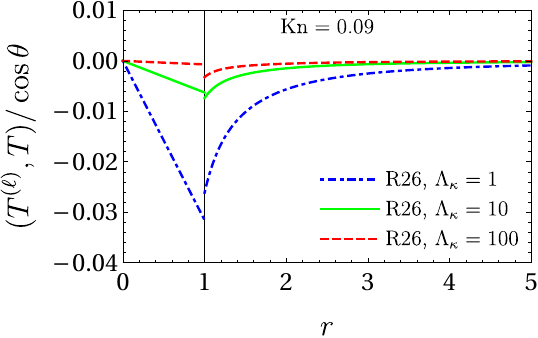}\hfill
      \includegraphics[width = 0.48\textwidth]{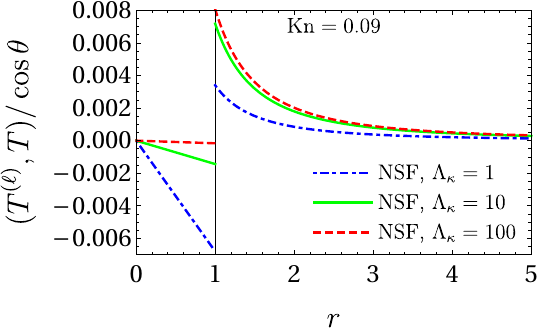}
\\
      \includegraphics[width = 0.48\textwidth]{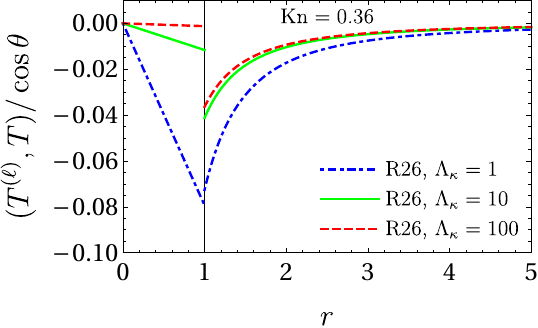}\hfill
      \includegraphics[width = 0.48\textwidth]{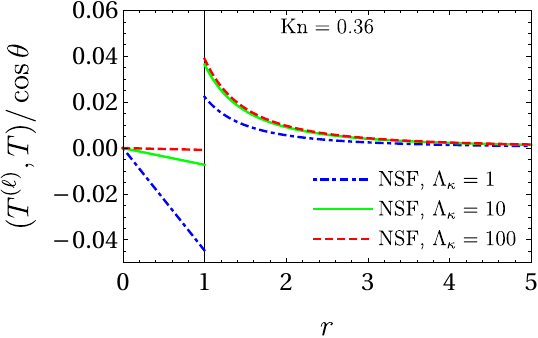}
\\
      \includegraphics[width = 0.48\textwidth]{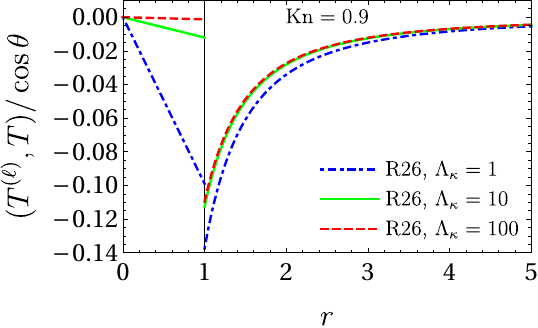}\hfill
      \includegraphics[width = 0.48\textwidth]{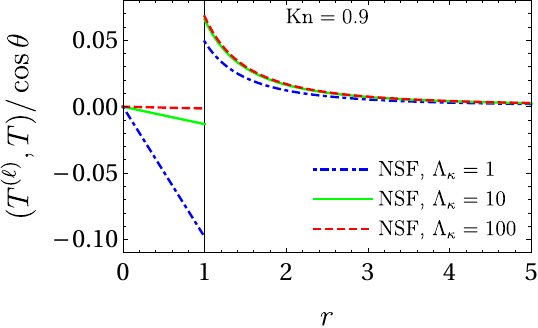}
\caption{\label{Fig:Temp}
Dimensionless deviations in the temperature (scaled with $\cos{\theta}$) as a function of position $r$ for a fixed viscosity ratio $\Lambda_\mu=100$ and for different values of the Knudsen number: $\mathrm{Kn} = 0.09$ (top row), $\mathrm{Kn} = 0.36$ (middle row) and $\mathrm{Kn} = 0.9$ (bottom row).
The vertical black line at $r=1$ demarcates the interface between the liquid and gas.
The results for the liquid phase (internal flow) have been computed with the NSF equations in all the cases while those for the gas phase (external flow) have been computed with the LR26 equations for the panels in the left column and with the NSF equations for the panels in the right column.
}
\end{figure}

\begin{figure}
     \centering
     \includegraphics[width = 0.48\textwidth]{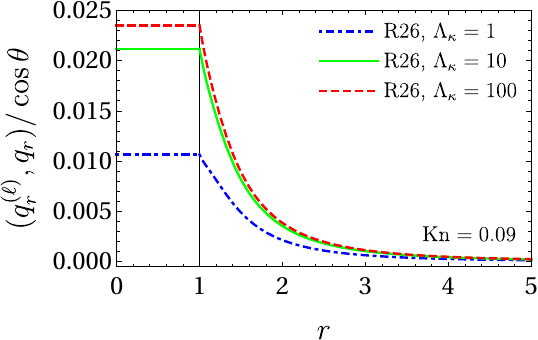}\hfill
      \includegraphics[width = 0.48\textwidth]{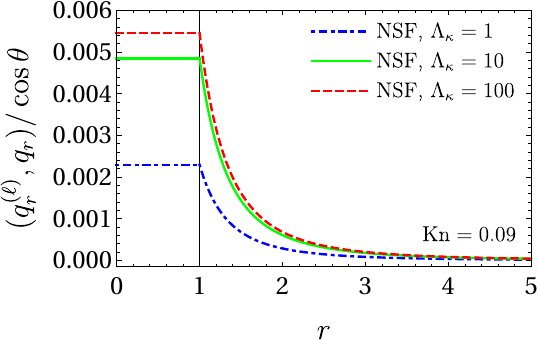}
\\
      \includegraphics[width = 0.48\textwidth]{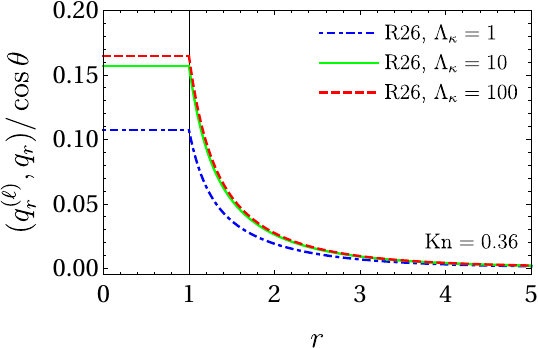}\hfill
      \includegraphics[width = 0.48\textwidth]{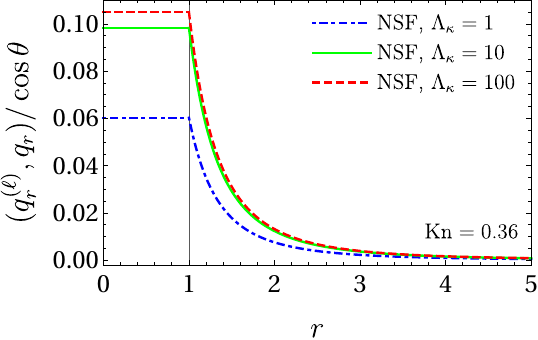}
\\
      \includegraphics[width = 0.48\textwidth]{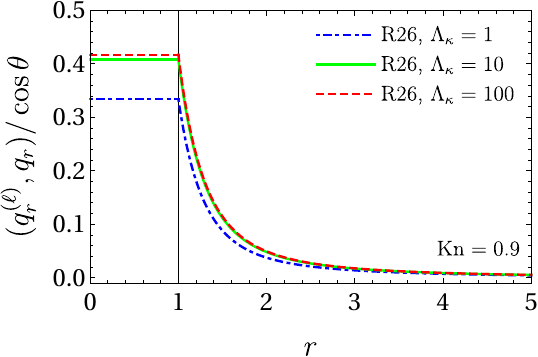}\hfill
      \includegraphics[width = 0.48\textwidth]{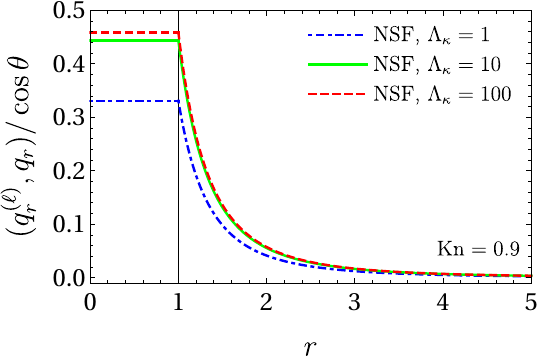}
\caption{\label{Fig:Hflux}
Dimensionless radial heat flux (scaled with $\cos{\theta}$) as a function of position $r$ for a fixed viscosity ratio $\Lambda_\mu=100$ and for different values of the Knudsen number: $\mathrm{Kn} = 0.09$ (top row), $\mathrm{Kn} = 0.36$ (middle row) and $\mathrm{Kn} = 0.9$ (bottom row).
The vertical black line at $r=1$ demarcates the interface between the liquid and gas.
The results for the liquid phase (internal flow) have been computed with the NSF equations in all the cases while those for the gas phase (external flow) have been computed with the LR26 equations for the panels in the left column and with the NSF equations for the panels in the right column.
}
\end{figure}

In order to get an insight of the temperature and heat flux at a given position, we plot the temperature and radial component of the heat flux (both divided by $\cos{\theta}$ to understand the results for any angle $\theta$) with respect to the position $r$ in figures \ref{Fig:Temp} and \ref{Fig:Hflux}, respectively, again  for a fixed viscosity ratio $\Lambda_\mu = 100$ and for the thermal conductivity ratios $\Lambda_\kappa=1$, $10$ and $100$.
The panels in top, middle and bottom rows in both figures display the results for the Knudsen numbers $\mathrm{Kn}=0.09, 0.36$ and $0.9$, respectively. 
The panels in the left column of each of figures \ref{Fig:Temp} and \ref{Fig:Hflux} illustrate the results obtained with the LR26 equations for the rarefied gas flow outside the droplet and with the linear NSF equations for the liquid inside the droplet while those in the right column of each of these figures display the results obtained with the linear NSF equations for both liquid inside the droplet and gas outside the droplet.
The droplet interface has been demarcated by a vertical black line  at $r=1$ in all panels of figures \ref{Fig:Temp} and \ref{Fig:Hflux}.
For the liquid phase, the analytic solution for the temperature and radial component of the heat flux can actually be written quite easily from \eqref{closure_ql}$_1$, \eqref{field_ansatz_liquid}$_4$ and \eqref{sol_l}$_3$, and they read
\begin{align}
T^{(\ell)} = b_3 r \cos{\theta}
\quad\text{and}\quad
q_r^{(\ell)} =- \frac{5}{2} \Lambda_\kappa \frac{\mathrm{Kn}}{\mathrm{Pr}} b_3 \cos{\theta}.
\end{align}
Therefore, $T^{(\ell)}/\cos{\theta}$ for the liquid phase is a linear function of $r$ while $q_r^{(\ell)}/\cos{\theta}$ for the liquid phase at a given Knudsen number and at a given thermal conductivity ratio is just a constant.
Consequently, the curves on the left of the vertical black lines in each panel of figure \ref{Fig:Temp} are straight lines and the curves on the left of the vertical black lines in each panel of figure \ref{Fig:Hflux} are horizontal lines.

The temperature contours depicted in figures \ref{fig:tempcontLambdakappa1}, \ref{fig:tempcontLambdakappa10} and \ref{fig:tempcontLambdakappa100} actually correspond to temperature profiles exhibited in figure \ref{Fig:Temp} by red, black and green lines, respectively.
To see this, let us, for example, fix $\theta = 0$ and let us focus on figure \ref{fig:tempcontLambdakappa1} and the red lines in figure \ref{Fig:Temp} (figures \ref{fig:tempcontLambdakappa10} and \ref{fig:tempcontLambdakappa100} and the lines of other colours in figure \ref{Fig:Temp} can be understood analogously).
From (the red lines in) figure \ref{Fig:Temp}, the temperature of the liquid droplet decreases when moving from its centre towards its interface. However, on comparing the left and right panels in figure \ref{Fig:Temp}, one finds that the temperature of the external gas predicted by the LR26 equations (left panels) increases on moving away from the interface whereas that predicted by the linear NSF equations (left panels) decreases on moving away from the interface.
These are in agreement with the temperature contours portrayed in figure \ref{fig:tempcontLambdakappa1} (notice the contours for $\theta=0$ when moving away from the centre of the droplet).
Furthermore, each panel of figure \ref{Fig:Temp} clearly shows a nonzero temperature jump $\mathcal{T} = T - T^{(\ell)}$ at the interface.
From the panels on the right column of figure \ref{Fig:Temp}, the temperature of the liquid (gas) predicted by the linear NSF equations is minimum (maximum and positive) at the interface, and hence the temperature jump $\mathcal{T} = T - T^{(\ell)}$ predicted by the NSF equations is always positive. 
This is not the case when the LR26 equations are used for the external flow (see panels on the left column of figure \ref{Fig:Temp}).
In fact, when the LR26 equations are used for the external flow, the temperature jump $\mathcal{T} = T - T^{(\ell)}$ is negative in most of the cases but positive in some cases.

\begin{figure}
     \centering
     \includegraphics[width = 0.48\textwidth]{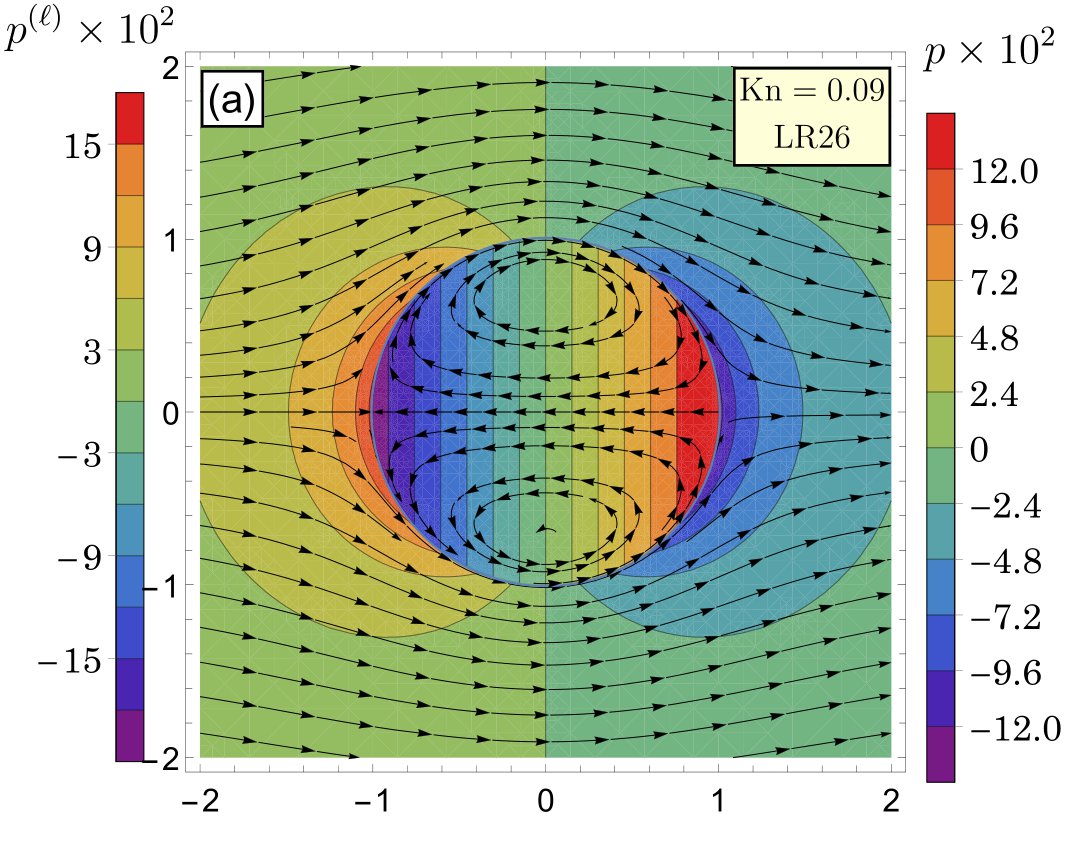}\hfill
      \includegraphics[width = 0.48\textwidth]{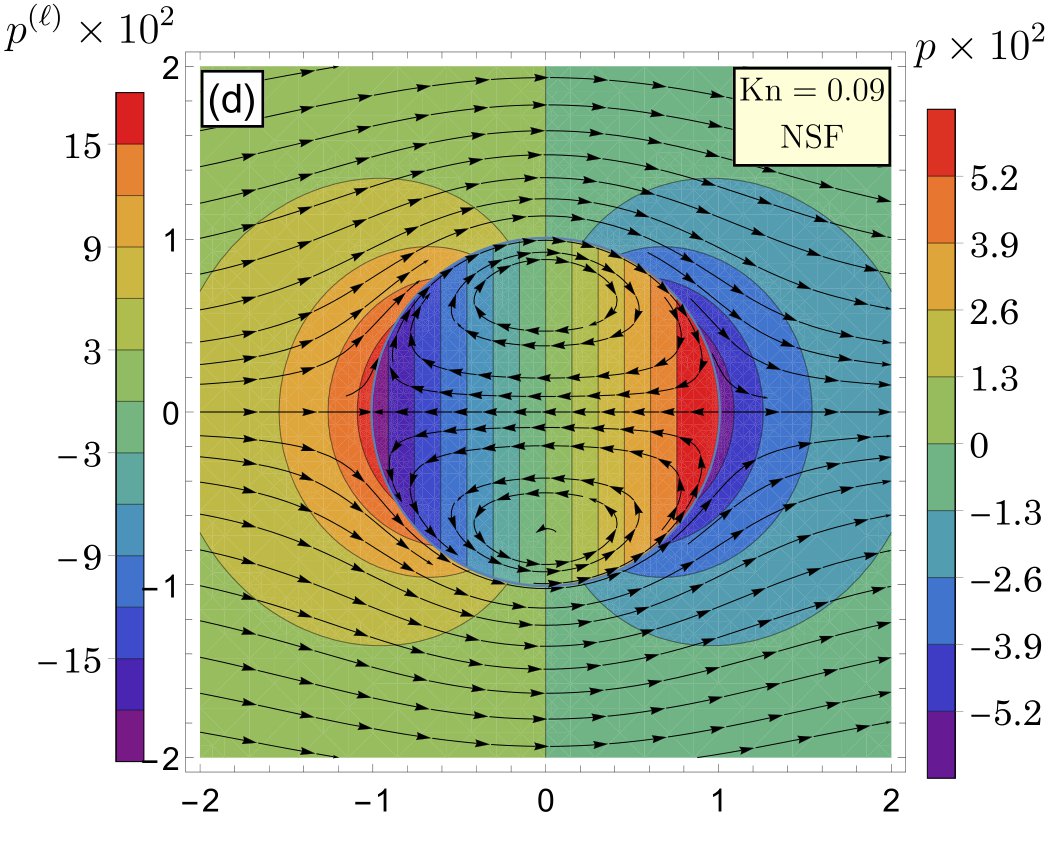}
      \includegraphics[width = 0.48\textwidth]{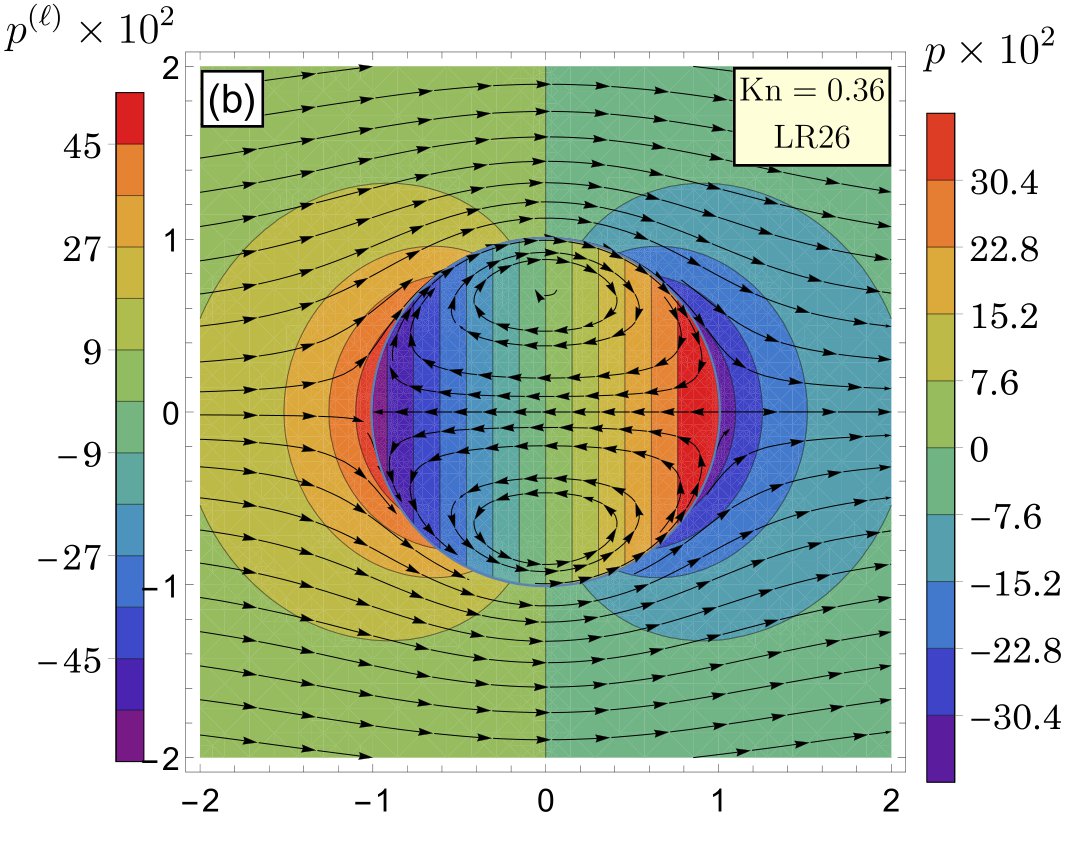}\hfill
      \includegraphics[width = 0.48\textwidth]{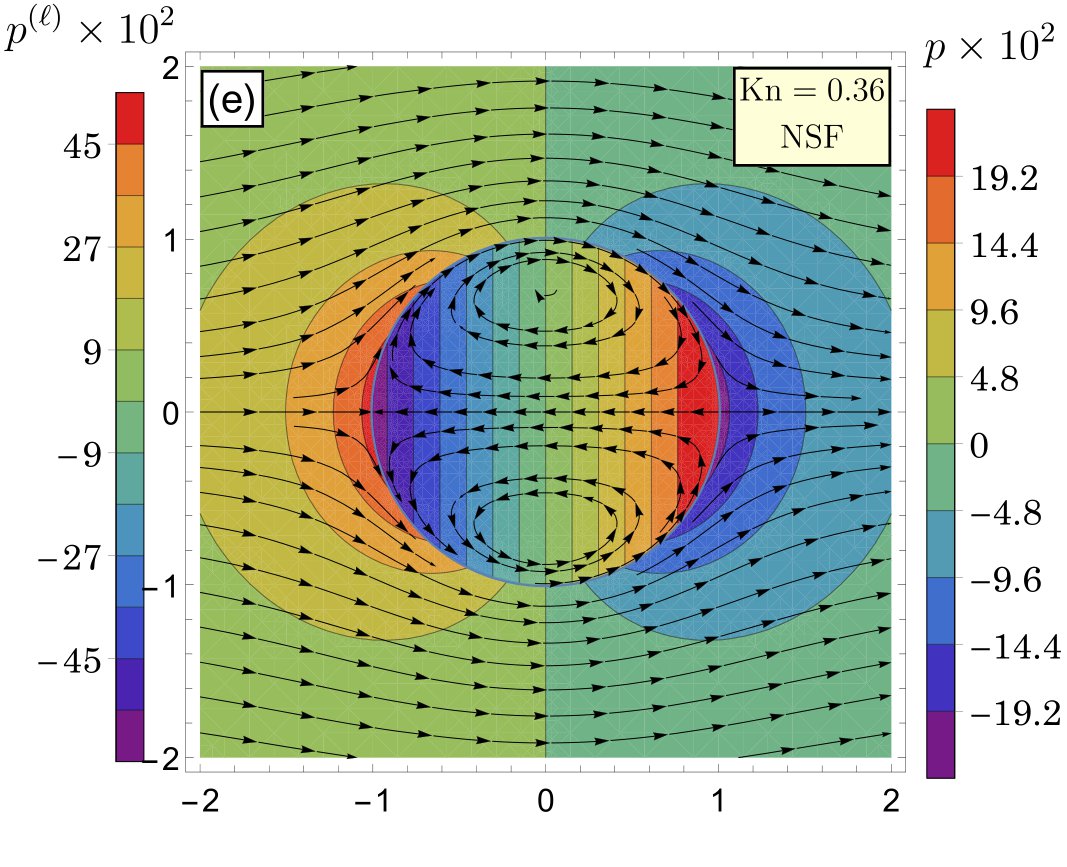}
      \includegraphics[width = 0.48\textwidth]{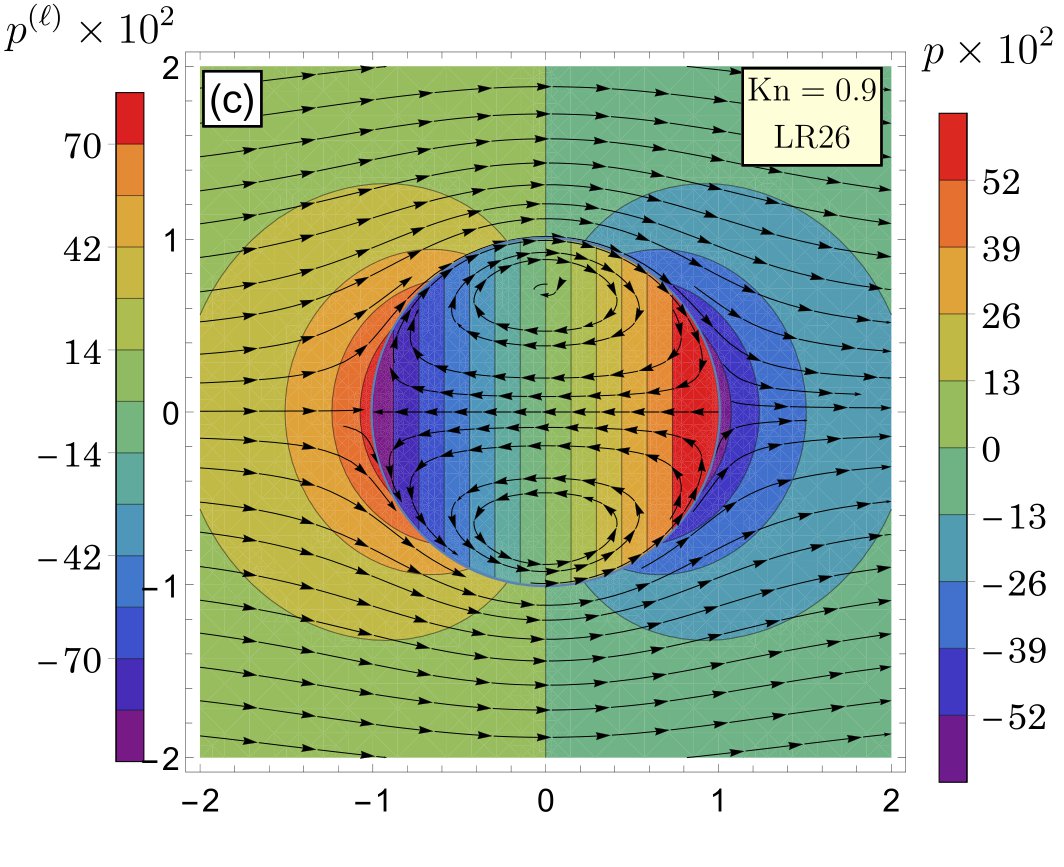}\hfill
      \includegraphics[width = 0.48\textwidth]{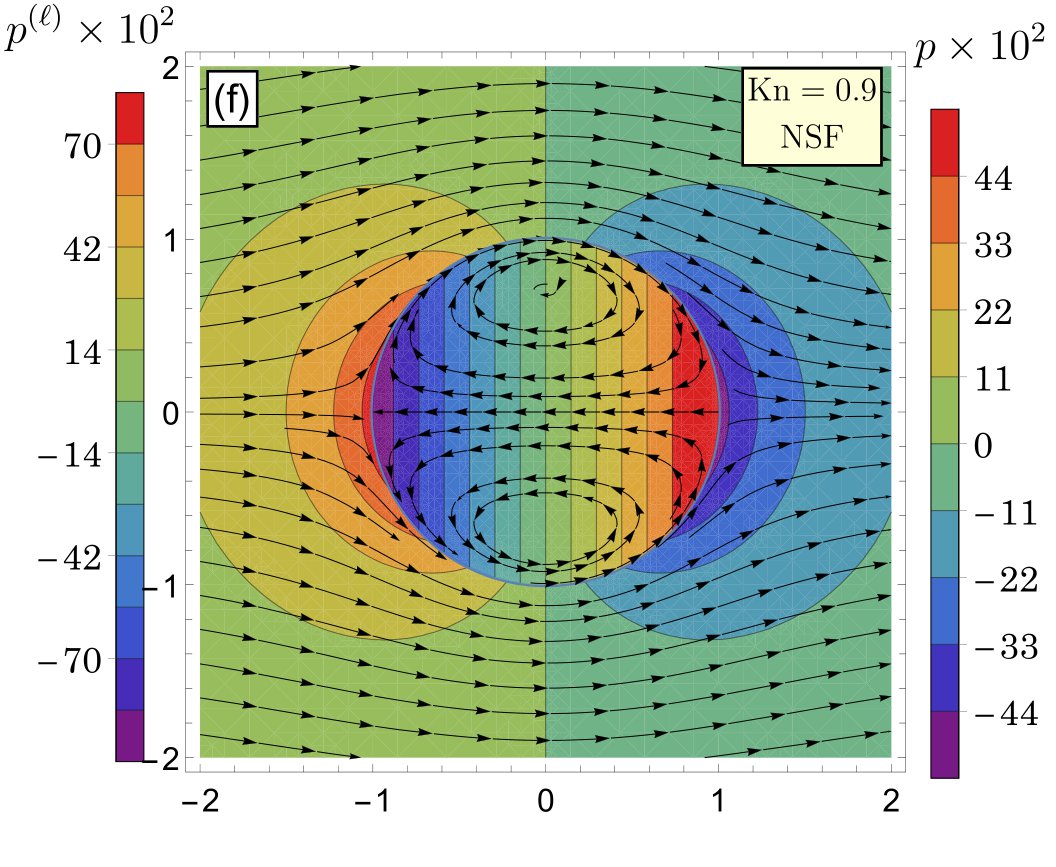}
 \caption{\label{fig:presscontLambdamu1}
Velocity streamlines plotted over the pressure contours for the thermal conductivity ratio $\Lambda_\kappa=100$ and for different values of the Knudsen number: $\mathrm{Kn} = 0.09$ (top row), $\mathrm{Kn} = 0.36$ (middle row) and $\mathrm{Kn} = 0.9$ (bottom row).
The results for the liquid phase (internal flow) have been computed with the NSF equations in all the cases while those for the gas phase (external flow) have been computed with the LR26 equations for the panels in the left column and with the NSF equations for the panels in the right column. The viscosity ratio is $\Lambda_\mu = 1$. }
\end{figure}

Figure \ref{Fig:Hflux} exhibits the radial component of the heat flux  (scaled with $\cos{\theta}$) with respect to the position $r$. 
As described above, the radial component of the heat flux for the liquid phase is constant (notice the horizontal lines on the left of the vertical black line in each panel of figure \ref{Fig:Hflux}).
On the contrary, the radial component of the heat flux is non-constant for the gas phase (see the curves on the right of the vertical black line in each panel of figure \ref{Fig:Hflux}), and is decreasing monotonically for $r>1$. 

%
\begin{figure}
     \centering
     \includegraphics[width = 0.48\textwidth]{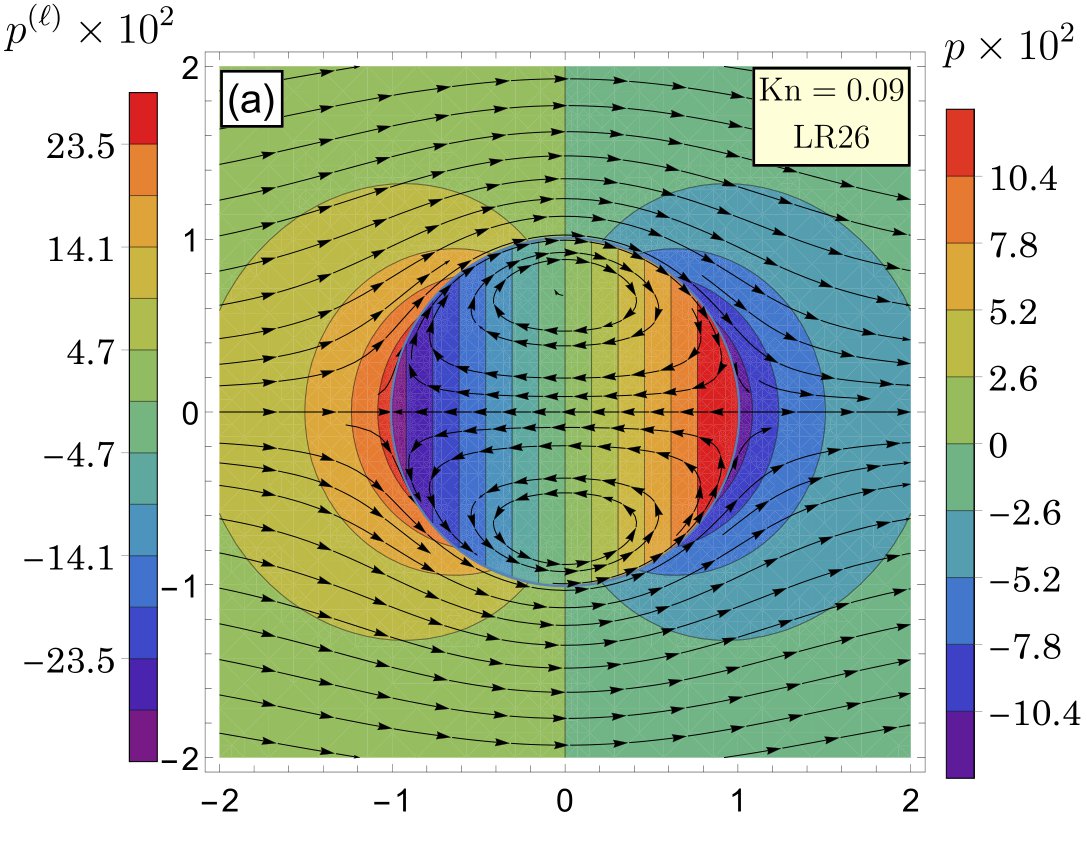}\hfill
      \includegraphics[width = 0.48\textwidth]{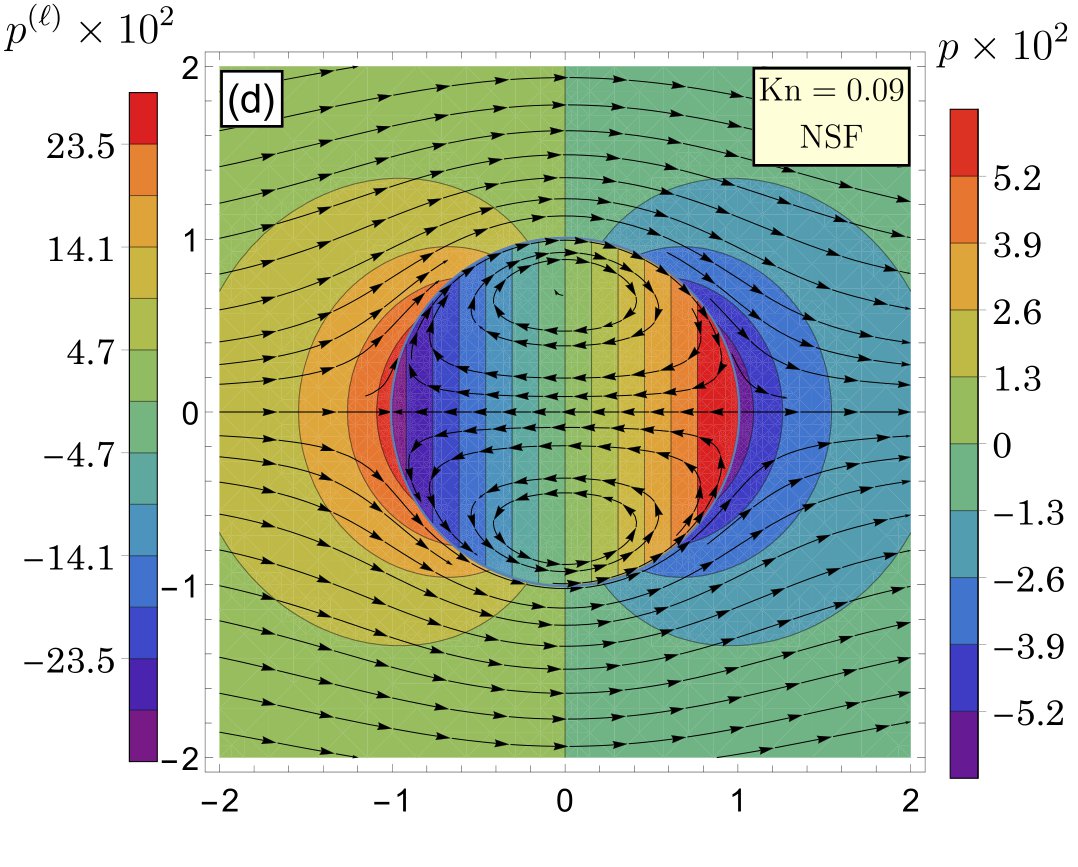}
      \includegraphics[width = 0.48\textwidth]{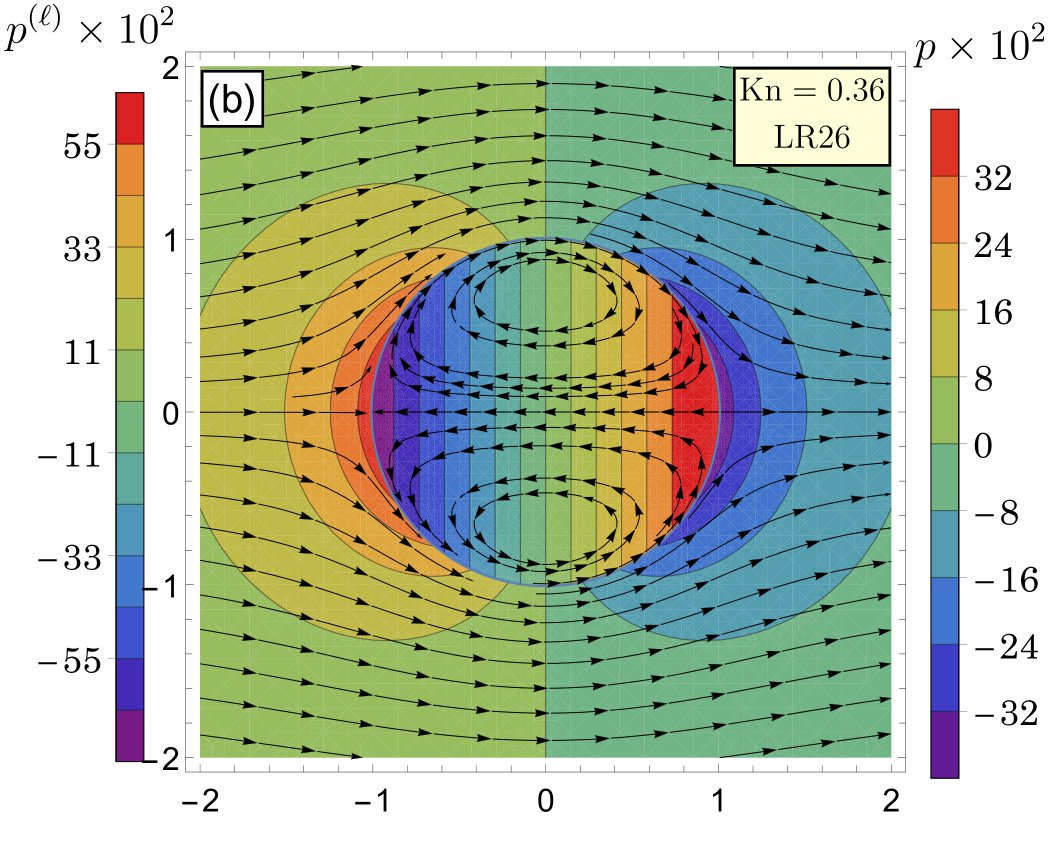}\hfill
      \includegraphics[width = 0.48\textwidth]{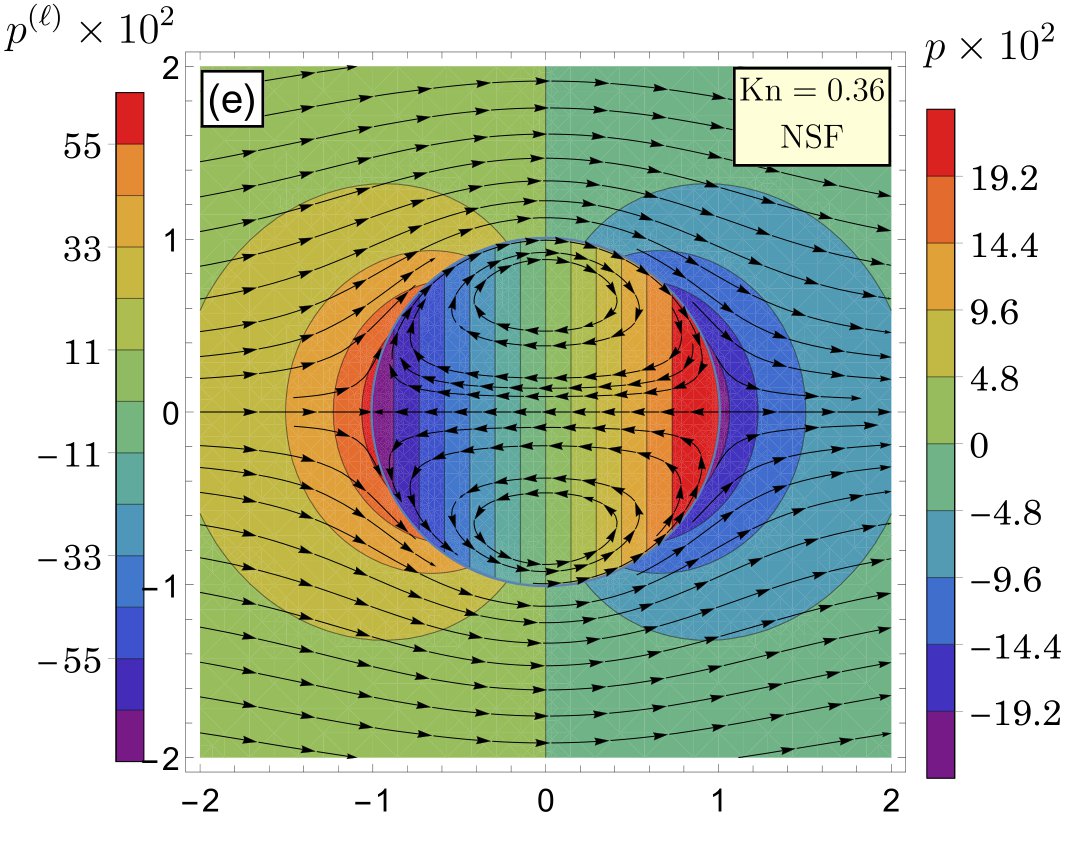}
      \includegraphics[width = 0.48\textwidth]{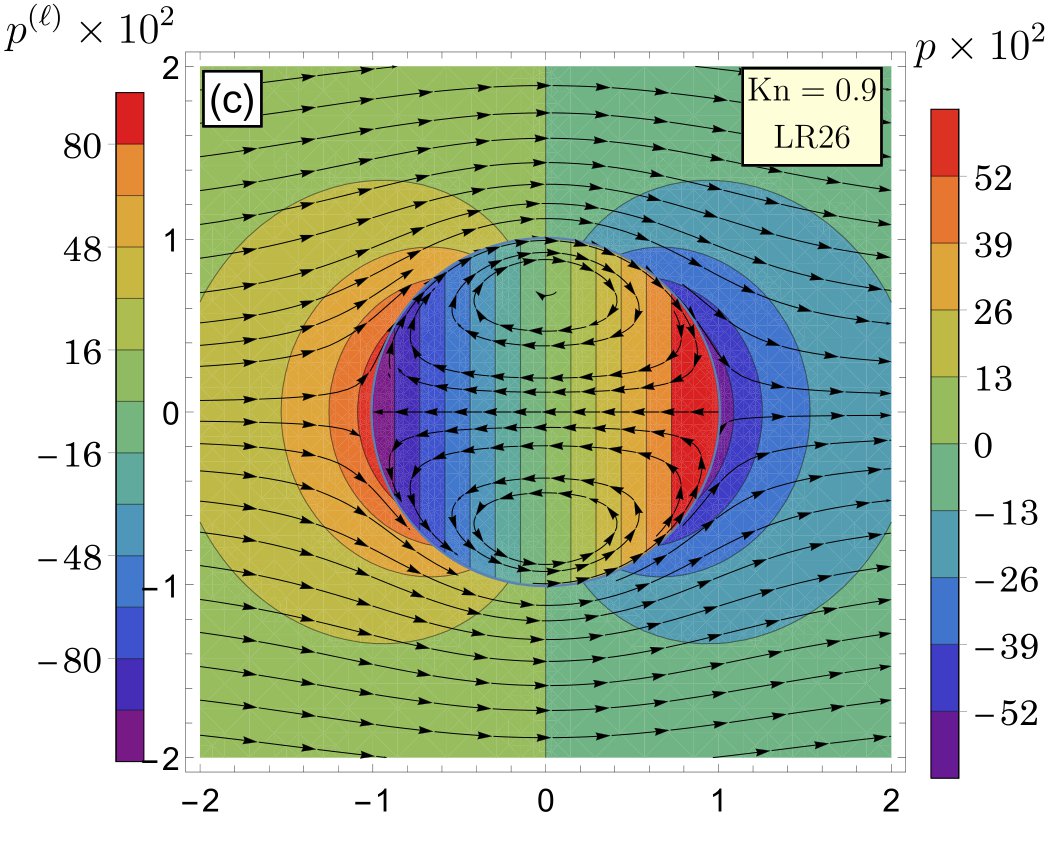}\hfill
      \includegraphics[width = 0.48\textwidth]{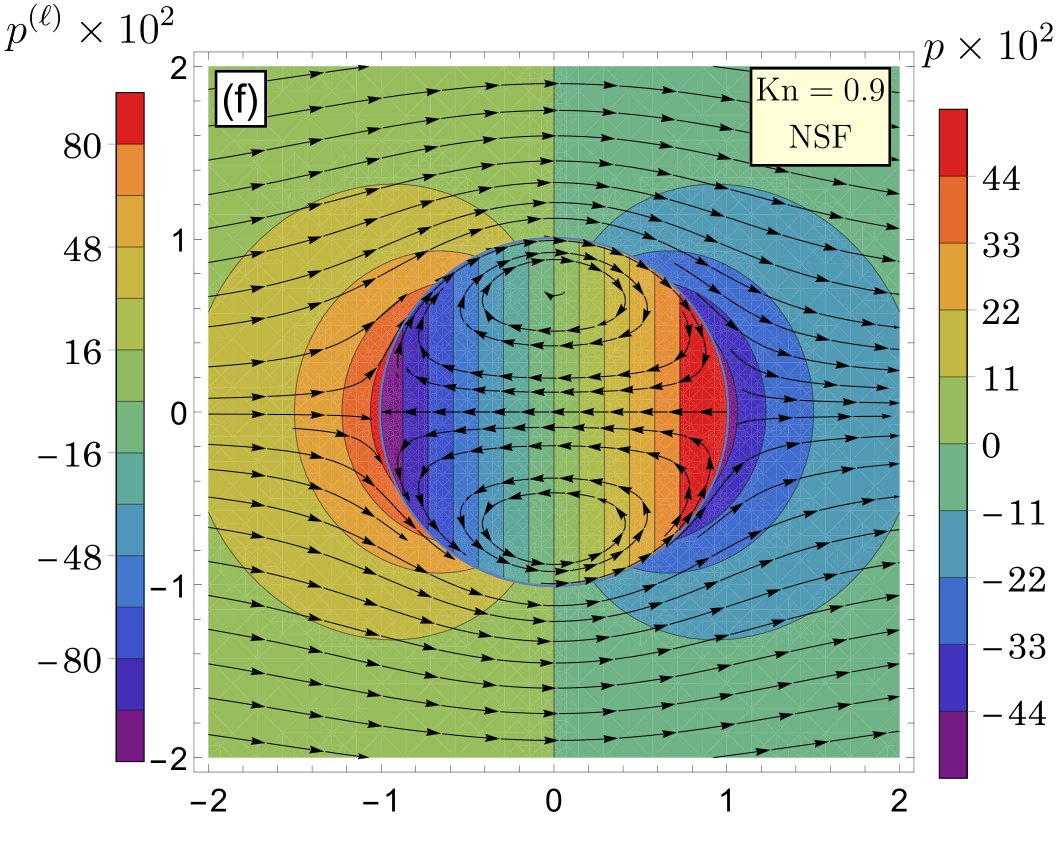}
 \caption{\label{fig:presscontLambdamu10}Same as figure \ref{fig:presscontLambdamu1} but for the viscosity ratio $\Lambda_\mu=10$.}
\end{figure}
\begin{figure}
     \centering
     \includegraphics[width = 0.48\textwidth]{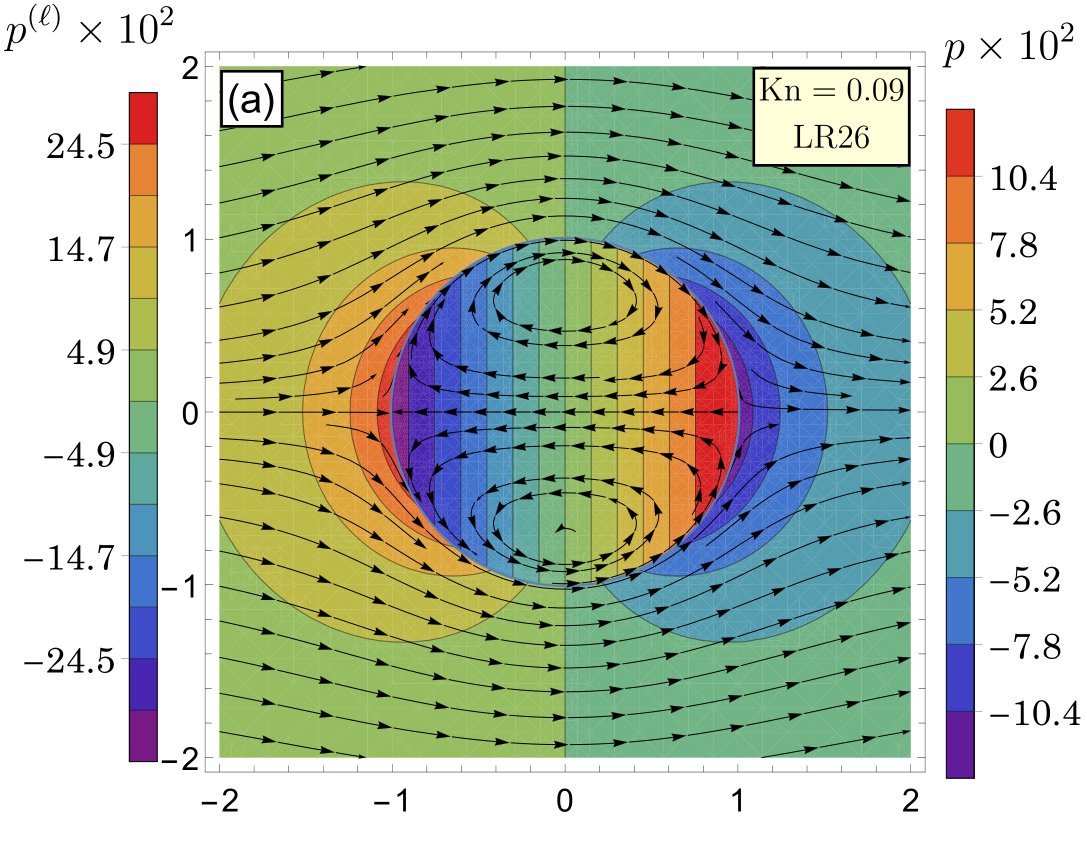}\hfill
      \includegraphics[width = 0.48\textwidth]{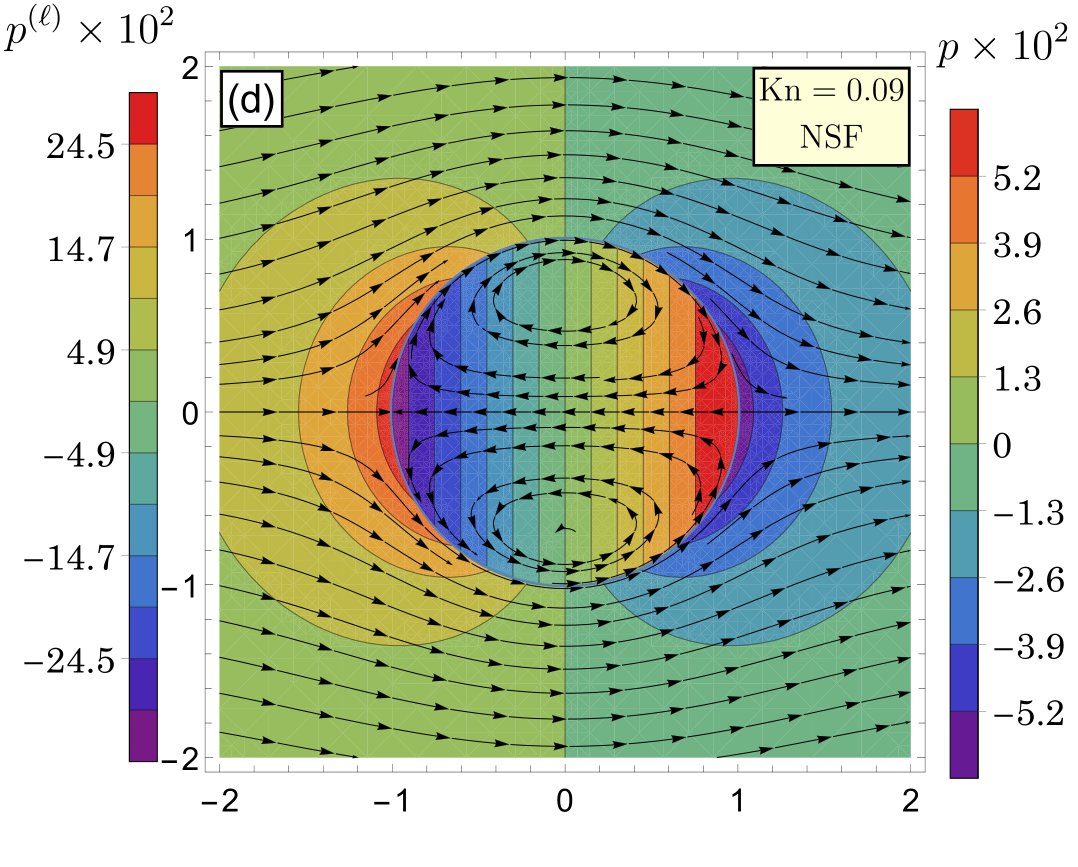}
      \includegraphics[width = 0.48\textwidth]{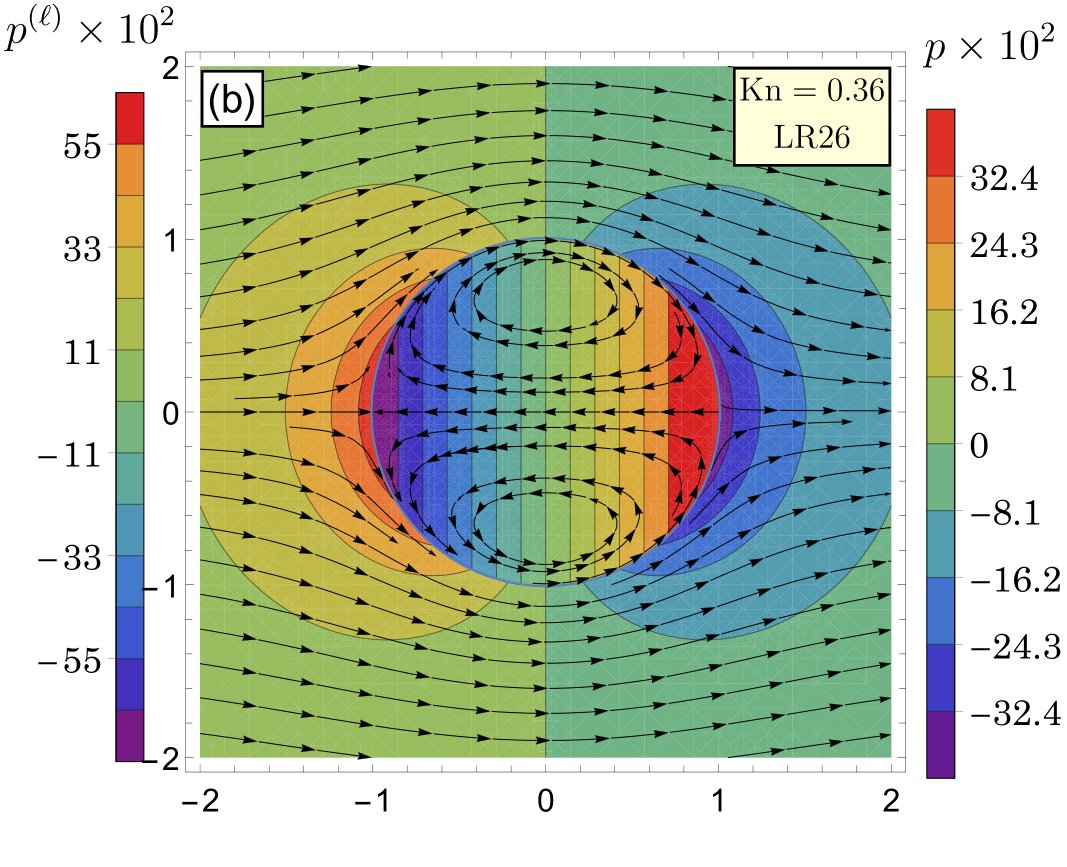}\hfill
      \includegraphics[width = 0.48\textwidth]{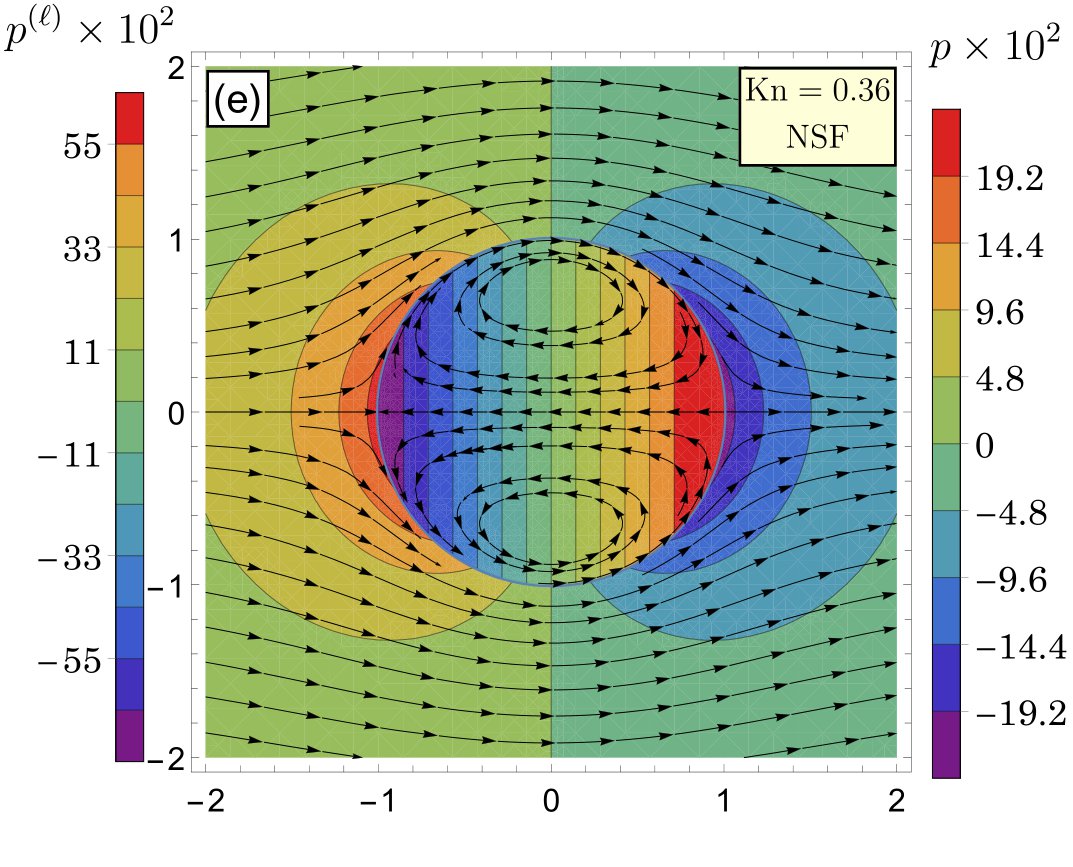}
      \includegraphics[width = 0.48\textwidth]{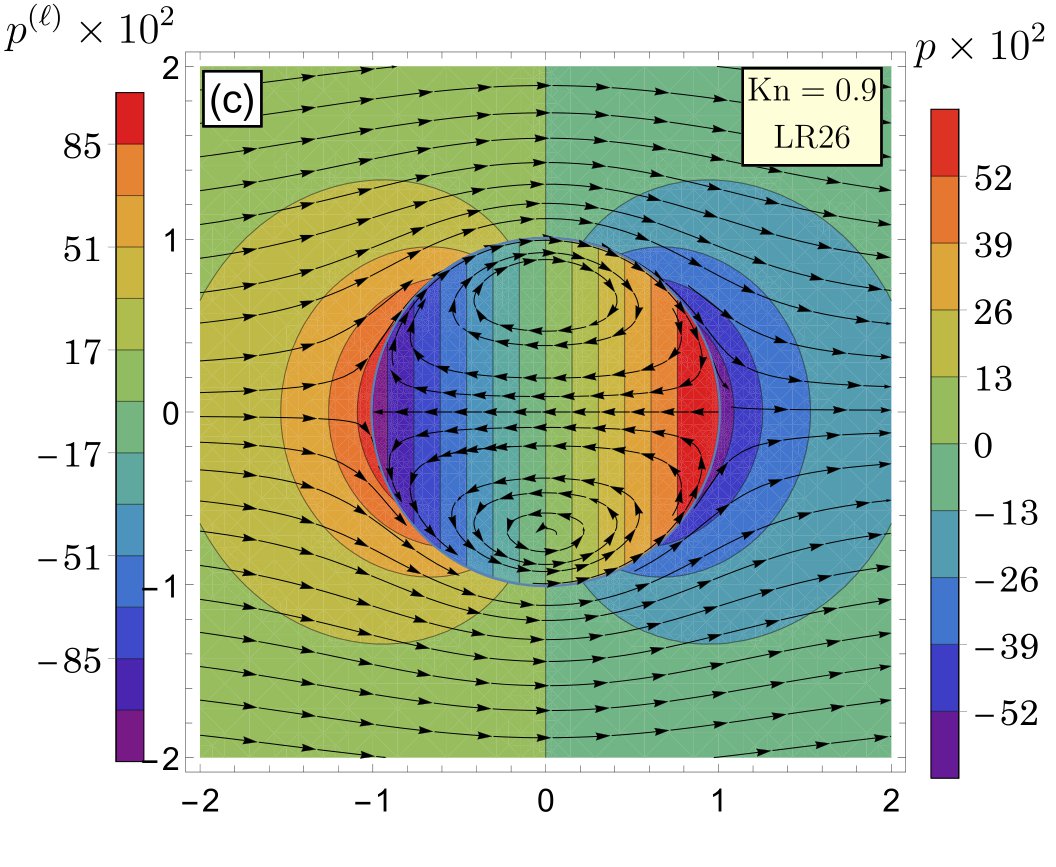}\hfill
      \includegraphics[width = 0.48\textwidth]{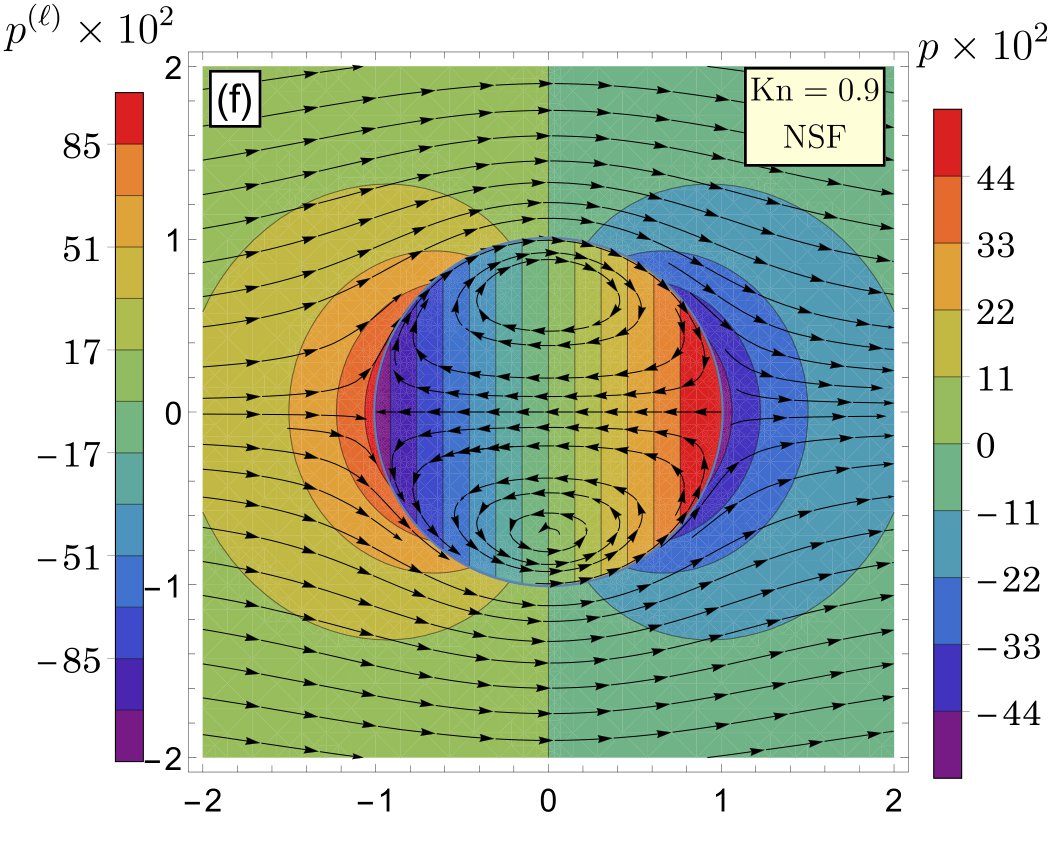}
 \caption{\label{fig:presscontLambdamu100}Same as figure \ref{fig:presscontLambdamu1} but for the viscosity ratio $\Lambda_\mu=100$.}
\end{figure}

\subsection{\label{Subsec:pv}Flow fields: pressure and velocity}
Figures \ref{fig:presscontLambdamu1}, \ref{fig:presscontLambdamu10} and \ref{fig:presscontLambdamu100} illustrate the velocity streamlines plotted over the pressure contours in the $\hat{y}=0$ plane for a fixed thermal conductivity ratio $\Lambda_\kappa = 100$ and for
the viscosity ratios $\Lambda_\mu = 1$, $10$ and $100$, respectively.
The panels in top, middle and bottom rows of each figure again denote the results for the Knudsen numbers $\mathrm{Kn}=0.09$, $0.36$ and $0.9$, respectively. 
The panels in the left columns of figures \ref{fig:presscontLambdamu1}, \ref{fig:presscontLambdamu10} and \ref{fig:presscontLambdamu100} illustrate the results obtained with the LR26 equations for the rarefied gas flow outside the droplet and with the linear NSF equations for the liquid inside the droplet while those on the right columns of figures \ref{fig:presscontLambdamu1}, \ref{fig:presscontLambdamu10} and \ref{fig:presscontLambdamu100} are obtained with the linear NSF equations for both liquid inside the droplet and gas flow outside the droplet. 
The streamlines in figures \ref{fig:presscontLambdamu1}, \ref{fig:presscontLambdamu10} and \ref{fig:presscontLambdamu100} exhibit that, owing to the motion of the gas in the $\hat{z}$-direction, the liquid inside the droplet near the top [close to $(r,\theta,\phi) = (1, \pi/2, 0)$] and bottom [close to $(r,\theta,\phi) = (1, \pi/2, \pi)$] surfaces of the droplet starts moving in direction of the gas flow but since the liquid cannot move outside of the droplet, it flows in the opposite directions near the centre of the droplet, hence forming two counter-rotating vortices inside the droplet.
The pressure contours in figures \ref{fig:presscontLambdamu1}, \ref{fig:presscontLambdamu10} and \ref{fig:presscontLambdamu100} show that the magnitude of the pressure at a point (inside or outside of the liquid droplet) increases with an increase in the Knudsen number.
Furthermore, a comparison of  the corresponding panels on the left and right columns of each of figures \ref{fig:presscontLambdamu1}, \ref{fig:presscontLambdamu10} and \ref{fig:presscontLambdamu100} reveals that there are practically no differences in the magnitudes of the pressure contours inside the liquid droplet but there are noticeable differences in the magnitudes of the pressure contours for the gas phase in the two cases [i.e.~when using the LR26 equations (panels in the left columns) and the linear NSF equations (panels in the right columns) for the gas phase].
Nonetheless, the differences in the magnitudes of the pressure contours for the gas phase decrease with an increase in the Knudsen number.

\begin{figure}
     \centering
     \includegraphics[width = 0.48\textwidth]{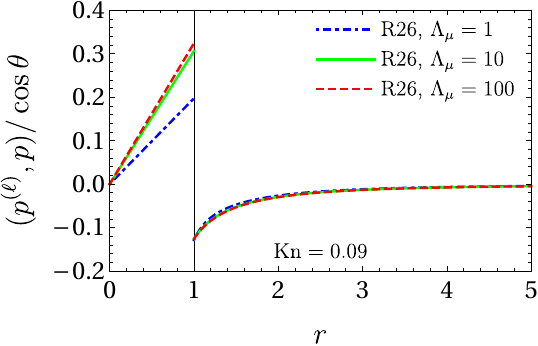}\hfill
      \includegraphics[width = 0.48\textwidth]{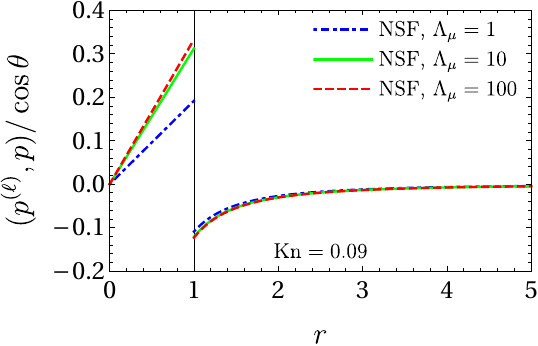}
\\
      \includegraphics[width = 0.48\textwidth]{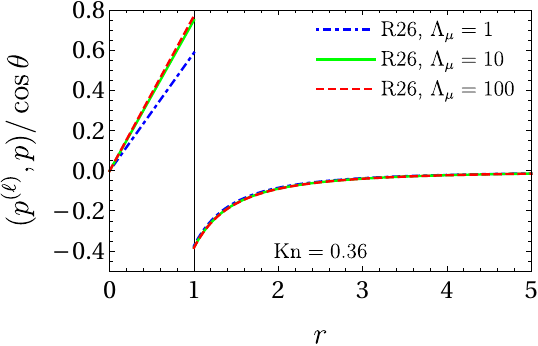}\hfill
      \includegraphics[width = 0.48\textwidth]{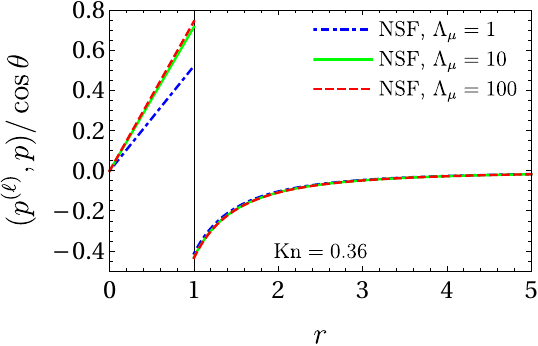}
\\
      \includegraphics[width = 0.48\textwidth]{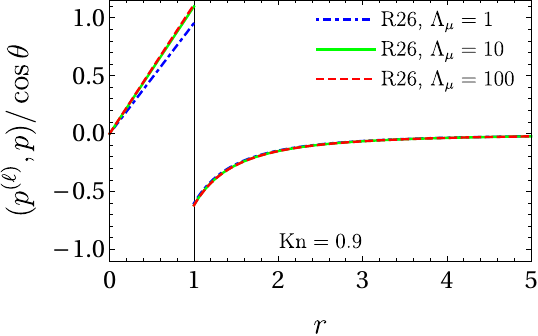}\hfill
      \includegraphics[width = 0.48\textwidth]{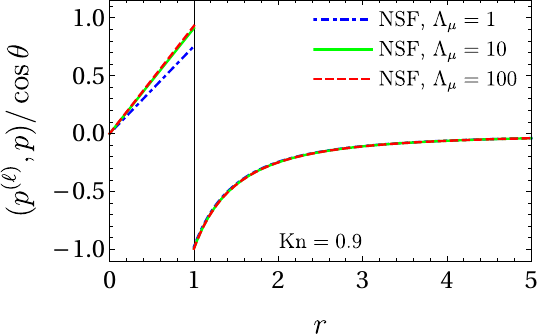}
\caption{\label{Fig:Pressure}
Dimensionless deviations in the pressure (scaled with $\cos{\theta}$) as a function of position $r$ for a fixed thermal conductivity ratio $\Lambda_\kappa=100$ and for different values of the Knudsen number: $\mathrm{Kn} = 0.09$ (top row), $\mathrm{Kn} = 0.36$ (middle row) and $\mathrm{Kn} = 0.9$ (bottom row).
The vertical black line at $r=1$ demarcates the interface between the liquid and gas.
The results for the liquid phase (internal flow) have been computed with the NSF equations in all the cases while those for the gas phase (external flow) have been computed with the LR26 equations for the panels in the left column and with the NSF equations for the panels in the right column.}
\end{figure}
\begin{figure}
     \centering
     \includegraphics[width = 0.48\textwidth]{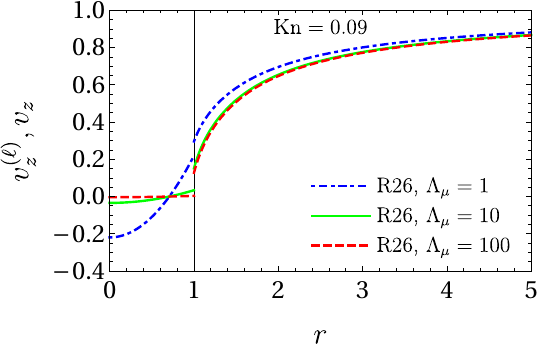}\hfill
      \includegraphics[width = 0.48\textwidth]{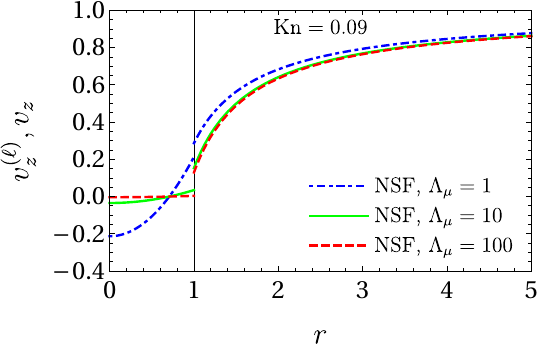}
\\
      \includegraphics[width = 0.48\textwidth]{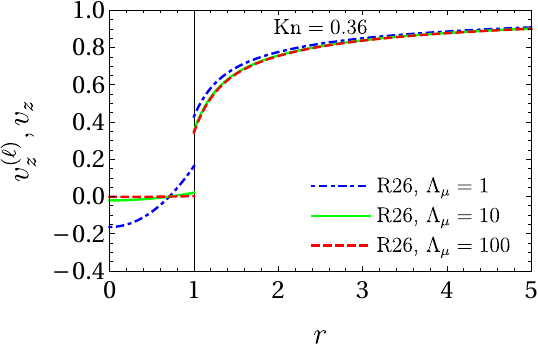}\hfill
      \includegraphics[width = 0.48\textwidth]{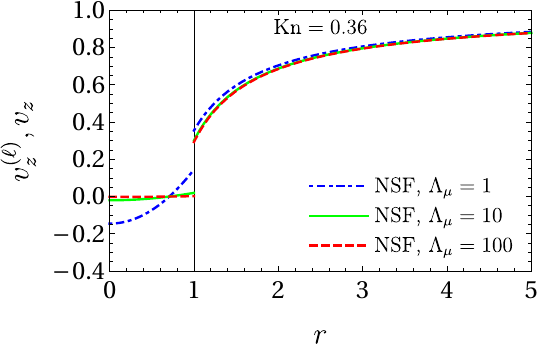}
\\      
      \includegraphics[width = 0.48\textwidth]{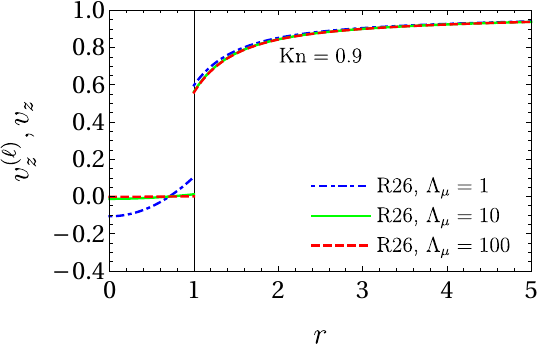}\hfill
      \includegraphics[width = 0.48\textwidth]{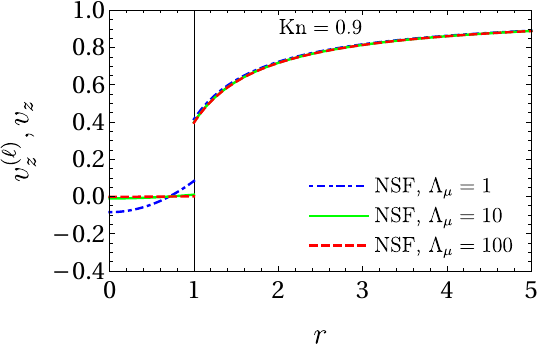}
\caption{\label{Fig:Velocity}
The $z$-component of the (dimensionless) velocity as a function of position $r$ in the plane $\theta=\pi/2$ for a fixed thermal conductivity ratio $\Lambda_\kappa=100$ and for different values of the Knudsen number: $\mathrm{Kn} = 0.09$ (top row), $\mathrm{Kn} = 0.36$ (middle row) and $\mathrm{Kn} = 0.9$ (bottom row).
The vertical black line at $r=1$ demarcates the interface between the liquid and gas.
The results for the liquid phase (internal flow) have been computed with the NSF equations in all the cases while those for the gas phase (external flow) have been computed with the LR26 equations for the panels in the left column and with the NSF equations for the panels in the right column.}
\end{figure}

Similarly to the above, in order to get an insight of the pressure and velocity at a given position, we plot the pressure (divided by $\cos{\theta}$ to understand the results for any angle $\theta$) with respect to the position $r$ in figure \ref{Fig:Pressure} and the $z$-component of the velocity for a fixed $\theta=\pi/2$ with respect to the position $r$ in figure \ref{Fig:Velocity}. 
For both figures \ref{Fig:Pressure} and \ref{Fig:Velocity}, the thermal conductivity ratio is again fixed to $\Lambda_\kappa = 100$ and the viscosity ratios are taken as $\Lambda_\mu=1$, $10$ and $100$.
The panels in top, middle and bottom rows in both figures again exhibit the results for the Knudsen numbers $\mathrm{Kn}=0.09, 0.36$ and $0.9$, respectively. 
The panels in the left column of each of figures \ref{Fig:Pressure} and \ref{Fig:Velocity} depict the results obtained with the LR26 equations for the rarefied gas flow outside the droplet and with the linear NSF equations for the liquid inside the droplet while those in the right column of each of these figures depict the results obtained with the linear NSF equations for both liquid inside the droplet and gas outside the droplet.
The droplet interface has again been demarcated by a vertical black line  at $r=1$ in all panels of figures \ref{Fig:Pressure} and \ref{Fig:Velocity}.
For the liquid phase, the analytic solution for the pressure and $z$-component of the velocity can also be written quite easily from \eqref{field_ansatz_liquid}$_{1,2,3}$ and \eqref{sol_l}$_{1,2,3}$, and they read
\begin{align}
p^{(\ell)} = 5 b_2 \Lambda_\mu \mathrm{Kn} \, r \cos{\theta}
\end{align}
and
\begin{align}
v_z^{(\ell)} 
= v_r^{(\ell)} \cos{\theta} - v_\theta^{(\ell)} \sin{\theta} 
= b_1 + b_2 r^2 - \frac{b_2 r^2}{2}\cos^2{\theta}.
\end{align}
Therefore, inside the droplet, $p^{(\ell)}/\cos{\theta}$ is a linear function of $r$ while $v_z^{(\ell)}$ is parabolic in $r$ for $\theta = \pi/2$.
Consequently, the curves on the left of the vertical black lines in each panel of figure \ref{Fig:Pressure} are straight lines and the curves on the left of the vertical black lines in each panel of figure \ref{Fig:Velocity} are parabolas.
Figure \ref{Fig:Pressure} delineates that the dependence of the pressure in the gas on the viscosity ratio is negligible and that the pressure in the liquid increases with increasing the viscosity ratio, although the increase in pressure in the liquid for large viscosity ratios is also insignificant. 
It is important to note that figure~\ref{Fig:Pressure} illustrates the deviations in the pressures in the liquid droplet and in the gas from the pressures in their respective equilibrium states, which are different for the liquid and gas due to the Laplace pressure. 
The dimensional pressure in the liquid from \eqref{p_liquid} is given by 
\begin{align}
\hat{p}^{(\ell)} = \hat{p}_0 
\left[ p^{(\ell)} + 1 + 2\gamma \mathrm{Kn} \right]
\end{align}
where 
\begin{align}
\label{surfacetension}
\gamma = \frac{\hat{\gamma}}{\hat{\mu}_0 \sqrt{\hat{R}\hat{T}_0}},
\end{align}
is the dimensionless surface tension. 
The dimensionless deviation in the liquid pressure $p^{(\ell)}$
can be made as small as desired in comparison with $1+2\gamma \mathrm{Kn}$ by taking $\hat{u}_\infty$ to be sufficiently small.
This keeps our assumptions of the shape of the droplet being spherical and its size being fixed justifiable. 

Figure \ref{Fig:Velocity} is actually a one-dimensional interpretation of the velocity streamlines shown in figures \ref{fig:presscontLambdamu1}, \ref{fig:presscontLambdamu10} and \ref{fig:presscontLambdamu100}.
To see the interpretation, notice that figures \ref{fig:presscontLambdamu1}, \ref{fig:presscontLambdamu10} and \ref{fig:presscontLambdamu100} show the results in the $\hat{y}=0$ plane while figure \ref{Fig:Velocity} displays the results in the $\hat{z}=0$ (or $\theta=\pi/2$) plane, and hence we can see the results in both figures along a common line, which is the $x$-axis.
Figure \ref{Fig:Velocity} reveals that as we start from the centre of the droplet along the positive $x$-axis, the $z$-component of the velocity of the liquid inside the droplet is initially negative, then at some point it becomes zero and then it turns positive as we approach towards the surface of the droplet, rendering a vortex in the upper half of the droplet. 
Further, on moving outside of the droplet (along the positive $x$-axis), the $z$-component of the velocity of the gas remains always positive and approaches $u_\infty=1$ as $r$ approaches $\infty$.
This is exactly what figures \ref{fig:presscontLambdamu1}, \ref{fig:presscontLambdamu10} and \ref{fig:presscontLambdamu100} show if we focus on the streamlines falling on the positive $x$-axis in figures \ref{fig:presscontLambdamu1}, \ref{fig:presscontLambdamu10} and \ref{fig:presscontLambdamu100}. 
It is evident from figure \ref{Fig:Velocity} that there is a discontinuity in the $z$-components of the velocities of the liquid and gas at the interface of the liquid droplet---referred to as the velocity slip---and that the magnitude of the discontinuity increases with increase in the Knudsen number.
Furthermore, figure \ref{Fig:Velocity} also reveals that the velocity of the liquid droplet approaches zero as the viscosity ratio $\Lambda_\mu$ becomes larger and larger, which is consistent with the case of a rarefied gas flow over a solid sphere.

\subsection{\label{Subsec:discussion}Discussion on results for some commonly used fluids}
In the above subsections, we have analysed the drag force on the droplet and flow profiles (streamlines, heat flux lines, velocity, temperature, pressure and heat flux) of the liquid and gas over a range of dimensionless parameters, namely, the Knudsen number $\mathrm{Kn}$,
the viscosity ratio $\Lambda_\mu$ and the thermal conductivity ratio $\Lambda_\kappa$.
In this section, we pick some commonly used fluids to gauge the applicability/limitations of the present work.

\begin{table}
\centering
\begin{tabular*}{0.8\textwidth}{@{\extracolsep{\fill}} c r r r c @{}}
Liquid & \multicolumn{1}{c}{$\Lambda _\mu$} & \multicolumn{1}{c}{$\Lambda _\kappa$} & \multicolumn{1}{c}{$\gamma$}
& \multicolumn{1}{c}{$\hat{p}_{0}$ (in bars)}  
\\ 
\midrule
Water & $37.6$ & $34.2$ & 
$12.625$ & $0.05$ \\ 
Propyl Alcohol & $84.5$ & $9.0$ & 
$4.175$ & $0.05$ \\
Methanol & $23.3$ & $11.2$ & 
$3.875$ & $0.2$ \\ 
Ammonia & $5.5$ & $26.2$ & 
$3.540$ & $11$ \\ 
Pentane & $9.5$ & $6.2$ & 
$2.685$ & $7$ \\
Propyne & $6.1$ & $6.5$ & 
$1.965$ & $7$\\ 
Isobutane & $6.5$ & $5.0$ & 
$1.725$ & $4$ \\ 
R134a & $8.4$ & $4.5$ & 
$1.370$ & $7.5$ \\ 
Propane & $4.2$ & $5.2$ & 
$1.200$ & $12.5$ \\ 
\bottomrule
\end{tabular*}%
\caption{\label{table:dataliquids}
Parameters for some common engineering fluids at $300$\,K. 
The viscosity and thermal conductivity ratios are with respect to the argon gas.}
\end{table}

Table~\ref{table:dataliquids} exhibits the parameters, namely the viscosity ratio $\Lambda_\mu$, the thermal conductivity ratio $\Lambda_\kappa$, the dimensionless surface tension $\gamma$ and the reference pressure $\hat{p}_0$, for some commonly used liquids at the reference temperature $\hat{T}_0 = 300$\,K.
The viscosity and thermal conductivity ratios are with respect to the argon gas, which has the viscosity $\hat{\mu}_0 = 2.2725\times 10^{-5}$\,Pa\,s at the reference temperature $\hat{T}_0 = 300$\,K.
The gas constant for argon is $\hat{R} = 208.1$\,J/(Kg K).
The parameters listed in table~\ref{table:dataliquids} have been determined using the data provided by the \cite{NISTwebsite}.
The reference pressure $\hat{p}_0$ is taken to be slightly above the saturation pressure of the fluid at the reference temperature $\hat{T}_{0}=300$\,K.

\begin{figure}
     \centering
     \includegraphics[width = 0.7\textwidth]{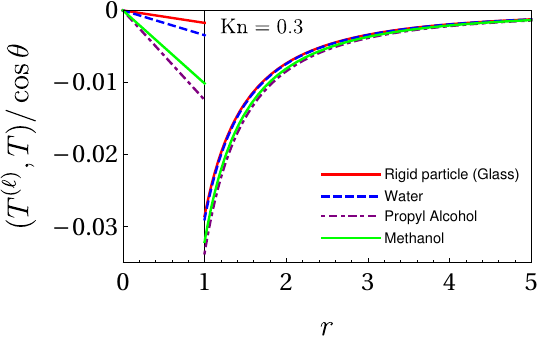}
\caption{\label{Fig:Tdiffliquids}
Dimensionless deviations in the temperature (scaled with $\cos{\theta}$) as a function of position $r$ for $\mathrm{Kn} = 0.3$ in the case of argon gas flow past a rigid spherical glass ball (solid red lines), in the cases of argon gas flow past spherical liquid droplets of water (dashed blue lines), propyl alcohol (purple dot-dashed lines) and methanol (green solid lines).
The vertical black line at $r=1$ demarcates the interface between the solid/liquid and gas.}
\end{figure}

Recall from \S\,\ref{Subsec:drag} that, for large viscosity ratios, the drag force on the liquid droplet approaches the drag force on a rigid sphere (the case of $\Lambda_\mu \to \infty$).
As a matter of fact, Happel \& Brenner's formula \eqref{HDrag} also deduces that the percentage error between the drag force on the liquid droplet with respect to the drag force in the rigid sphere case is less than $10\%$ for all $\Lambda_\mu \geq 2.5$.
For most common engineering fluids, the viscosity ratio $\Lambda_\mu$ is apparently bigger than $2.5$ (see table~\ref{table:dataliquids}). 
Hence, for many practical applications (such as, a water droplet moving through air), the viscosity ratio is high enough to treat the liquid droplet to be a rigid sphere---as far as one is only concerned about the drag force. 
Notwithstanding, the coupled dynamics of the liquid droplet and gas leads to some interesting flow features in the temperature and heat profiles, as shown in \S\,\ref{Subsec:Tq} and \ref{Subsec:pv}, that could potentially play important roles in processes involving phase change \citep{RLS2018}.
To emphasise the point,
we plot in figure~\ref{Fig:Tdiffliquids} the temperature profile in the case of argon gas flow past a rigid spherical ball made of glass, which has the thermal conductivity about twice of that of water, at $\mathrm{Kn} = 0.3$ by solid red lines.
The corresponding temperature profiles in the cases of argon gas flow past spherical liquid droplets of water (dashed blue lines),  propyl alcohol (purple dot-dashed lines) and methanol (green solid lines) are also shown in figure~\ref{Fig:Tdiffliquids}.
The vertical black line at $r=1$ demarcates the interface between the solid/liquid and gas. 
Evidently, there are significant differences in temperature profiles in all above cases. 
The differences are primarily due to their different thermal conductivity ratios and are more pronounced within the solid sphere or liquid droplet. 
Smaller heat conductivity ratio apparently leads to larger temperature gradients within and outside the liquid droplet or solid sphere. 

\begin{figure}
     \centering
     \includegraphics[width = 0.7\textwidth]{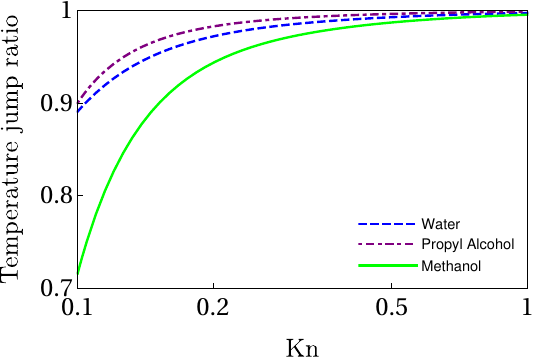}
\caption{\label{Fig:jumpratio}
The ratio of the temperature jump at the interface of a liquid droplet (made of water, propyl alcohol and methanol) to the temperature jump at the interface of a corresponding solid sphere ($\Lambda_\mu\to \infty$) having the same thermal conductivity as that of the liquid against plotted over the Knudsen number.}
\end{figure}


In order to highlight the effects of the internal motion on the temperature profiles, we plot in figure~\ref{Fig:jumpratio} the ratio of the temperature jump at the interface of a liquid droplet to the temperature jump at the interface of a corresponding solid sphere ($\Lambda_\mu\to \infty$) having the same thermal conductivity as that of the liquid against the Knudsen number.
We choose the liquid droplets to be made of water (dashed blue lines in figure~\ref{Fig:jumpratio}), propyl alcohol (dot-dashed purple lines in figure~\ref{Fig:jumpratio}) and methanol (solid green lines in figure~\ref{Fig:jumpratio}) so that our assumption of the liquid droplet being spherical holds good.
The differences in the temperature jump ratios for different liquid droplets with respect to its solid counterpart (having the same thermal conductivity) are evident in figure~\ref{Fig:jumpratio}, especially for relatively smaller Knudsen numbers, although the differences decrease with the increasing Knudsen number. 
Methanol having the viscosity ratio about $23$ shows the largest deviation (of about $30\%$).


An important assumption made in this work is that the liquid droplet remains spherical throughout. 
This assumption, in other words, means that the surface tension force on the droplet is assumed to be larger than the pressure difference between inside and outside of the droplet at the interface, i.e.
\begin{align}
\label{LaplaceCond}
\hat{p}^{(\ell)}-\hat{p} < \frac{2\hat{\gamma}}{\hat{R}_0}
\end{align}
where $\hat{\gamma}$ is the surface tension and $\hat{R}_0$ is the radius of the droplet. 
Condition \eqref{LaplaceCond} in the dimensionless form reads
\begin{align}
\label{LaplaceConddimless}
\frac{p^{(\ell)}-p}{\mathrm{Kn}} < 2\gamma.
\end{align}
From figure \ref{Fig:Pressure}, $(p^{(\ell)}-p)/\mathrm{Kn}$ at the interface is about $4.44$ (for the top row), about $3.33$ (for the middle row) and about $2$ (for the bottom row). 
Comparing these numbers with the twice of the dimensionless surface tension given in table~\ref{table:dataliquids}, one can see that condition \eqref{LaplaceConddimless} holds true for water, propyl alcohol, methanol, ammonia and pentane but does not hold for isobutane, R134a and propane.
Therefore, there are some liquids (e.g.~water, propyl alcohol, methanol, etc.) for which our assumptions of the droplet being spherical is valid.
Nevertheless, there are also many liquids for which the surface tension forces are not strong enough to maintain the liquid droplet spherical.
For such liquids, it will be necessary to account for the effect of surface tension forces and to extend the present work. 
\section{\label{Sec:concl}Conclusion}
In this article, we have studied the effects of internal motion within a spherical liquid droplet with its diameter being comparable to the mean free path of the surrounding (monatomic) gas under the assumptions of no phase change involved and the surface tension force being strong enough to maintain the liquid droplet spherical throughout. 
The rarefied gas phase in this article has been modelled with the LR26 equations while the flow in the liquid phase has been modelled with the incompressible NSF equations.
Owing to our assumptions, the problem has become somewhat simplified---allowing for the analytic solution, which was still not so easy to obtain.
The analytic solution plays a vital role in understanding the kinetic effects in rarefied gases and also helps in developing a deeper understanding on the effects, which the internal motion in a liquid droplet has on the flow pattern of the rarefied gas flowing past it. 
With the analytic solution, the effects of the liquid to gas viscosity ratio and the liquid to gas thermal conductivity ratio on the drag force and the overall flow dynamics have been investigated. 

To validate the analytic solution obtained in the present work, the analytic results for the drag force obtained from the LR26 theory in the present work have been compared with   the limited experimental measurements available in the literature.
It turns out that the analytic results for the drag force obtained from the LR26 theory agree closely with experimental data even for significantly large values of the Knudsen number.
After validating the analytic results on the drag force, the physical field variables, which are usually difficult to measure in experiments, have been presented. 
The analytic results obtained in the present work show a clear discontinuity in the velocities of the liquid and gas at the interface of the liquid droplet (figure~\ref{Fig:Velocity}).
It is worthwhile noting that the viscosity ratio of the liquid to the gas in practical applications is usually large. 
Thus, from the findings of figure~\ref{Fig:Drag1}, it is acceptable in such applications to treat the liquid droplet as a rigid sphere of the same size if one is concerned only about the drag force.
Nevertheless, \S\,\ref{Subsec:discussion} shows that there could be significant differences in the other quantities (e.g., the temperature) when treating the liquid droplet as a rigid sphere.
In addition, for some liquids and gases used in some other applications, the viscosity ratio could be close to unity even at standard conditions (at $25$\textdegree C and $1$ atmospheric pressure); for instance, the viscosity ratio between water (liquid) and ammonia (gas) at standard conditions is about 3.6.
For such applications, the presented theory would yield meaningful results.

It is important to note that, owing to our assumptions, the density ratio does not appear in our analysis and hence does not affect the results in the present work. 
Nevertheless, the density ratio as well as the surface tension are the two most important parameters, whose effects are significant in problems of rarefied gas flow past liquid droplets and cannot be ignored in practical applications.
The present work may be treated as a first step to achieving a good mathematical understanding of the (simplified) problem. 
More realistic problems, involving
surface tension and/or phase change, will be subjects of future research.
%
\bigskip

\noindent
\paragraph{\bf Acknowledgments.}
V.~K.~Gupta acknowledges the facilities of the Bhaskaracharya Mathematics Laboratory and  Brahmagupta Mathematics Library of IIT Indore supported by the DST FIST project (SR/FST/MS-I/2018/26), which have been used to carry out this work.
\bigskip

\noindent
\paragraph{\bf Funding.}
A.~S.~Rana gratefully acknowledges the financial supports through ``SRG" (SRG/2021/000790) and ``MATRICS" (MTR/2021/000417) schemes funded by the Science \& Engineering Research Board, India.
\bigskip

\noindent
\paragraph{\bf Declaration of interests.}
The authors report no conflict of interest.

\appendix
\section{Analytic expressions for the unknowns in ansatzes \eqref{field_ansatz_gas} and \eqref{field_ansatz_liquid}}
\label{app:sol}
\noindent
The analytic solution for the unknowns appearing in \eqref{field_ansatz_gas} and \eqref{field_ansatz_liquid} are as follows.
\begingroup
\allowdisplaybreaks
\begin{fleqn}
\begin{align}
\label{v1gas}
\mathbbm{v}_{1}(r) ={}& \frac{C_1}{2r}+\frac{C_2}{3r^3}-K_1 \Theta_1(r) \left(\frac{0.975701 \mathrm{Kn}^3}{r^3}+\frac{0.497885 \mathrm{Kn}^2}{r^2}\right)
\nonumber\\
& - K_2 \Theta_2(r) \left(\frac{1.4893 \mathrm{Kn}^3}{r^3}+\frac{1.88528 \mathrm{Kn}^2}{r^2}\right),
\\
\label{v2gas}
\mathbbm{v}_{2}(r) ={}& \frac{C_1}{4r}-\frac{C_2}{6r^3}+K_1 \Theta_1(r) \left(\frac{0.487851 \mathrm{Kn}^3}{r^3}+\frac{0.248943 \mathrm{Kn}^2}{r^2}+\frac{0.127032 \mathrm{Kn}}{r}\right)
\nonumber\\
& + K_2 \Theta_2(r) \left(\frac{0.74465 \mathrm{Kn}^3}{r^3}+\frac{0.942641 \mathrm{Kn}^2}{r^2}+\frac{1.19327 \mathrm{Kn}}{r}\right),
\\[2ex]
\label{pgas}
\mathbbm{p}(r) ={}&\frac{C_1\mathrm{Kn}}{2r^2}-K_3\Theta_3(r) \left(\frac{0.931108 \mathrm{Kn}^2}{r^2}+\frac{1.08307 \mathrm{Kn}}{r}\right)
\nonumber\\
&-K_4 \Theta_4(r) \left(\frac{0.254173 \mathrm{Kn}^2}{r^2}+\frac{0.172163 \mathrm{Kn}}{r}\right)
\nonumber\\
&-K_5 \Theta_5(r) \left(\frac{0.0501203 \mathrm{Kn}^2}{r^2}+\frac{0.0226838 \mathrm{Kn}}{r}\right),
\\
\label{Tgas}
\mathbbm{T}(r) ={}& \frac{C_3}{45 r^2}-K_3 \Theta_3(r) \left(\frac{0.0172036 \mathrm{Kn}^2}{r^2}+\frac{0.0200113 \mathrm{Kn}}{r}\right)
\nonumber\\
&-K_4 \Theta_4(r) \left(\frac{0.0629047 \mathrm{Kn}^2}{r^2}+\frac{0.0426083 \mathrm{Kn}}{r}\right)
\nonumber\\
&-K_5 \Theta_5(r) \left(\frac{0.443094 \mathrm{Kn}^2}{r^2}+\frac{0.200538 \mathrm{Kn}}{r}\right),
\end{align}
\begin{align}
\label{s1gas}
\mathbbm{s}_1(r) ={}& \frac{C_1 \mathrm{Kn}}{r^2}+\frac{\mathrm{Kn}^4}{r^4}\left(-\frac{10 C_1}{\mathrm{Kn}}+\frac{2 C_2}{\mathrm{Kn}^3}+\frac{2 C_3}{5\mathrm{Kn}^2}\right)
\nonumber\\
&+K_3 \Theta_3(r) \left(\frac{6.19341 \mathrm{Kn}^4}{r^4}+\frac{7.20421 \mathrm{Kn}^3}{r^3}+\frac{3.72443 \mathrm{Kn}^2}{r^2}+\frac{1.08307 \mathrm{Kn}}{r}\right)
\nonumber\\
&+K_4 \Theta_4(r) \left(\frac{4.98597 \mathrm{Kn}^4}{r^4}+\frac{3.37723 \mathrm{Kn}^3}{r^3}+\frac{1.01669 \mathrm{Kn}^2}{r^2}+\frac{0.172163 \mathrm{Kn}}{r}\right)
\nonumber\\
&+K_5 \Theta_5(r) \left(\frac{2.20218 \mathrm{Kn}^4}{r^4}+\frac{0.996676 \mathrm{Kn}^3}{r^3}+\frac{0.200481 \mathrm{Kn}^2}{r^2}+\frac{0.0226838 \mathrm{Kn}}{r}\right),
\\
\label{s2gas}
\mathbbm{s}_2(r) ={}& \frac{\mathrm{Kn}^4}{r^4}\left(-\frac{5 C_1}{\mathrm{Kn}}+\frac{C_2}{\mathrm{Kn}^3}+\frac{C_3}{5\mathrm{Kn}^2}\right)
\nonumber\\
&+K_3 \Theta_3(r)\left(\frac{3.09671 \mathrm{Kn}^4}{r^4}+\frac{3.6021 \mathrm{Kn}^3}{r^3}+\frac{1.39666 \mathrm{Kn}^2}{r^2}\right)
\nonumber\\
&+K_4 \Theta_4(r) \left(\frac{2.49299 \mathrm{Kn}^4}{r^4}+\frac{1.68862 \mathrm{Kn}^3}{r^3}+\frac{0.381259 \mathrm{Kn}^2}{r^2}\right)
\nonumber\\
&+K_5 \Theta_5(r) \left(\frac{1.10109 \mathrm{Kn}^4}{r^4}+\frac{0.498338 \mathrm{Kn}^3}{r^3}+\frac{0.0751805 \mathrm{Kn}^2}{r^2}\right),
\end{align}
\begin{align}
\label{q1gas}
\mathbbm{q}_{1}(r) ={}& \frac{\mathrm{Kn}^2 }{r^3}\left(\frac{C_3}{6\mathrm{Kn}}-\frac{3 C_1}{2}\right)
+K_1 \Theta_1(r) \left(\frac{1.50825 \mathrm{Kn}^3}{r^3}+\frac{0.769634 \mathrm{Kn}^2}{r^2}\right)
\nonumber\\
& -K_2 \Theta_2(r) \left(\frac{0.80299 \mathrm{Kn}^3}{r^3}+\frac{1.01649 \mathrm{Kn}^2}{r^2}\right),
\\
\label{q2gas}
\mathbbm{q}_{2}(r) ={}& \frac{\mathrm{Kn}^2 }{r^3}\left(\frac{3C_1}{4}-\frac{C_3}{12 \mathrm{Kn}}\right)
\nonumber\\
&-K_1 \Theta_1(r) \left(\frac{0.754123 \mathrm{Kn}^3}{r^3}+\frac{0.384817 \mathrm{Kn}^2}{r^2}+\frac{0.196366 \mathrm{Kn}}{r}\right)
\nonumber\\
&+ K_2 \Theta_2(r)  \left(\frac{0.401495 \mathrm{Kn}^3}{r^3}+\frac{0.508246 \mathrm{Kn}^2}{r^2}+\frac{0.643381 \mathrm{Kn}}{r}\right),
\end{align}
\begin{align}
\label{m1gas}
\mathbbm{m}_1(r) ={}&\frac{24 C_1 \mathrm{Kn}^2}{5 r^3}+\frac{\mathrm{Kn}^5}{r^5}\left(-\frac{7344 C_1}{49 \mathrm{Kn}}+\frac{16 C_2}{\mathrm{Kn}^3}+\frac{208 C_3}{35 \mathrm{Kn}^2}\right)
\nonumber\\
&+K_1\Theta_1(r) \left(\frac{64.3578 \mathrm{Kn}^5}{r^5}+\frac{32.8408 \mathrm{Kn}^4}{r^4}+\frac{6.70326 \mathrm{Kn}^3}{r^3}+\frac{0.570095 \mathrm{Kn}^2}{r^2}\right)
\nonumber\\
&+K_2 \Theta_2(r)\left(\frac{50.8419 \mathrm{Kn}^5}{r^5}+\frac{64.3599 \mathrm{Kn}^4}{r^4}+\frac{32.5889 \mathrm{Kn}^3}{r^3}+\frac{6.87563 \mathrm{Kn}^2}{r^2}\right)
\nonumber\\
&+K_3\Theta_3(r)\left(\frac{30.5159 \mathrm{Kn}^5}{r^5}+\frac{35.4962 \mathrm{Kn}^4}{r^4}+\frac{18.5802 \mathrm{Kn}^3}{r^3}+\frac{5.60327 \mathrm{Kn}^2}{r^2}+\frac{0.931108 \mathrm{Kn}}{r}\right)
\nonumber\\
&+K_4 \Theta_4(r)\left(\frac{72.4497 \mathrm{Kn}^5}{r^5}+\frac{49.0736 \mathrm{Kn}^4}{r^4}+\frac{14.9579 \mathrm{Kn}^3}{r^3}+\frac{2.62674 \mathrm{Kn}^2}{r^2}+\frac{0.254173 \mathrm{Kn}}{r}\right)
\nonumber\\
&+K_5 \Theta_5(r)\left(\frac{71.6731 \mathrm{Kn}^5}{r^5}+\frac{32.4383 \mathrm{Kn}^4}{r^4}+\frac{6.60653 \mathrm{Kn}^3}{r^3}+\frac{0.775193 \mathrm{Kn}^2}{r^2}+\frac{0.0501203 \mathrm{Kn}}{r}\right),
\\
\label{m2gas}
\mathbbm{m}_2(r) ={}&\frac{4 C_1 \mathrm{Kn}^2}{5 r^3}+\frac{\mathrm{Kn}^5}{r^5}\left(-\frac{3672 C_1}{49 \mathrm{Kn}}+\frac{8 C_2}{\mathrm{Kn}^3}+\frac{104 C_3}{35 \mathrm{Kn}^2}\right)
\nonumber\\
&+K_1 \Theta_1(r)\left(\frac{32.1789 \mathrm{Kn}^5}{r^5}+\frac{16.4204 \mathrm{Kn}^4}{r^4}+\frac{3.91023 \mathrm{Kn}^3}{r^3}+\frac{0.570095 \mathrm{Kn}^2}{r^2}+\frac{0.0484851 \mathrm{Kn}}{r}\right)
\nonumber\\
&+K_2\Theta_2(r)\left(\frac{25.4209 \mathrm{Kn}^5}{r^5}+\frac{32.18 \mathrm{Kn}^4}{r^4}+\frac{19.0102 \mathrm{Kn}^3}{r^3}+\frac{6.87563 \mathrm{Kn}^2}{r^2}+\frac{1.45063 \mathrm{Kn}}{r}\right)
\nonumber\\
&+K_3\Theta_3(r)\left(\frac{15.2579 \mathrm{Kn}^5}{r^5}+\frac{17.7481 \mathrm{Kn}^4}{r^4}+\frac{8.25788 \mathrm{Kn}^3}{r^3}+\frac{1.60094 \mathrm{Kn}^2}{r^2}\right)
\nonumber\\
&+K_4\Theta_4(r)\left(\frac{36.2249 \mathrm{Kn}^5}{r^5}+\frac{24.5368 \mathrm{Kn}^4}{r^4}+\frac{6.64796 \mathrm{Kn}^3}{r^3}+\frac{0.750496 \mathrm{Kn}^2}{r^2}\right)
\nonumber\\
&+K_5\Theta_5(r)\left(\frac{35.8366 \mathrm{Kn}^5}{r^5}+\frac{16.2192 \mathrm{Kn}^4}{r^4}+\frac{2.93623 \mathrm{Kn}^3}{r^3}+\frac{0.221484 \mathrm{Kn}^2}{r^2}\right).
\end{align} 
\begin{align}
\label{R1gas}
\mathbbm{R}_1(r) ={}&\frac{\mathrm{Kn}^4}{r^4}\left(\frac{12C_3}{5\mathrm{Kn}^2}-\frac{228 C_1}{7\mathrm{Kn}}\right)
\nonumber\\
&+K_1 \Theta_1(r) \left(\frac{46.338 \mathrm{Kn}^4}{r^4}+\frac{23.6455 \mathrm{Kn}^3}{r^3}+\frac{4.02199 \mathrm{Kn}^2}{r^2}\right)
\nonumber\\
&-K_2 \Theta_2(r) \left(\frac{4.00877 \mathrm{Kn}^4}{r^4}+\frac{5.07465 \mathrm{Kn}^3}{r^3}+\frac{2.14131 \mathrm{Kn}^2}{r^2}\right)
\nonumber\\
&-K_3 \Theta_3(r)\left(\frac{42.868 \mathrm{Kn}^4}{r^4}+\frac{49.8643 \mathrm{Kn}^3}{r^3}+\frac{25.7788 \mathrm{Kn}^2}{r^2}+\frac{7.49652 \mathrm{Kn}}{r}\right)
\nonumber\\
&+K_4\Theta_4(r) \left(\frac{7.38704 \mathrm{Kn}^4}{r^4}+\frac{5.00359 \mathrm{Kn}^3}{r^3}+\frac{1.5063 \mathrm{Kn}^2}{r^2}+\frac{0.255071 \mathrm{Kn}}{r}\right)
\nonumber\\
&+K_5\Theta_5(r) \left(\frac{37.9774 \mathrm{Kn}^4}{r^4}+\frac{17.1881 \mathrm{Kn}^3}{r^3}+\frac{3.45738 \mathrm{Kn}^2}{r^2}+\frac{0.391191 \mathrm{Kn}}{r}\right),
\\
\label{R2gas}
\mathbbm{R}_2(r) ={}&\frac{\mathrm{Kn}^4}{r^4}\left(\frac{6C_3}{5\mathrm{Kn}^2}-\frac{114 C_1}{7\mathrm{Kn}}\right)
\nonumber\\
&+K_1\Theta_1(r) \left(\frac{23.169 \mathrm{Kn}^4}{r^4}+\frac{11.8228 \mathrm{Kn}^3}{r^3}+\frac{3.01649 \mathrm{Kn}^2}{r^2}+\frac{0.51309 \mathrm{Kn}}{r}\right)
\nonumber\\
&-K_2\Theta_2(r)\left(\frac{2.00439 \mathrm{Kn}^4}{r^4}+\frac{2.53732 \mathrm{Kn}^3}{r^3}+\frac{1.60598 \mathrm{Kn}^2}{r^2}+\frac{0.677662 \mathrm{Kn}}{r}\right)
\nonumber\\
&-K_3 \Theta_3(r)\left(\frac{21.434 \mathrm{Kn}^4}{r^4}+\frac{24.9321 \mathrm{Kn}^3}{r^3}+\frac{9.66706 \mathrm{Kn}^2}{r^2}\right)
\nonumber\\
&+K_4\Theta_4(r)\left(\frac{3.69352 \mathrm{Kn}^4}{r^4}+\frac{2.50179 \mathrm{Kn}^3}{r^3}+\frac{0.564861 \mathrm{Kn}^2}{r^2}\right)
\nonumber\\
&+K_5\Theta_5(r) \left(\frac{18.9887 \mathrm{Kn}^4}{r^4}+\frac{8.59403 \mathrm{Kn}^3}{r^3}+\frac{1.29652 \mathrm{Kn}^2}{r^2}\right),
\end{align}
\begin{align}
\label{dgas}
\mathbbm{d}(r) ={}& K_3 \Theta_3(r)  \left(\frac{14.0055 \mathrm{Kn}^2}{r^2}+\frac{16.2913 \mathrm{Kn}}{r}\right)
-K_4 \Theta_4(r) \left(\frac{1.71119 \mathrm{Kn}^2}{r^2}+\frac{1.15907 \mathrm{Kn}}{r}\right)
\nonumber\\
&+K_5 \Theta_5(r)\left(\frac{3.75265 \mathrm{Kn}^2}{r^2}+\frac{1.6984 \mathrm{Kn}}{r}\right).
\end{align}
\end{fleqn}
\endgroup
In \eqref{v1gas}--\eqref{dgas}, the coefficients 
\begin{align}
\left.
\begin{gathered}
\Theta_1(r)  =  \mathrm{e}^{-\frac{0.510285 (r-1)}{\mathrm{Kn}}},
\quad
\Theta_2(r)=\mathrm{e}^{-\frac{1.26588 (r-1)}{\mathrm{Kn}}},
\quad
\Theta_3(r)  =  \mathrm{e}^{-\frac{1.16321 (r-1)}{\mathrm{Kn}}},
\\[2ex]
\Theta_4(r)  =  \mathrm{e}^{-\frac{0.677347 (r-1)}{\mathrm{Kn}}}
\quad\textrm{and}\quad
\Theta_5(r)  =  \mathrm{e}^{-\frac{0.452587 (r-1)}{\mathrm{Kn}}}
\end{gathered}
\right\}
\end{align}
describe the Knudsen layer functions that vanish as $r \to \infty$.



\bibliography{refer}

\begin{thebibliography}{56}
\expandafter\ifx\csname natexlab\endcsname\relax\def\natexlab#1{#1}\fi
\def\au#1{#1} \def\ed#1{#1} \def\yr#1{#1}\def\at#1{#1}\def\jt#1{\textit{#1}}
  \def\bt#1{#1}\def\bvol#1{\textbf{#1}} \def\vol#1{#1} \def\pg#1{#1}
  \def\publ#1{#1}\def\arxiv#1{#1}\def\org#1{#1}\def\st#1{\textit{#1}}

\bibitem[Abdel-Alim \& Hamielec(1975)]{AH1975}
{\sc \au{Abdel-Alim, A.~H.} \& \au{Hamielec, A.~E.}} \yr{1975}  \at{A
  theoretical and experimental investigation of the effect of internal
  circulation on the drag of spherical droplets falling at terminal velocity in
  liquid media}. \href{http://dx.doi.org/10.1021/i160056a004}{ \jt{Ind. Eng.
  Chem. Fundam.}} \href{http://dx.doi.org/10.1021/i160056a004}{ \bvol{14},
  \pg{308--312}}.

\bibitem[Allen \& Raabe(1982)]{AR1982}
{\sc \au{Allen, M.~D.} \& \au{Raabe, O.~G.}} \yr{1982}  \at{Re-evaluation of
  {M}illikan's oil drop data for the motion of small particles in air}.
  \href{http://dx.doi.org/10.1016/0021-8502(82)90019-2}{ \jt{J. Aerosol Sci.}}
  \href{http://dx.doi.org/10.1016/0021-8502(82)90019-2}{ \bvol{13},
  \pg{537--547}}.

\bibitem[Allen \& Raabe(1985)]{AR1985}
{\sc \au{Allen, M.~D.} \& \au{Raabe, O.~G.}} \yr{1985}  \at{Slip correction
  measurements of spherical solid aerosol particles in an improved {Millikan}
  apparatus}. \href{http://dx.doi.org/10.1080/02786828508959055}{ \jt{Aerosol
  Sci. Technol.}} \href{http://dx.doi.org/10.1080/02786828508959055}{ \bvol{4},
   \pg{269--286}}.

\bibitem[Batchelor(1967)]{Batchelor1967}
{\sc \au{Batchelor, G.~K.}} \yr{1967} {\em An Introduction to Fluid
  Dynamics\/}.  \publ{Cambridge: Cambridge University Press}.

\bibitem[Beckmann {\em et~al.\/}(2018)Beckmann, Rana, Torrilhon \&
  Struchtrup]{BRTS2018}
{\sc \au{Beckmann, A.~F.}, \au{Rana, A.~S.}, \au{Torrilhon, M.} \&
  \au{Struchtrup, H.}} \yr{2018}  \at{Evaporation boundary conditions for the
  linear {R13} equations based on the {O}nsager theory}.
  \href{http://dx.doi.org/10.3390/e20090680}{ \jt{Entropy}}
  \href{http://dx.doi.org/10.3390/e20090680}{ \bvol{20},  \pg{680}}.

\bibitem[Bhatnagar {\em et~al.\/}(1954)Bhatnagar, Gross \& Krook]{BGK1954}
{\sc \au{Bhatnagar, P.~L.}, \au{Gross, E.~P.} \& \au{Krook, M.}} \yr{1954}
  \at{A model for collision processes in gases.~{I}.~{S}mall amplitude
  processes in charged and neutral one-component systems}.
  \href{http://dx.doi.org/10.1103/PhysRev.94.511}{ \jt{Phys. Rev.}}
  \href{http://dx.doi.org/10.1103/PhysRev.94.511}{ \bvol{94},  \pg{511--525}}.

\bibitem[Bird(1994)]{Bird1994}
{\sc \au{Bird, G.~A.}} \yr{1994} {\em Molecular Gas Dynamics and the Direct
  Simulation of Gas Flows\/}.  \publ{Oxford: Clarendon Press}.

\bibitem[Chakraborty(2019)]{C2019}
{\sc \au{Chakraborty, I.}} \yr{2019}  \at{Numerical modeling of the dynamics of
  bubble oscillations subjected to fast variations in the ambient pressure with
  a coupled level set and volume of fluid method}.
  \href{http://dx.doi.org/10.1103/PhysRevE.99.043107}{ \jt{Phys. Rev. E}}
  \href{http://dx.doi.org/10.1103/PhysRevE.99.043107}{ \bvol{99},
  \pg{043107}}.

\bibitem[Chakraborty {\em et~al.\/}(2019)Chakraborty, Ricouvier, Yazhgur,
  Tabeling \& Leshansky]{CRYTL2019}
{\sc \au{Chakraborty, I.}, \au{Ricouvier, J.}, \au{Yazhgur, P.}, \au{Tabeling,
  P.} \& \au{Leshansky, A.~M.}} \yr{2019}  \at{Droplet generation at {Hele-Shaw
  microfluidic T-junction}}. \href{http://dx.doi.org/10.1063/1.5086808}{
  \jt{Phys. Fluids}} \href{http://dx.doi.org/10.1063/1.5086808}{ \bvol{31},
  \pg{022010}}.

\bibitem[Chapman \& Cowling(1970)]{CC1970}
{\sc \au{Chapman, S.} \& \au{Cowling, T.~G.}} \yr{1970} {\em The Mathematical
  Theory of Non-Uniform Gases\/}.  \publ{Cambridge: Cambridge University
  Press}.

\bibitem[Clift {\em et~al.\/}(1978)Clift, Grace \& Weber]{CGW1978}
{\sc \au{Clift, R.}, \au{Grace, J.~R.} \& \au{Weber, M.~E.}} \yr{1978} {\em
  Bubbles, Drops, and Particles\/}.  \publ{New York: Dover}.

\bibitem[{De Fraja} {\em et~al.\/}(2022){De Fraja}, Rana, Enright, Cooper,
  Lockerby \& Sprittles]{de2022efficient}
{\sc \au{{De Fraja}, T.~C.}, \au{Rana, A.~S.}, \au{Enright, R.}, \au{Cooper,
  L.~J.}, \au{Lockerby, D.~A.} \& \au{Sprittles, J.~E.}} \yr{2022}
  \at{Efficient moment method for modeling nanoporous evaporation}.
  \href{http://dx.doi.org/10.1103/PhysRevFluids.7.024201}{ \jt{Phys. Rev.
  Fluids}} \href{http://dx.doi.org/10.1103/PhysRevFluids.7.024201}{ \bvol{7},
  \pg{024201}}.

\bibitem[Garner \& Lane(1959)]{GL1959}
{\sc \au{Garner, F.~H.} \& \au{Lane, J.~J.}} \yr{1959}  \at{Mass transfer to
  drops of liquid suspended in a gas stream. {Part II: E}xperimental work and
  results}.  \jt{Trans. Inst. Chem. Eng.}  \bvol{37},  \pg{162--172}.

\bibitem[Grad(1949)]{Grad1949b}
{\sc \au{Grad, H.}} \yr{1949}  \at{On the kinetic theory of rarefied gases}.
  \href{http://dx.doi.org/10.1002/cpa.3160020403}{ \jt{Comm. Pure Appl. Math.}}
  \href{http://dx.doi.org/10.1002/cpa.3160020403}{ \bvol{2},  \pg{331--407}}.

\bibitem[Gu \& Emerson(2007)]{Gu&Emerson2007}
{\sc \au{Gu, X.~J.} \& \au{Emerson, D.~R.}} \yr{2007}  \at{A computational
  strategy for the regularized 13 moment equations with enhanced wall-boundary
  conditions}. \href{http://dx.doi.org/10.1016/j.jcp.2006.11.032}{ \jt{J.
  Comput. Phys.}} \href{http://dx.doi.org/10.1016/j.jcp.2006.11.032}{
  \bvol{225},  \pg{263--283}}.

\bibitem[Gu \& Emerson(2009)]{GuEmerson2009}
{\sc \au{Gu, X.~J.} \& \au{Emerson, D.~R.}} \yr{2009}  \at{A high-order moment
  approach for capturing non equilibrium phenomena in the transition regime}.
  \href{http://dx.doi.org/10.1017/S002211200900768X}{ \jt{J. Fluid Mech.}}
  \href{http://dx.doi.org/10.1017/S002211200900768X}{ \bvol{636},
  \pg{177--216}}.

\bibitem[Happel \& Brenner(1965)]{Happel1965}
{\sc \au{Happel, J.} \& \au{Brenner, H.}} \yr{1965} {\em Low Reynolds number
  hydrodynamics prentice-hall\/}, ,  \vol{vol. 331}.

\bibitem[Holway(1966)]{Holway1966}
{\sc \au{Holway, L.~H.}} \yr{1966}  \at{New statistical models for kinetic
  theory: Methods of construction}. \href{http://dx.doi.org/10.1063/1.1761920}{
  \jt{Phys. Fluids}} \href{http://dx.doi.org/10.1063/1.1761920}{ \bvol{9},
  \pg{1658--1673}}.

\bibitem[Hutchins {\em et~al.\/}(1995)Hutchins, Harper \& Felder]{HHF1995}
{\sc \au{Hutchins, D.~K.}, \au{Harper, M.~H.} \& \au{Felder, R.~L.}} \yr{1995}
  \at{Slip correction measurements for solid spherical particles by modulated
  dynamic light scattering}.
  \href{http://dx.doi.org/10.1080/02786829408959741}{ \jt{Aerosol Sci.
  Technol.}} \href{http://dx.doi.org/10.1080/02786829408959741}{ \bvol{22},
  \pg{202--218}}.

\bibitem[Kennard(1938)]{Kennard1938}
{\sc \au{Kennard, E.~H.}} \yr{1938} {\em Kinetic Theory of Gases\/}.  \publ{New
  York: McGraw-Hill Book Company Inc.}

\bibitem[Knudsen \& Weber(1911)]{KW1911}
{\sc \au{Knudsen, M.} \& \au{Weber, S.}} \yr{1911}  \at{{Luftwiderstand gegen
  die langsame Bewegung kleiner Kugeln}}.
  \href{http://dx.doi.org/10.1002/andp.19113411506}{ \jt{Ann. Phys.}}
  \href{http://dx.doi.org/10.1002/andp.19113411506}{ \bvol{341},
  \pg{981--994}}.

\bibitem[Landau \& Lifshitz(1959)]{LL1959}
{\sc \au{Landau, L.~D.} \& \au{Lifshitz, E.~M.}} \yr{1959} {\em Fluid
  Mechanics\/}.  \publ{Oxford: Pergamon Press}.

\bibitem[LeClair {\em et~al.\/}(1972)LeClair, Hamielec, Pruppacher \&
  Hall]{leclair1972theoretical}
{\sc \au{LeClair, B.~P.}, \au{Hamielec, A.~E.}, \au{Pruppacher, H.~R.} \&
  \au{Hall, W.~D.}} \yr{1972}  \at{A theoretical and experimental study of the
  internal circulation in water drops falling at terminal velocity in air}.
  \href{http://dx.doi.org/10.1175/1520-0469(1972)029<0728:ATAESO>2.0.CO;2}{
  \jt{J. Atmos. Sci.}}
  \href{http://dx.doi.org/10.1175/1520-0469(1972)029<0728:ATAESO>2.0.CO;2}{
  \bvol{29},  \pg{728--740}}.

\bibitem[Lenard(1904)]{Lenard1904}
{\sc \au{Lenard, P.}} \yr{1904}  \at{{\"U}ber regen}.  \jt{Meteor. Z.}
  \bvol{21},  \pg{249--260}.

\bibitem[Malekzadeh \& Roohi(2015)]{MR2015}
{\sc \au{Malekzadeh, S.} \& \au{Roohi, E.}} \yr{2015}  \at{Investigation of
  different droplet formation regimes in a {T}-junction microchannel using the
  {VOF} technique in {OpenFOAM}}.
  \href{http://dx.doi.org/10.1007/s12217-015-9440-2}{ \jt{Microgravity Sci.
  Technol.}} \href{http://dx.doi.org/10.1007/s12217-015-9440-2}{ \bvol{27},
  \pg{231--243}}.

\bibitem[McDonald(1954)]{mcdonald1954shape}
{\sc \au{McDonald, J.~E.}} \yr{1954}  \at{The shape and aerodynamics of large
  raindrops}.
  \href{http://dx.doi.org/10.1175/1520-0469(1954)011<0478:TSAAOL>2.0.CO;2}{
  \jt{J. Atmos. Sci.}}
  \href{http://dx.doi.org/10.1175/1520-0469(1954)011<0478:TSAAOL>2.0.CO;2}{
  \bvol{11},  \pg{478--494}}.

\bibitem[Millikan(1923)]{Millikan1923}
{\sc \au{Millikan, R.~A.}} \yr{1923}  \at{The general law of fall of a small
  spherical body through a gas, and its bearing upon the nature of molecular
  reflection from surfaces}. \href{http://dx.doi.org/10.1103/PhysRev.22.1}{
  \jt{Phys. Rev.}} \href{http://dx.doi.org/10.1103/PhysRev.22.1}{ \bvol{22},
  \pg{1--23}}.

\bibitem[{National Institute of Standards and Technology}(2023)]{NISTwebsite}
{\sc \au{{National Institute of Standards and Technology}}} \yr{2023} {NIST
  Chemistry WebBook}. \url{https://webbook.nist.gov/chemistry/}.

\bibitem[Oliver \& Chung(1985)]{oliver1985steady}
{\sc \au{Oliver, D. L.~R.} \& \au{Chung, J.~N.}} \yr{1985}  \at{Steady flows
  inside and around a fluid sphere at low {R}eynolds numbers}.
  \href{http://dx.doi.org/10.1017/S0022112085001495}{ \jt{J. Fluid Mech.}}
  \href{http://dx.doi.org/10.1017/S0022112085001495}{ \bvol{154},
  \pg{215--230}}.

\bibitem[Oliver \& Chung(1987)]{oliver1987flow}
{\sc \au{Oliver, D. L.~R.} \& \au{Chung, J.~N.}} \yr{1987}  \at{Flow about a
  fluid sphere at low to moderate {R}eynolds numbers}.
  \href{http://dx.doi.org/10.1017/S002211208700082X}{ \jt{J. Fluid Mech.}}
  \href{http://dx.doi.org/10.1017/S002211208700082X}{ \bvol{177},  \pg{1--18}}.

\bibitem[Onsager(1931{\natexlab{{\em a\/}}})]{Onsager37}
{\sc \au{Onsager, L.}} \yr{1931{\natexlab{{\em a\/}}}}  \at{Reciprocal
  relations in irreversible processes. {I}.}
  \href{http://dx.doi.org/10.1103/PhysRev.37.405}{ \jt{Phys. Rev.}}
  \href{http://dx.doi.org/10.1103/PhysRev.37.405}{ \bvol{37},  \pg{405--426}}.

\bibitem[Onsager(1931{\natexlab{{\em b\/}}})]{Onsager38}
{\sc \au{Onsager, L.}} \yr{1931{\natexlab{{\em b\/}}}}  \at{Reciprocal
  relations in irreversible processes. {II}.}
  \href{http://dx.doi.org/10.1103/PhysRev.38.2265}{ \jt{Phys. Rev.}}
  \href{http://dx.doi.org/10.1103/PhysRev.38.2265}{ \bvol{38},
  \pg{2265--2279}}.

\bibitem[Pozrikidis(1989)]{Pozrikidis1989}
{\sc \au{Pozrikidis, C.}} \yr{1989}  \at{Inviscid drops with internal
  circulation}. \href{http://dx.doi.org/10.1017/S0022112089003046}{ \jt{J.
  Fluid Mech.}} \href{http://dx.doi.org/10.1017/S0022112089003046}{ \bvol{209},
   \pg{77--92}}.

\bibitem[Pruppacher \& Beard(1970)]{PB1970}
{\sc \au{Pruppacher, H.~R.} \& \au{Beard, K.~V.}} \yr{1970}  \at{A wind tunnel
  investigation of the internal circulation and shape of water drops falling at
  terminal velocity in air}. \href{http://dx.doi.org/10.1002/qj.49709640807}{
  \jt{Q. J. R. Meteorol. Soc.}}
  \href{http://dx.doi.org/10.1002/qj.49709640807}{ \bvol{96},  \pg{247--256}}.

\bibitem[Rana {\em et~al.\/}(2021{\natexlab{{\em a\/}}})Rana, Gupta, Sprittles
  \& Torrilhon]{RGST2021}
{\sc \au{Rana, A.~S.}, \au{Gupta, V.~K.}, \au{Sprittles, J.~E.} \&
  \au{Torrilhon, M.}} \yr{2021{\natexlab{{\em a\/}}}}  \at{{$H$}-theorem and
  boundary conditions for the linear {R26} equations: application to flow past
  an evaporating droplet}. \href{http://dx.doi.org/10.1017/jfm.2021.622}{
  \jt{J. Fluid Mech.}} \href{http://dx.doi.org/10.1017/jfm.2021.622}{
  \bvol{924},  \pg{A16}}.

\bibitem[Rana {\em et~al.\/}(2018{\natexlab{{\em a\/}}})Rana, Gupta \&
  Struchtrup]{RGS2018}
{\sc \au{Rana, A.~S.}, \au{Gupta, V.~K.} \& \au{Struchtrup, H.}}
  \yr{2018{\natexlab{{\em a\/}}}}  \at{Coupled constitutive relations: a second
  law based higher-order closure for hydrodynamics}.
  \href{http://dx.doi.org/10.1098/rspa.2018.0323}{ \jt{Proc. Roy. Soc. A}}
  \href{http://dx.doi.org/10.1098/rspa.2018.0323}{ \bvol{474},  \pg{20180323}}.

\bibitem[Rana {\em et~al.\/}(2018{\natexlab{{\em b\/}}})Rana, Lockerby \&
  Sprittles]{RLS2018}
{\sc \au{Rana, A.~S.}, \au{Lockerby, D.~A.} \& \au{Sprittles, J.~E.}}
  \yr{2018{\natexlab{{\em b\/}}}}  \at{Evaporation-driven vapour microflows:
  analytical solutions from moment methods}.
  \href{http://dx.doi.org/10.1017/jfm.2018.85}{ \jt{J. Fluid Mech.}}
  \href{http://dx.doi.org/10.1017/jfm.2018.85}{ \bvol{841},  \pg{962--988}}.

\bibitem[Rana {\em et~al.\/}(2019)Rana, Lockerby \&
  Sprittles]{rana2019lifetime}
{\sc \au{Rana, A.~S.}, \au{Lockerby, D.~A.} \& \au{Sprittles, J.~E.}} \yr{2019}
   \at{Lifetime of a nanodroplet: Kinetic effects and regime transitions}.
  \href{http://dx.doi.org/10.1103/PhysRevLett.123.154501}{ \jt{Phys. Rev.
  Lett.}} \href{http://dx.doi.org/10.1103/PhysRevLett.123.154501}{ \bvol{123},
  \pg{154501}}.

\bibitem[Rana {\em et~al.\/}(2015)Rana, Mohammadzadeh \& Struchtrup]{RMS2015}
{\sc \au{Rana, A.~S.}, \au{Mohammadzadeh, A.} \& \au{Struchtrup, H.}} \yr{2015}
   \at{A numerical study of the heat transfer through a rarefied gas confined
  in a microcavity}. \href{http://dx.doi.org/10.1007/s00161-014-0371-8}{
  \jt{Continuum Mech. Thermodyn.}}
  \href{http://dx.doi.org/10.1007/s00161-014-0371-8}{ \bvol{27},
  \pg{433--446}}.

\bibitem[Rana {\em et~al.\/}(2021{\natexlab{{\em b\/}}})Rana, Saini,
  Chakraborty, Lockerby \& Sprittles]{RSCLS2021}
{\sc \au{Rana, A.~S.}, \au{Saini, S.}, \au{Chakraborty, S.}, \au{Lockerby,
  D.~A.} \& \au{Sprittles, J.~E.}} \yr{2021{\natexlab{{\em b\/}}}}
  \at{Efficient simulation of non-classical liquid--vapour phase-transition
  flows: a method of fundamental solutions}.
  \href{http://dx.doi.org/10.1017/jfm.2021.405}{ \jt{J. Fluid Mech.}}
  \href{http://dx.doi.org/10.1017/jfm.2021.405}{ \bvol{919},  \pg{A35}}.

\bibitem[Rana \& Struchtrup(2016)]{RanaStruchtrup2016}
{\sc \au{Rana, A.~S.} \& \au{Struchtrup, H.}} \yr{2016}  \at{Thermodynamically
  admissible boundary conditions for the regularized 13 moment equations}.
  \href{http://dx.doi.org/10.1063/1.4941293}{ \jt{Phys. Fluids}}
  \href{http://dx.doi.org/10.1063/1.4941293}{ \bvol{28},  \pg{027105}}.

\bibitem[Rivkind \& Ryskin(1976)]{RR1976}
{\sc \au{Rivkind, V.~{\relax Ya}.} \& \au{Ryskin, G.~M.}} \yr{1976}  \at{Flow
  structure in motion of a spherical drop in a fluid medium at intermediate
  {R}eynolds numbers}. \href{http://dx.doi.org/10.1007/BF01023387}{ \jt{Fluid
  Dyn.}} \href{http://dx.doi.org/10.1007/BF01023387}{ \bvol{11},  \pg{5--12}}.

\bibitem[Rivkind {\em et~al.\/}(1976)Rivkind, Ryskin \&
  Fishbein]{rivkind1976flow}
{\sc \au{Rivkind, V.~{\relax Ya}.}, \au{Ryskin, G.~M.} \& \au{Fishbein, G.~A.}}
  \yr{1976}  \at{Flow around a spherical drop at intermediate {R}eynolds
  numbers}. \href{http://dx.doi.org/10.1016/0021-8928(76)90181-7}{ \jt{J. Appl.
  Math. Mech.}} \href{http://dx.doi.org/10.1016/0021-8928(76)90181-7}{
  \bvol{40},  \pg{687--691}}.

\bibitem[Sadr \& Hadjiconstantinou(2023)]{SH2023}
{\sc \au{Sadr, M.} \& \au{Hadjiconstantinou, N.~G.}} \yr{2023}  \at{A
  variance-reduced direct {Monte Carlo} simulation method for solving the
  boltzmann equation over a wide range of rarefaction}.
  \href{http://dx.doi.org/10.1016/j.jcp.2022.111677}{ \jt{J. Comput. Phys.}}
  \href{http://dx.doi.org/10.1016/j.jcp.2022.111677}{ \bvol{472},
  \pg{111677}}.

\bibitem[Shakhov(1968)]{Shakhov1968}
{\sc \au{Shakhov, E.~M.}} \yr{1968}  \at{Generalization of the {K}rook kinetic
  relaxation equation}. \href{http://dx.doi.org/10.1007/BF01029546}{ \jt{Fluid
  Dyn.}} \href{http://dx.doi.org/10.1007/BF01029546}{ \bvol{3},  \pg{95--96}}.

\bibitem[Sone(2002)]{Sone2002}
{\sc \au{Sone, Y.}} \yr{2002} {\em Kinetic Theory and Fluid Dynamics\/}.
  \publ{Boston: Birkh{\"a}user}.

\bibitem[Stefanov {\em et~al.\/}(2022)Stefanov, Roohi \& Shoja-Sani]{SRS2022}
{\sc \au{Stefanov, S.}, \au{Roohi, E.} \& \au{Shoja-Sani, A.}} \yr{2022}  \at{A
  novel transient-adaptive subcell algorithm with a hybrid application of
  different collision techniques in direct simulation {Monte Carlo (DSMC)}}.
  \href{http://dx.doi.org/10.1063/5.0104613}{ \jt{Phys. Fluids}}
  \href{http://dx.doi.org/10.1063/5.0104613}{ \bvol{34},  \pg{092003}}.

\bibitem[Struchtrup(2004)]{Struchtrup2004}
{\sc \au{Struchtrup, H.}} \yr{2004}  \at{Stable transport equations for
  rarefied gases at high orders in the {K}nudsen number}.
  \href{http://dx.doi.org/10.1063/1.1782751}{ \jt{Phys. Fluids}}
  \href{http://dx.doi.org/10.1063/1.1782751}{ \bvol{16},  \pg{3921--3934}}.

\bibitem[Struchtrup(2005)]{Struchtrup2005}
{\sc \au{Struchtrup, H.}} \yr{2005} {\em Macroscopic Transport Equations for
  Rarefied Gas Flows\/}.  \publ{Berlin: Springer}.

\bibitem[Struchtrup \& Torrilhon(2003)]{StruchtrupTorrilhon2003}
{\sc \au{Struchtrup, H.} \& \au{Torrilhon, M.}} \yr{2003}  \at{Regularization
  of {G}rad's 13 moment equations: {D}erivation and linear analysis}.
  \href{http://dx.doi.org/10.1063/1.1597472}{ \jt{Phys. Fluids}}
  \href{http://dx.doi.org/10.1063/1.1597472}{ \bvol{15},  \pg{2668--2680}}.

\bibitem[Taheri {\em et~al.\/}(2022)Taheri, Roohi \& Stefanov]{TRS2022}
{\sc \au{Taheri, E.}, \au{Roohi, E.} \& \au{Stefanov, S.}} \yr{2022}  \at{A
  symmetrized and simplified {Bernoulli} trial collision scheme in direct
  simulation {Monte Carlo}}. \href{http://dx.doi.org/10.1063/5.0076025}{
  \jt{Phys. Fluids}} \href{http://dx.doi.org/10.1063/5.0076025}{ \bvol{34},
  \pg{012010}}.

\bibitem[Tiwari {\em et~al.\/}(2021)Tiwari, Klar \& Russo]{TKR2021}
{\sc \au{Tiwari, S.}, \au{Klar, A.} \& \au{Russo, G.}} \yr{2021}  \at{Modelling
  and simulations of moving droplet in a rarefied gas}.
  \href{http://dx.doi.org/10.1080/10618562.2021.2024520}{ \jt{Int. J. Comput.
  Fluid Dyn.}} \href{http://dx.doi.org/10.1080/10618562.2021.2024520}{
  \bvol{35},  \pg{666--684}}.

\bibitem[Torrilhon(2010)]{Torrilhon2010}
{\sc \au{Torrilhon, M.}} \yr{2010}  \at{Slow gas microflow past a sphere:
  Analytical solution based on moment equations}.
  \href{http://dx.doi.org/10.1063/1.3453707}{ \jt{Phys. Fluids}}
  \href{http://dx.doi.org/10.1063/1.3453707}{ \bvol{22},  \pg{072001}}.

\bibitem[Torrilhon(2016)]{TorrilhonARFM}
{\sc \au{Torrilhon, M.}} \yr{2016}  \at{Modeling nonequilibrium gas flow based
  on moment equations}.
  \href{http://dx.doi.org/10.1146/annurev-fluid-122414-034259}{ \jt{Annu. Rev.
  Fluid Mech.}} \href{http://dx.doi.org/10.1146/annurev-fluid-122414-034259}{
  \bvol{48},  \pg{429--458}}.

\bibitem[Torrilhon \& Struchtrup(2008)]{ST2008}
{\sc \au{Torrilhon, M.} \& \au{Struchtrup, H.}} \yr{2008}  \at{Boundary
  conditions for regularized 13-moment equations for micro-channel-flows}.
  \href{http://dx.doi.org/10.1016/j.jcp.2007.10.006}{ \jt{J. Comput. Phys.}}
  \href{http://dx.doi.org/10.1016/j.jcp.2007.10.006}{ \bvol{227},
  \pg{1982--2011}}.

\bibitem[Yang {\em et~al.\/}(2020)Yang, Gu, Wu, Emerson, Zhang \&
  Tang]{YGWEZT2020}
{\sc \au{Yang, W.}, \au{Gu, X.-J.}, \au{Wu, L.}, \au{Emerson, D.~R.},
  \au{Zhang, Y.} \& \au{Tang, S.}} \yr{2020}  \at{A hybrid approach to couple
  the discrete velocity method and {Method of Moments} for rarefied gas flows}.
  \href{http://dx.doi.org/10.1016/j.jcp.2020.109397}{ \jt{J. Comput. Phys.}}
  \href{http://dx.doi.org/10.1016/j.jcp.2020.109397}{ \bvol{410},
  \pg{109397}}.

\end{thebibliography}
\bibliographystyle{jfm_doi}

\end{document}